\documentclass[11pt]{iopart}
\usepackage[margin=3cm]{geometry}

\newcommand{\volume}{{\ooalign{\hfil$V$\hfil\cr\kern0.08em--\hfil\cr}}}
\usepackage{threeparttable}    
\usepackage{dcolumn}           
\newcolumntype{d}{D{.}{.}{-1}}
\usepackage{nomencl}           
\makenomenclature
\usepackage{graphicx}          
\usepackage{caption}          
\usepackage{fancyvrb}          
\fvset{fontsize=\footnotesize,xleftmargin=2em}  
\usepackage{lettrine}          
\expandafter\let\csname equation*\endcsname\relax
\expandafter\let\csname endequation*\endcsname\relax
\usepackage{amsmath}           
\usepackage{hyperref}          
\graphicspath{{figs/}}         
\usepackage{float}             
\usepackage{upquote}
\usepackage{placeins}
\usepackage{enumitem}
\usepackage[retainorgcmds]{IEEEtrantools}
\usepackage{subfig}
\usepackage{siunitx}
\usepackage[export]{adjustbox}
\usepackage[title]{appendix}
\newcommand{\be}{\begin{equation}}
\newcommand{\ee}{\end{equation}} 

\usepackage{cite}

\usepackage[nameinlink,capitalise]{cleveref}
\usepackage{tikz}
\usetikzlibrary{positioning,shadows,arrows,shapes.multipart,decorations.pathmorphing}

\newcommand{\colormarkerempty}[1]{\raisebox{0.5pt}{\protect\tikz{\protect\node[draw,scale=0.3,circle,fill=white](){};}}}

\begin{document}

\title[]{Numerical analysis of three-dimensional magnetohydrodynamic effects in an inductively coupled plasma wind tunnel}

\author{Sanjeev Kumar, Alessandro Munaf\`{o}, Daniel J. Bodony and Marco Panesi\footnote{\label{footnote_2}Corresponding author (mpanesi@illinois.edu).}}

\address{Center for Hypersonics and Entry Systems Studies (CHESS), \\
Department of Aerospace Engineering, \\
University of Illinois Urbana-Champaign, Urbana, IL 61801, USA}
\ead{mpanesi@illinois.edu}
\vspace{10pt}


\begin{abstract}
This paper introduces a three-dimensional model for the \SI{350}{\kilo\watt} Plasmatron X inductively coupled plasma facility at the University of Illinois Urbana-Champaign, designed for testing high-temperature materials. Simulations of the facility have been performed using a three-dimensional, multiphysics computational framework, which reveals pronounced three-dimensional characteristics within the facility. The analysis of the plasma and electromagnetic field in the torch region reveals the influence of the helical coils, which cause a non-axisymmetric distribution of the plasma discharge. Additionally, simulations of the torch-chamber configuration at two operating pressures have been conducted to examine the impact of plasma asymmetry in the torch on jet characteristics in the chamber. The results indicate an unsteady, three-dimensional behavior of the plasma jet at high pressure. Spectral Proper Orthogonal Decomposition (SPOD) has been performed on the unsteady flow field to identify the dominant modes and their associated frequencies. At low pressure, a steady, supersonic, nearly axisymmetric plasma jet forms with consistent flow properties, such as temperature and velocity. However, strong non-equilibrium effects at low pressures lead to substantial deviations in species concentrations from axial symmetry despite having an almost axisymmetric distribution for quantities such as velocity and temperatures.  

\end{abstract}
\section{Introduction}\label{sec:intro}
When an atmospheric entry vehicle travels across the upper layers of a planetary atmosphere, it reaches extremely high speeds. This kinetic energy is dissipated in the strong bow shock ahead of the vehicle and further in the boundary layer, resulting in intense heat loads. To protect the spacecraft in this extreme environment, its surface must be shielded by a heat-resistant barrier made of advanced thermal protection materials (TPMs). Since in-flight testing of TPMs is expensive, the thermal loads experienced during hypersonic entry may be instead simulated by exposing a TPM sample to a plasma jet replicating the conditions behind the shock. A key type of plasma wind tunnel, the ICP (inductively coupled plasma) facility, produces a contamination-free plasma flow over an extended period without the need for electrodes. Consequently, ICP facilities are widely used to test the thermal protection systems of entry vehicles. The ease and long testing times along with their versatility in terms of sizes have made ICPs widely used not only for testing of TPMs \cite{chazot2008tps,wu2022characterization,chinnaraj2020ablation,matveev2022inductively} but also for other industrial applications such as plasma spray processes\cite{fauchais2004understanding,meillot2015numerical}, synthesis of nano-particles\cite{shigeta2011thermal}, electric propulsion for low earth orbit satellites\cite{zheng2023atmosphere}, studying the behavior of metal-oxides injected in thermal plasmas \cite{borgianni1969behaviour,capitelli1970decomposition}, \emph{etc.}

Numerical simulations of ICPs have proven highly valuable for elucidating the magnetohydrodynamics within these facilities, thereby enabling improved experimental designs and advancements in the development and design of future facilities. A significant amount of work has been done in modeling ICPs with one- and two-dimensional descriptions \cite{Boulos_1976,mostaghimi1984parametric,mostaghimi1990effect,chen1991modeling,panesi2007analysis,abeele2000efficient,utyuzhnikov2004simulation,munafo2015tightly} which describe the main features of the plasma in ICPs. However, these models are unable to account for the three-dimensional effects of actual facilities. The advancements in computational methods and high-performance computing have heightened interest in the three-dimensional modeling of ICPs. Recent numerical analyses of ICP facilities have demonstrated the importance of taking into account three-dimensional effects resulting from complex geometrical configurations, such as the realistic coil shapes \cite{colombo2008three,bernardi2003three,bernardi2003threetwo,bernardi20043,colombo2010three,bernardi20053}, the detailed inlet gas region often incorporating swirl dynamics\cite{colombo20073,tsivilskiy2020experimentally}, and the transverse injection of cold gas jets \cite{njah1993mathematical}, among others. Moreover, under certain operating conditions, ICPs have been found to show unstable perturbations resulting from electromagnetic interactions as well as the hydrodynamic instabilities arising due to shearing between the hot plasma and the cold ambient gas \cite{cipullo2014investigation,shigeta2012time}. These instabilities often exhibit a three-dimensional nature, further contributing to the non-axisymmetric distribution of quantities of interest (\emph{e.g.}, temperature). Several works in the literature present the linear stability analysis of two-dimensional axisymmetric plasma jets in ICPs \cite{anfuso2021multiscale,demange2020local,balestra2015absolute,demange2020high,shigeta2016turbulence}, underlining the unstable nature of the plasma jets produced in these facilities. To accurately model plasma behavior and explore its three-dimensional characteristics in ICP facilities, time-dependent three-dimensional simulations are essential. These simulations capture the complex interaction between the dynamics of the plasma discharge and the three-dimensional electromagnetic fields generated by the coils. Such a detailed understanding of a specific ICP setup not only provides support in optimizing experimental operating conditions but also reveals critical geometric parameters to mitigate asymmetries in future ICP systems. 

The majority of the aforementioned three-dimensional studies are based on the assumption of local thermodynamic equilibrium (LTE), primarily because they focus on ICP facilities operating at atmospheric pressure. Moreover, modeling plasmas in non-local thermodynamic equilibrium (NLTE) is associated with significant computational cost and increased complexity. To the best of the authors' knowledge, there is a lack of three-dimensional studies on ICP facilities incorporating the NLTE effect. A distinctive characteristic of ICPs used for testing TPMs is their operation at extremely low sub-atmospheric pressures to simulate planetary re-entry conditions, which can induce significant non-local thermodynamic equilibrium (NLTE) effects in the plasma. Consequently, accurately capturing NLTE effects becomes a critical aspect of modeling such environments, in addition to incorporating three-dimensional effects. 

The most accurate methodology to account for non-equilibrium is the state-to-state (StS) approach, which treats each internal state (\emph{e.g.}, electronic) as a separate pseudo-species, thereby allowing for non-Boltzmann distributions   \cite{laux2012state,munafo2015tightly,panesi2013rovibrational,kumar2024investigation,colonna2015non,laporta2016electron,capitelli2013plasma}. However, this approach, though accurate, is computationally intensive and requires extensive atomic and molecular data (\emph{e.g.}, cross-sections) which are not always available. To circumvent these issues, simplified models that assume Maxwell-Boltzmann distributions for internal states have been developed \cite{park1989nonequilibrium,park1993review,gnoffo1989conservation}. Among these, the two-temperature formulation is most commonly employed for simulating ICP discharges \cite{park1989nonequilibrium,park1993review,park2001chemical}. This model presumes rapid equilibration between translational and rotational energy modes of heavy particles (\emph{i.e.}, atoms and molecules) while assuming that electronic and vibrational energy modes are in equilibrium with the translation of free electrons. Recent studies \cite{atsuchi2006modeling,panesi2007analysis,yu2019effects,munafo2022multi,munafo2024self-consistent,kumar2024electronic,sun2023validation} have reported simulations of ICP discharges using two-temperature NLTE models, highlighting the importance of non-equilibrium distributions of quantities of interest such as temperatures and chemical composition. The same modeling approach is employed in the present work to explore three-dimensional NLTE effects within an ICP wind tunnel.
\\

The preceding discussion highlights two significant gaps in the current literature on the numerical modeling of high-enthalpy ICP wind tunnels:
\begin{itemize}
    \item A scarcity of studies investigating the unsteady behavior of high-enthalpy ICP wind tunnels under fully three-dimensional configurations.
    \item A limited understanding of the potential effects of thermochemical non-equilibrium on the three-dimensional distribution of key flow properties, such as species concentrations and temperature fields, within ICP wind tunnels.
\end{itemize}

Hence, this work aims to examine the three-dimensional behavior of plasma discharges in the \SI{350}{\kilo\watt} Plasmatron X facility at the University of Illinois Urbana-Champaign (UIUC), employing an in-house multi-physics computational framework developed by the authors with the purpose of filling the gap with the works published in the literature so far. 

The paper is organized as follows: \cref{sec:physical_modeling} discusses the physical model for the plasma and the electromagnetic field. \cref{sec:numerical_framework} describes the numerical framework used for the simulations. \cref{sec:problem_desc} describes the Plasmatron X facility and simulation setup. \cref{sec:results} begins with the presentation of results for the torch-only configuration of the Plasmatron X facility, followed by an analysis of the torch-chamber layout to examine the behavior of the plasma jet within the chamber region. Conclusions are summarized in \cref{sec:conclusions}.

\section{Physical Modeling}\label{sec:physical_modeling}
The plasma and the electromagnetic field in this work are modeled under the following assumptions:
\begin{enumerate}
    \item The plasma is a collection of neutral and charged components, each behaving macroscopically as an ideal gas \cite{boulos1994thermal}. 
    \item The plasma is quasi-neutral, un-magnetized, and collision-dominated, such that the use of a hydrodynamic description is appropriate. \label{assum:unmagnetized}
    \item The plasma is laminar\cite{Mitchner_book,mostaghimi1987two,mostaghimi1990effect}. 
    \item Low magnetic Reynolds number \cite{abeele2000efficient}. \label{assum:mag_reynolds}
    \item Low-frequency approximation (\emph{i.e.,} the inductor frequency, $f$, is much smaller than the plasma frequency), which allows one to rule out both electrostatic and electromagnetic waves \cite{abeele2000efficient}. \label{assum:waves}
    \item Negligible effect of radiation transport and radiative processes (\emph{e.g.}, line emission) within the plasma \cite{abeele2000efficient,asinovsky1971experimental,devoto1978air}.
\end{enumerate}

\subsection{\label{sec:plasma Field}Plasma} 
The plasma is described based on either a two-temperature (2T) NLTE formulation or an LTE model. Here, only a summary of the main formulas and concepts is provided. For the details, the reader is referred to previous works from the authors and their co-workers \cite{munafo2023plato,munafo2024self-consistent,kumar2024electronic,kumar2024investigation}. 

The chemical components (\emph{i.e.}, species) are stored in the $\mathcal{S}$ set. The heavy-particles (\emph{i.e.}, atoms and molecules) form the heavy subset $\mathcal{H}$, such that $\mathcal{S} = \left\{ \mathrm{e} \right\} \cup \mathrm{H}$, where e denotes free-electrons. 

For the 2T model,  the internal energy levels of a given component are assumed to follow a Maxwell-Boltzmann distribution at a specific temperature. Here, thermal equilibrium is assumed between the rotational and translational degrees of freedom ($T_{\mathrm{tr}} = T_{\mathrm{r}}$) of heavy particles, and between free-electrons, electronic, and vibrational energy modes ($T_{\mathrm{e}} = T_{\mathrm{el}} = T_{\mathrm{v}}$). In light of this, the NLTE state of the plasma may be described in terms of the number densities, $n_s$, and the two temperatures: $T_{\mathrm{h}}$ and $T_{\mathrm{ve}}$, where the heavy-particle temperature, $T_{\mathrm{h}}$, refers to the roto-translational degrees of freedom of heavy-particles, whereas the \emph{vibronic} temperature, $T_{\mathrm{ve}}$, characterizes the energy of free-electrons and that of vibrational and electronic states of heavy-particles. To further simplify the formulation, rotational and vibrational energy modes are treated in a decoupled fashion by assuming that molecules execute rigid rotations and harmonic oscillations \cite{bottin1999thermodynamic}.

Under the above assumptions, the NLTE plasma dynamics are governed by the set of mass continuity, global momentum and energy, and vibronic energy equations \cite{Munafo_JCP_2020,Mitchner_book,gnoffo1989conservation,capitelli2013fundamantal,magin2004transport}:
\begin{IEEEeqnarray}{rCl}
\frac{\partial \rho_{s}}{\partial t}+ {\nabla}  \cdot\left[\rho_{s}\left(\mathbf{v}+ \mathbf{U}_s \right)\right] &=& \dot{\omega}_{s}, \quad s \in \mathcal{S}, \label{eq:cont} \\
\frac{\partial \rho \mathbf{v}}{\partial t}+ {\nabla} \cdot(\rho \mathbf{v} \mathbf{v} + p \mathsf{I}) &=&  {\nabla} \cdot \mathsf{\tau} + \mathbf{J} \times \mathbf{B}, \label{eq:momentum}\\
\frac{\partial \rho \mathcal{E}}{\partial t}+ {\nabla} \cdot(\rho H \mathbf{v}) &=& {\nabla} \cdot \left( \mathsf{\tau} \mathbf{v} - \mathbf{q}\right) + \mathbf{J} \cdot \mathbf{E^{\prime}}, \label{eq:global_E}\\
  \frac{\partial \rho e_{\mathrm{ve}}}{\partial t}+ \nabla \cdot\left(\rho e_{\mathrm{ve}} \mathbf{v} \right) &=&   - \nabla \cdot \mathbf{q}_{\mathrm{ve}}  -p_{\mathrm{e}} \nabla \cdot \mathbf{v} + \Omega_{\mathrm{ve}}^{\textsc{c}} + \mathbf{J} \cdot \mathbf{E^{\prime}},\label{eq:ve_eq}
\end{IEEEeqnarray}
Here, the subscript ve refers specifically to contributions from vibronic degrees of freedom. The species set $\mathcal{S}$ includes both free electrons and all macroscopic species present in the plasma. The symbols used in the governing equations (\ref{eq:cont})-(\ref{eq:ve_eq}) are defined as follows:
\begin{itemize}
    \item $t$: time,
    \item $\rho_s = m_s n_s$: the partial densities, with $m_s$ being the masses,
    \item $\mathbf{U}_s$: diffusion velocities of species $s$,
    \item $\rho = \sum_{s \in \mathcal{s}} \rho_s$: total mass density,
    \item $\mathbf{v}$: mass-averaged velocity of the plasma mixture,
    \item $p$: total plasma pressure,
    \item $p_{\mathrm{e}}$: partial pressure of free-electrons,
    \item $\mathcal{E} = e + \mathbf{v} \cdot$: total energy per unit mass, where $e$ is the mixture thermal energy per unit mass,
    \item $H = \mathcal{E} + p/\rho$: total enthalpy per unit mass,
    \item $\mathsf{\tau}$: viscous stress tensor,
    \item $\mathbf{q}$: heat flux vector,
    \item $\dot{\omega}_s$: mass production rate of species $s$ due to collisional processes,
    \item $\Omega_{\mathrm{ve}}^{\textsc{c}}$: volumetric energy exchange rate due to elastic, inelastic, and reactive interactions between electrons and heavy particles,
    \item $\mathbf{J}$: electric current density,
    \item $\mathbf{E}$ and $\mathbf{B}$: electric and magnetic fields, respectively,
    \item $\mathbf{E^{\prime}} = \mathbf{E} + \mathbf{v} \times \mathbf{B}$: electric field observed in the hydrodynamic (moving) frame.
    
\end{itemize}

More details on the computation of $\dot{\omega}_s$, $\Omega_{\mathrm{ve}}^{\textsc{c}}$, as well as on the evaluation of thermodynamic and transport properties, can be found in \cite{kumar2024investigation}.

This paper also presents simulations conducted under LTE assumptions, which is the most efficient method for modeling chemically reacting flows. This approach assumes that the time scales of kinetic processes are significantly shorter than those characterizing the flow (\emph{e.g.}, diffusion across a boundary layer). Under these circumstances, when elemental demixing is ignored \cite{panesi2007analysis,rini2005closed}, the resulting governing equations (\ref{eq:cont}-\ref{eq:ve_eq}) are mathematically equivalent to the Navier-Stokes equations for a simple non-reacting gas, with the effect of chemical reactions \emph{embedded} in the temperature and pressure dependence of thermodynamic and transport properties. The direct calculation of LTE properties such as the electrical conductivity during a simulation may be computationally demanding, as it requires solving a non-linear set of equations to determine the equilibrium composition \cite{bottin1999thermodynamic,vincenti1965introduction}. To address this issue, two-dimensional lookup tables are pre-computed and loaded at the start of a simulation.

\subsection{\label{sec:EM field} Electromagnetic field}
In the context of classical electrodynamics, electromagnetic phenomena are all governed by Maxwell's equations \cite{Mitchner_book}. Here, the latter equations are simplified by introducing the assumptions outlined in \cref{assum:unmagnetized,assum:mag_reynolds,assum:waves}  given at the beginning of this section. While a detailed derivation of the governing equations for the electromagnetic field is provided in \cite{kumar2024investigation}, the key expressions are summarized here for completeness.

The governing equation for the induced electric field is derived under the assumption that, in steady-state, all variables exhibit harmonic temporal behavior (\emph{e.g.,} $\mathbf{E} (\mathbf{r}, \, t)  =  \mathbf{\tilde{E} (\mathbf{r})} \exp\left({\imath \omega t}\right)$), which gives a Helmholtz-like vector equation for the electric field phasor \cite{kumar2024investigation}:
\begin{equation}\label{eq:Helm}
{\nabla} \times {\nabla} \times \mathbf{\tilde{E}} = - \imath \mu_0 \,   \omega \, (\mathbf{\tilde{J}} + \mathbf{\tilde{J_s}}),    
\end{equation}
where the angular frequency of the current running through the inductor is $\omega = 2\pi f$, whereas $\imath = \sqrt{-1}$ denotes the imaginary unit. $\mu_0$ is the vacuum magnetic permeability. In the above equation, amplitudes are taken complex (\emph{e.g.}, $\mathbf{\tilde{E}} = \mathbf{\tilde{E}}_{\mathrm{R}} + \imath \mathbf{\tilde{E}}_{\mathrm{I}}$) to account for the phase differences between electric and magnetic fields. The plasma contribution to the current density phasor in the torch is given by $\mathbf{\tilde{J}} = \sigma \mathbf{\tilde{E}}$ as derived in \cite{kumar2024investigation}, and zero elsewhere, whereas that of the inductor coils, $\mathbf{\tilde{J_s}}$, is prescribed as discussed in \cref{sec:problem_desc}. Once $\mathbf{\tilde{E}}$ is found, the magnetic induction is retrieved from Faraday's law:
\begin{equation}\label{eq:Bfield}
\mathbf{\tilde{B}} = \left(\dfrac{\imath}{\omega}\right) \nabla \times \mathbf{\tilde{E}}.
\end{equation}

ICP facilities typically operate at radio frequencies on the order of \si{\mega\hertz}. Given this, it is appropriate to consider that, over a single inductor cycle $1/f$, the plasma responds to the time-averaged effects of both the Lorentz force and Joule heating:
\begin{IEEEeqnarray}{rCl}\label{Lorentz_Joule_terms}
\left<\mathbf{J}\times \mathbf{B}\right>&= & \dfrac{1}{2} \left(\dfrac{\sigma}{\omega}\right) \Re{\left[
\mathbf{\tilde{E}} \times\left( \imath {\nabla}_{\mathbf{r}} \times\mathbf{ \tilde{E}}\right)^{*}\right]}, \\
\left<\mathbf{J}\cdot \mathbf{E}\right>&= &
    \frac{1}{2}\sigma\mathbf{\tilde{E}}\cdot \mathbf{\tilde{E}^*},
\end{IEEEeqnarray}
where the $z^*$ and  $\Re{(z)}$ denote, respectively, the complex conjugate and the real part of $z$.

As described by Boulos \cite{Boulos_1976}, the inductor coil current is dynamically adjusted during the simulation to ensure that the power dissipated through Joule heating matches a specified target value:
\be
P_t = \!\! \int \!\! \left<  \mathbf{J}\cdot \mathbf{E^{\prime}} \right> d V \simeq \!\! \int \!\! \left<  \mathbf{J}\cdot \mathbf{E} \right> d V, \label{eq:target_power}
\ee
where the volume integration is restricted to the plasma region.

\section{Numerical framework}\label{sec:numerical_framework}
The simulations presented in this work use a multi-solver coupled computational framework for ICPs described in ref. \cite{kumar2024investigation}, where the flow governing equations are solved in a 3D block-structured finite volume (FV) solver,
\textsc{hegel} (High fidElity tool for maGnEtogas-dynamic appLications) \cite{munafo2024hegel} and the electromagnetic (EM) field is solved in a mixed finite element solver,
\textsc{flux} (Finite-element soLver for Unsteady electromagnetic) \cite{kumar2022self}. In this work, the three-dimensional module of \textsc{flux} has been used, where the H(Curl) finite element space is used for the electric field. The H(Curl) finite element space contains the 1-form vector basis functions, which have well-defined curl. The magnetic field lies in the H(Div) finite element space, which contains the 2-form vector basis functions having well-defined divergence. \textsc{flux} solves for the steady-state electromagnetic field based on the given plasma electrical conductivity distribution. The resulting linear system is solved employing the Flexible Generalized Minimal Residual (FGMRES) Method \cite{saad1993flexible}, leveraging the Auxiliary space Maxwell Solver (AMS) \cite{hiptmair2007nodal} as a preconditioner to enhance convergence.

\begin{figure}[hbt!]
\centering
\includegraphics[scale=0.75]{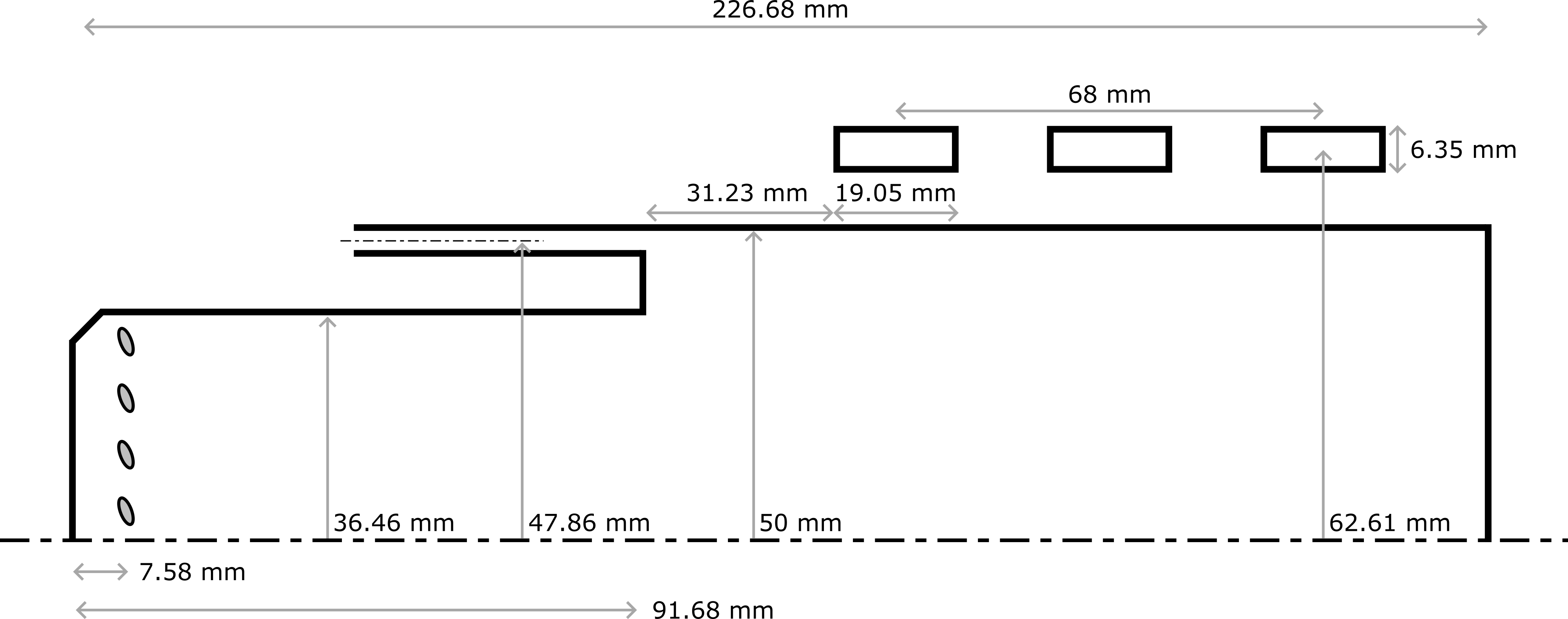}
\caption{Schematic of the torch of the Plasmatron X facility.}
\label{fig:pX_torch}
\end{figure}

The plasma and the electromagnetic field are coupled through the Lorentz force and Joule heating in the fluid momentum equation (\ref{eq:momentum}) and energy equations (\ref{eq:global_E})-(\ref{eq:ve_eq}), respectively, and the electrical conductivity in the induction equation (\ref{eq:Helm}). The communication of the exchanged quantities (\emph{e.g.}, $\sigma$) between the flow solver grid and the EM solver grid occurs only in the overlap region, which consists of the torch. To minimize communication-related slowdowns, the communication time window is set so that the FV fluid solver calls the EM solver every 100 fluid time steps. During each coupling interval, the exchanged source terms remain constant. The communication of data is achieved using preCICE \cite{bungartz2016precice}, an open-source coupling library for partitioned multi-physics simulations.

\section{\label{sec:problem_desc}Plasmatron X facility: problem description and simulation setup}

The multi-physics computational framework discussed above is used to simulate the \SI{350}{\kilo\watt} Plasmatron X facility at UIUC. \cref{fig:pX_torch} illustrates the schematic of the torch section \cite{oruganti2023modeling,kumar2024modeling}. The torch incorporates two types of gas injection systems: a central injector with 15 angled holes (inclined downward at \SI{15}{\degree} and imparting a \SI{24}{\degree} swirl), and a sheath injector composed of 72 holes aligned along the axial direction. For modeling purposes, both injection systems are idealized as continuous annular inlets. Throughout all simulations, the mass flow rates are held constant at \SI{0.86}{\gram/\second} for the central injector and \SI{7.13}{\gram/\second} for the sheath gas. The working fluid in all cases is an 11-species air mixture: \textbf{$\mathcal{S} = \left\{\mathrm{e}^-, \, \mathrm{N}_2, \, \mathrm{O}_2, \, \mathrm{NO},\mathrm{N},\, \mathrm{O}, \, \mathrm{N^+_2}, \, \mathrm{O^+_2},\, \mathrm{NO^+},\, \mathrm{N^+},\, \mathrm{O^+}\right\}$}. Among the three available nozzle configurations in the Plasmatron X facility, one straight and two of the converging-diverging type, this study focuses exclusively on the straight nozzle. This nozzle measures \SI{129.54}{mm} in length and has a diameter of \SI{100}{mm}, positioned downstream of the torch. The induction system comprises a three-turn coil with a rectangular cross-section. All simulations are conducted at a fixed coil operating frequency of \SI{2.1}{\mega\hertz}. Gravity has been applied in the negative $y$ direction, where the axis of the torch is along the $x$ direction. It is to be highlighted that during the operation of the facility, only a part of the operating power goes to the plasma due to several inefficiencies (such as the coupling efficiency between the inductor coils and the plasma, power lost in the inductor coils and matching networks, \emph{etc.}). Hence, the powers reported in this paper for all the simulations are accompanied by the corresponding overall efficiencies $\eta$ given as:
\begin{equation}
    \eta = (P - P_{loss})/P
\end{equation}
where $P$ is the facility operating (or input) power and $P_{loss}$ is the sum of all the power losses. Subsequently, the target power used for simulations $P_t$ as described in \cref{eq:target_power} is defined as $P \times \eta$.

\subsection{Boundary conditions}
The fluid governing equations are solved by imposing the following boundary conditions:
\begin{itemize}
\item injectors (subsonic inflow):
$$
\rho u=\frac{\dot{m_x}}{A}, \quad \rho v=\frac{\dot{m_y}}{A}, \quad \rho w=\frac{\dot{m_z}}{A}, \quad  y_s = y_{s\, \mathrm{in}}, \quad \frac{\partial p}{\partial n}=0 \quad \text{and} \quad T_{\mathrm{h}}= T_{\mathrm{ve}} = T_{\mathrm{in}},
$$
where the $\mathrm{in}$ lower-script denotes the injection conditions. In the above formulas, $\dot{m}$ is the mass flow, while $u$, $v$, and $w$ denote the velocity components $x$, $y$, and $z$, respectively. The $A$ symbol stands for the injector annular area. 
\item walls (isothermal and non-catalytic):
$$
u=v=w=0 \quad \text{and} \quad T_{\mathrm{h}}=T_{\mathrm{ve}} = T_{\mathrm{w}},
$$
where $T_{\mathrm{w}}$ denotes the wall temperature. 

\item outlet (subsonic/supersonic outflow): if the local normal Mach number at the boundary is less than 1 (\emph{i.e.} subsonic outflow), a constant ambient pressure is prescribed, $p=p_{\mathrm{a}}$. On the other hand, if the local Mach number is greater than 1 (\emph{i.e.} supersonic inflow), the flow properties are extrapolated from the interior. For the torch-chamber simulation, the outlet boundary condition is applied on the right-most as well as the side face of the chamber.
\end{itemize} 
In all simulations, both the wall and the ambient temperature are set to \SI{300}{\kelvin}.
\\

For the electric field, a far-field boundary condition is used at all the outer boundaries by imposing a Dirichlet boundary condition $\hat{n} \times \textbf{E} = 0$. Although it is not a truly absorbing boundary condition, it is sufficient if the outer boundaries are far from the coils such that the reflections are negligible. Moreover, for the variational formulation used in this work, this condition makes the problem well-posed without adding additional complexities. 

\subsection{Modeling of the helical coil}
The inductor coil of the Plasmatron X facility has three turns with a rectangular cross-section (see \cref{fig:pX_torch}). However, almost all of the current gets accumulated at the inner two corners of the rectangular cross-section due to the well-known skin-effect \cite{abeele2000efficient,belevitch1971lateral,blackwell2020demonstration}. To simplify the grid generation, it is assumed that the current flows through the midpoint of the line connecting the bottom two corners of the rectangular cross-section. This allows for meshing a single helical coil passing through the midpoint of the inner surface of the coil. The thickness of the helical coil is kept small relative to the length of the coil, and a uniform current is applied throughout the cross-section to simulate a thin current-carrying helical loop.

\begin{figure}[hbt!]
\centering
\includegraphics[trim={0 3cm 0 3cm},clip,scale=0.3]{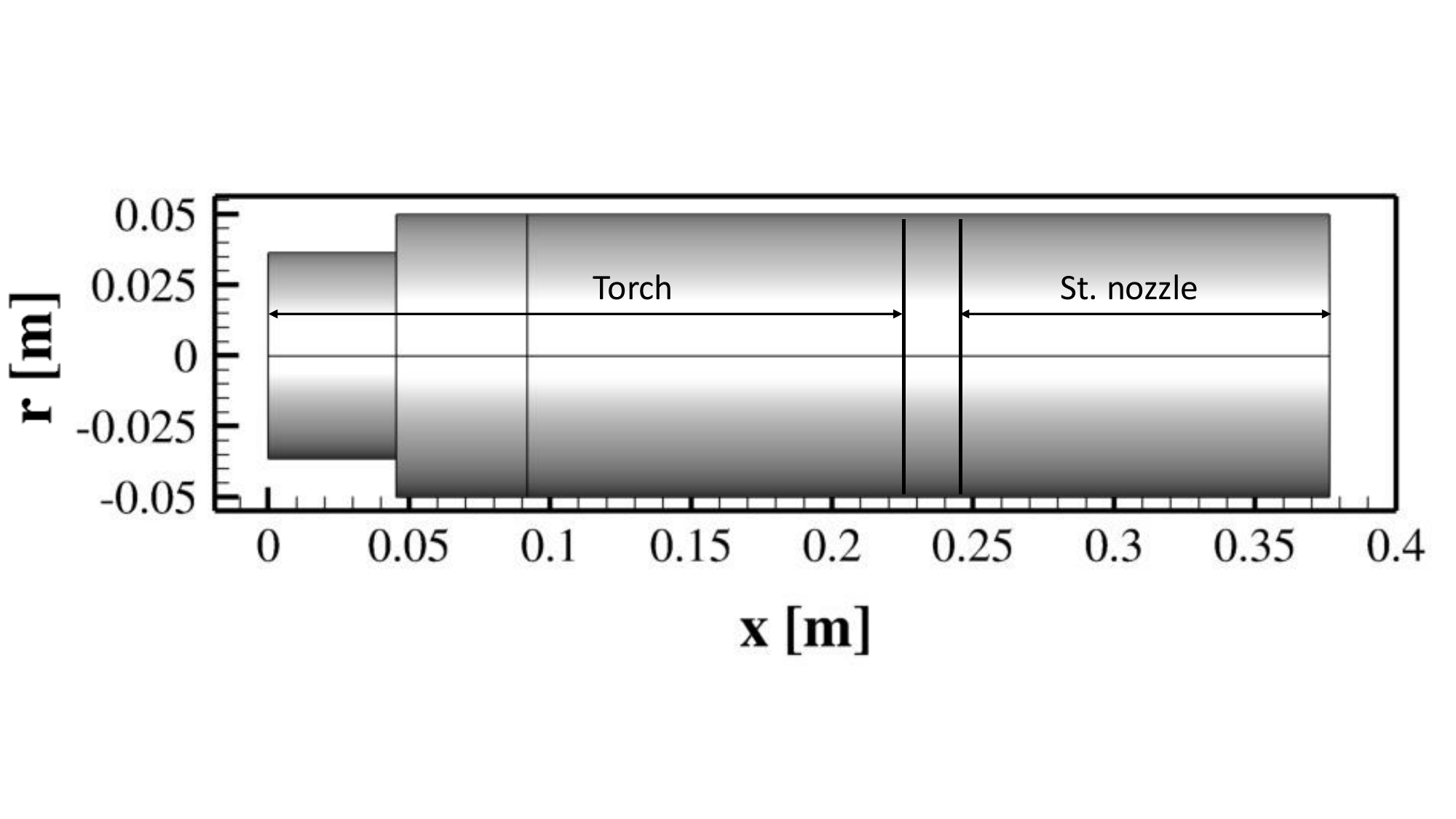}
\caption{Simplified geometry adopted for the torch-nozzle assembly.}
\label{fig:pX_torch_domain}
\end{figure}

\begin{figure}[hbt!]
\centering
\subfloat[][]{\includegraphics[scale=0.2]{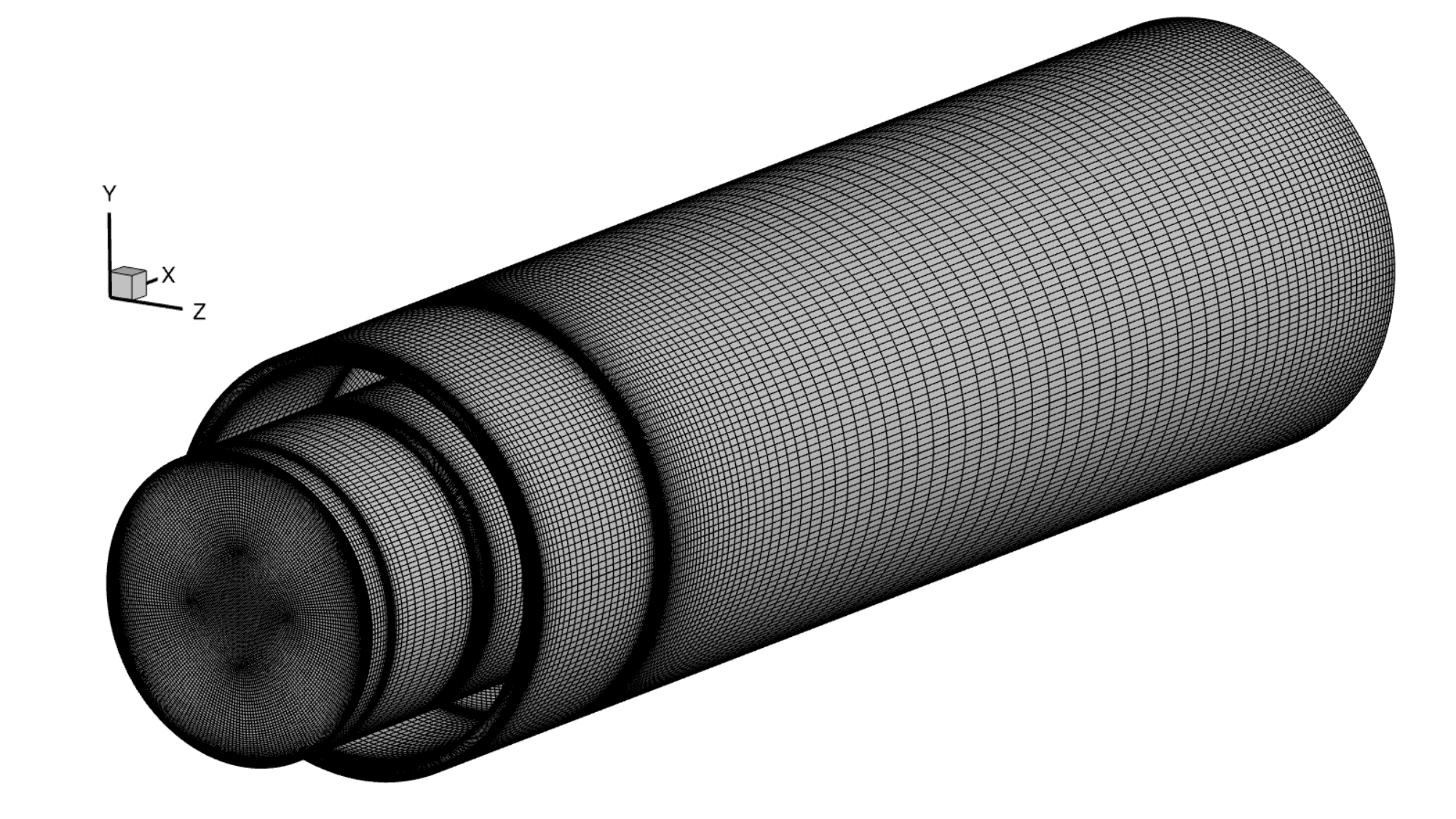}\label{fig:pX_torch_mesha}} 
\hspace{1cm}
\subfloat[][]{\includegraphics[scale=0.2]{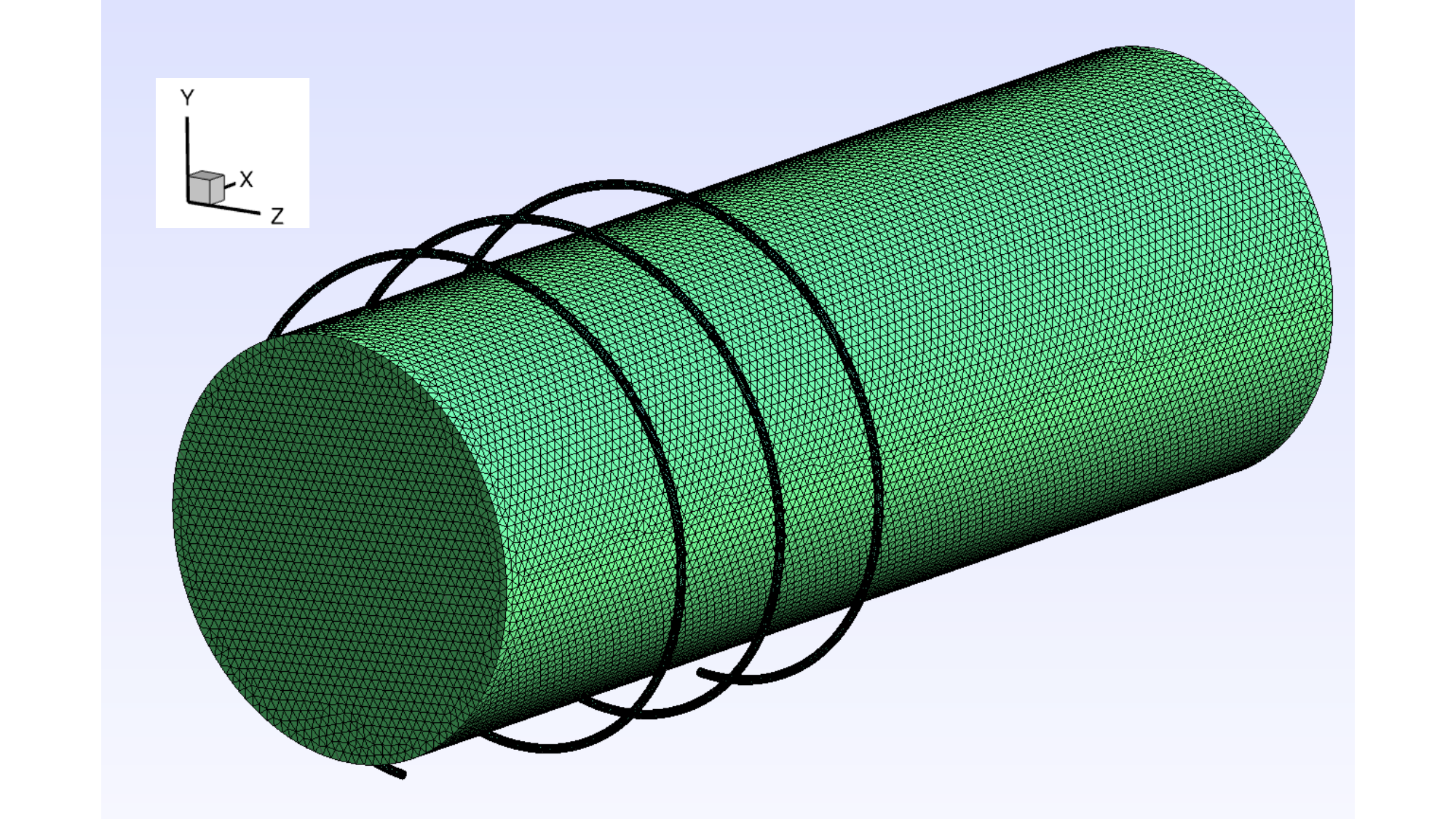}\label{fig:pX_torch_meshb}}
\caption{Mesh for the torch-nozzle geometry used in: \protect\subref{fig:pX_torch_mesha} the plasma solver (structured) and \protect\subref{fig:pX_torch_meshb} the electromagnetic solver (unstructured) including the helical coils.} 
\label{fig:pX_torch_mesh}
\end{figure}

\subsection{Computational domain and mesh}
\cref{fig:pX_torch_domain} shows the computational domain for plasma used for the simulation, which consists of the torch and the straight nozzle. It is to be noted that the term torch will refer to the torch along with the straight nozzle throughout this paper. A block-structured mesh has been used for the plasma solver, consisting of approximately \num{4.2e+6} cells, while that used in the electromagnetic solver (which includes the far field region) is fully unstructured, and consists of \num{1.9e+6} elements (see \cref{fig:pX_torch_mesh}). These mesh sizes were found to give a grid-converged solution for the torch simulation.

\begin{figure}[hbt!]
\centering
\includegraphics[scale=0.3]{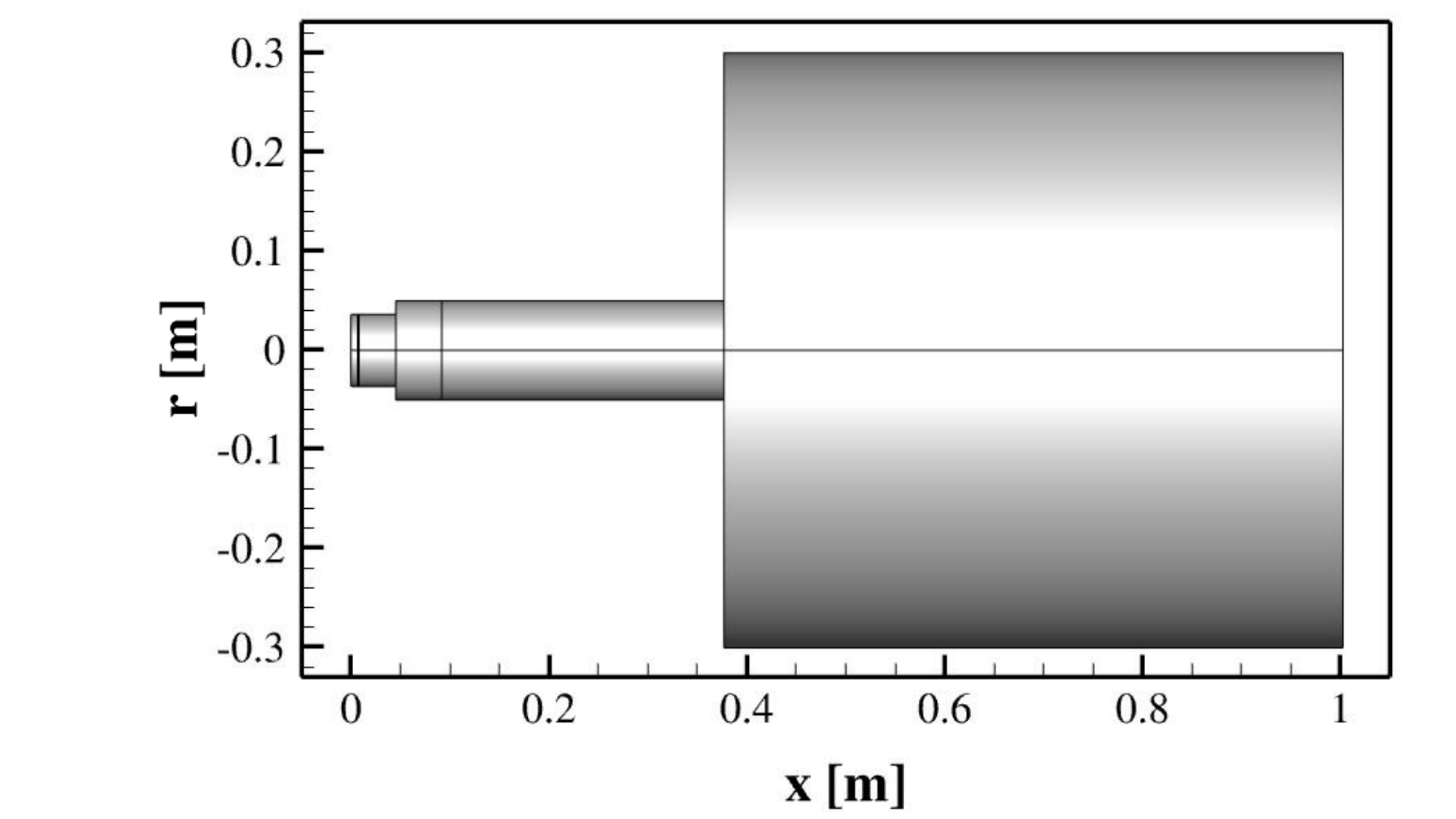}
\caption{Plasma computational domain for the torch-chamber geometry.}
\label{fig:pX_torch_chamber_domain}
\end{figure}

\cref{fig:pX_torch_chamber_domain} shows the plasma computational domain for the torch-chamber system considered in \cref{sec:torch_chamber}, with the mesh consisting of approximately \SI{9.5}{} million cells. This mesh size was found to be adequate to resolve the main flow structures in the chamber. The geometry of the former is here simplified to facilitate the multi-block grid generation process. The mesh for the EM solver remains the same.

\section{Results}\label{sec:results}

\subsection{\label{sec:torch} Three-dimensional analysis of the torch}
A steady-state simulation of the torch (along with the straight nozzle) has been conducted to assess possible three-dimensional effects in the torch region. The fluid governing equations are advanced in time using the backward Euler method with local CFL-based time-stepping to accelerate convergence to steady-state \cite{hirsch1990numerical}. It is important to note that under certain operating conditions, the solution may not achieve a fully steady state due to minor oscillations in the cold region near the inlet and within the recirculation zone. However, since these fluctuations are minimal and do not compromise the stability of the highly viscous plasma core, convergence is assessed by monitoring the residuals within the hot plasma core in the torch. For cases exhibiting significant unsteadiness, particularly in torch-chamber simulations as discussed later, time-accurate simulations are required. The operating conditions for the torch simulation are: $p_{\mathrm{a}} = \SI{10}{\kilo\pascal}$, $P =\SI{180}{\kilo\watt}$ and $\eta = \SI{55.5}{}\%$. The simulation has been performed assuming LTE, which was found suitable for such a large ambient pressure. 

\begin{figure}[hbt!]
\hspace{-1cm}
\subfloat[][]{\includegraphics[trim={1.5cm 0 2cm 0},clip,scale=0.3]{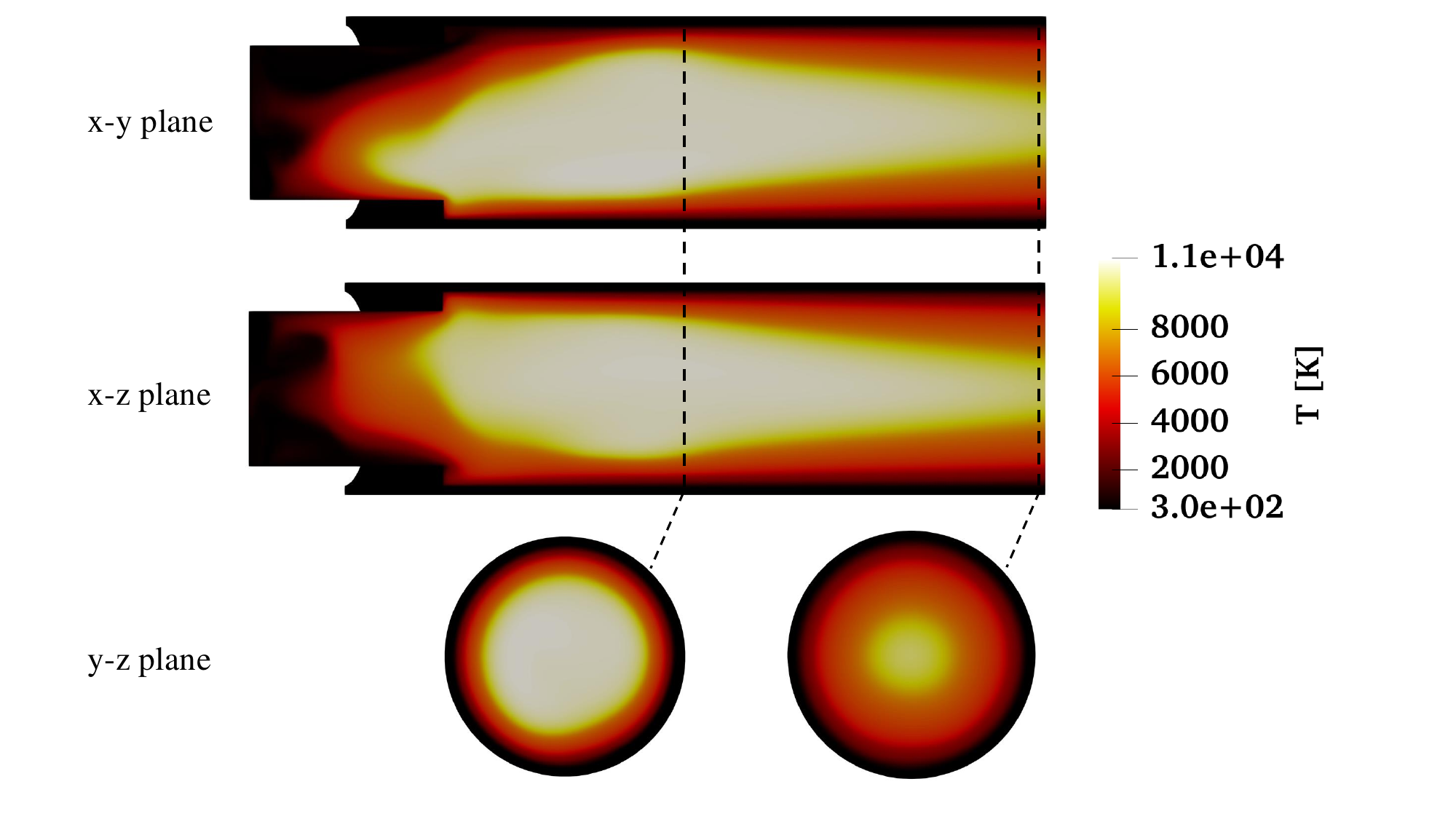}\label{fig:torch_contoursa}} 
\subfloat[][]{\includegraphics[trim={1.5cm 0 2cm 0},clip,scale=0.3]{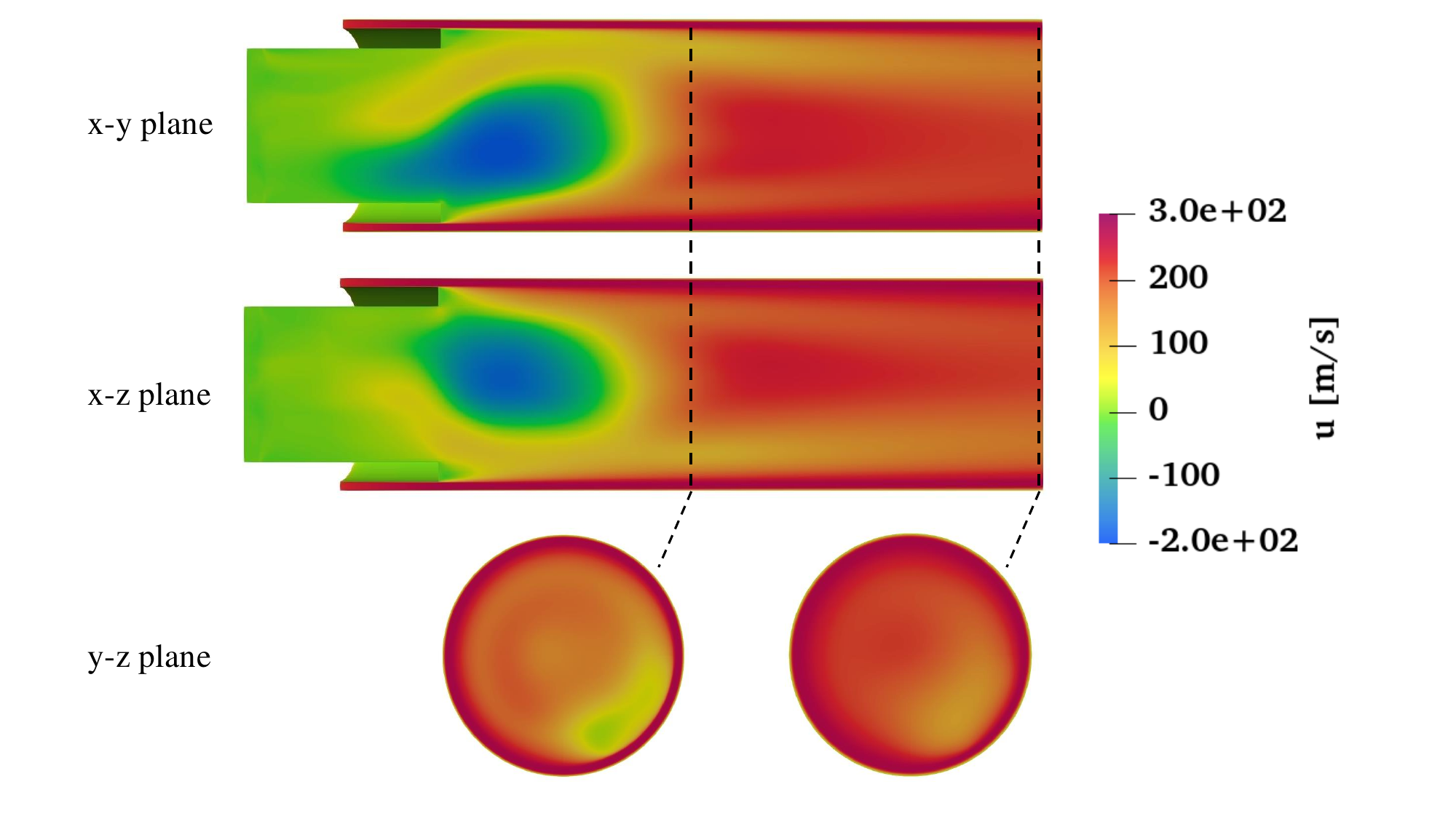}\label{fig:torch_contoursb}}
\caption{Plasma field inside the torch across various planes. The $y$-$z$ sections have been shown at two axial locations: $x = \SI{0.2}{m}$ (coil region) and $x = \SI{0.375}{m}$ (nozzle outlet).  \protect\subref{fig:torch_contoursa} Temperature, and \protect\subref{fig:torch_contoursb} axial velocity. ($p_{\mathrm{a}} = \SI{10}{\kilo\pascal}$,  $P = \SI{180}{\kilo\watt}$, $\eta = 55.5\%$).} 
\label{fig:torch_contours}
\end{figure}

\begin{figure}[hbt!]
\centering
\subfloat[][]{\includegraphics[scale=0.5]{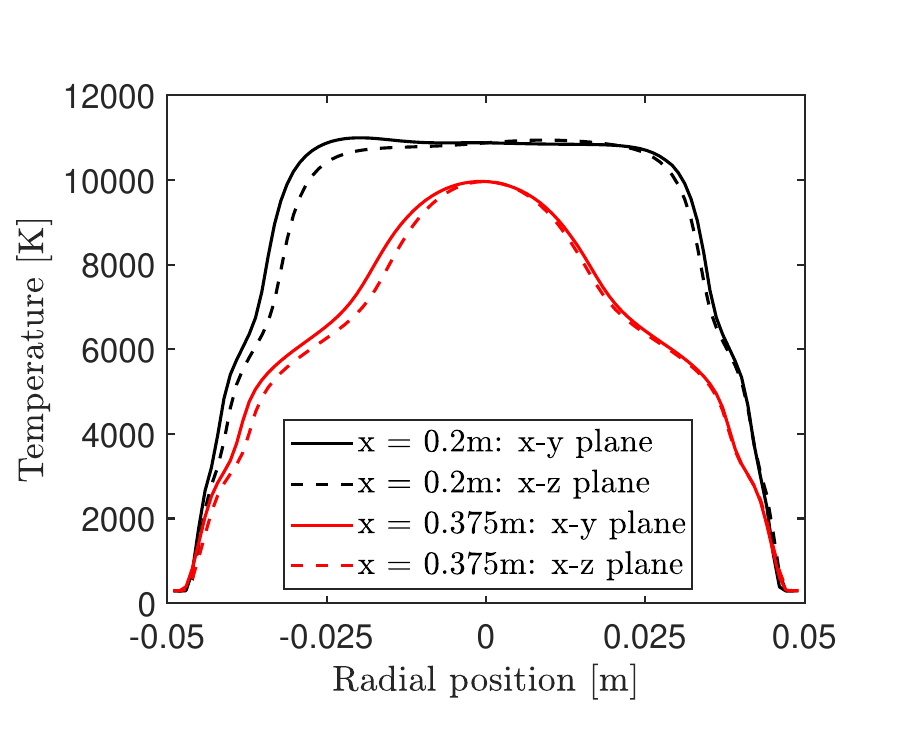}\label{fig:profiles_torcha}} 
\subfloat[][]{\includegraphics[scale=0.5]{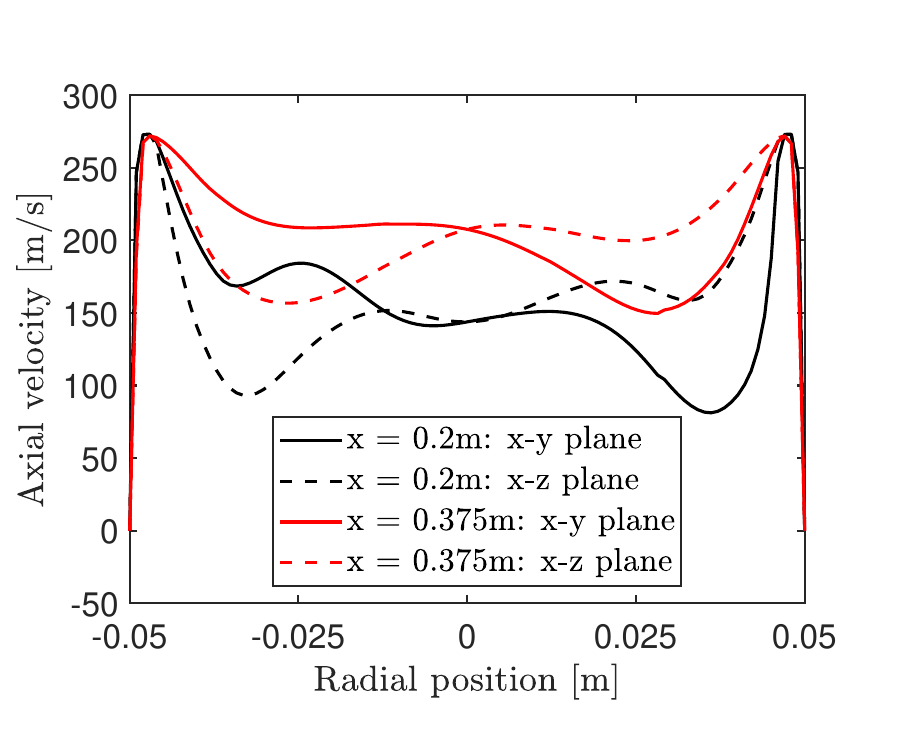}\label{fig:profiles_torchb}}
\caption{Radial profiles across the $x$-$y$ and $x$-$z$ planes at axial locations: $x= \SI{0.2}{m}$ (coil region) and $x = \SI{0.375}{m}$ (nozzle outlet). \protect\subref{fig:profiles_torcha} Temperature, and \protect\subref{fig:profiles_torchb} axial velocity. ($p_{\mathrm{a}} = \SI{10}{\kilo\pascal}$, $P = \SI{180}{\kilo\watt}$, $\eta = 55.5\%$). } 
\label{fig:profiles_torch}
\end{figure}

\begin{figure}[hbt!]
\centering
\subfloat[][]{\includegraphics[trim={11cm 0 12cm 0},clip,scale=0.3]{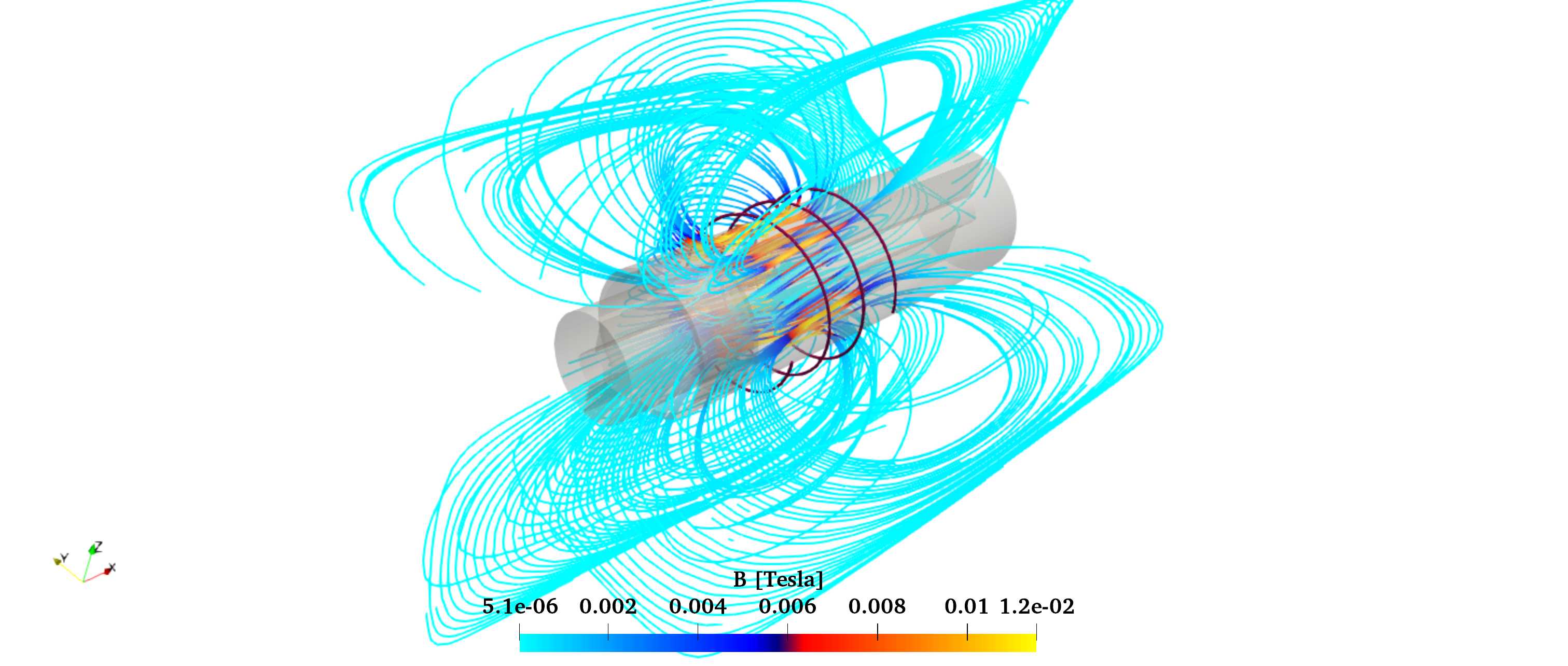}\label{fig:torch_E_B_linesa}} 
\subfloat[][]{\includegraphics[trim={12cm 0 12cm 0},clip,scale=0.3]{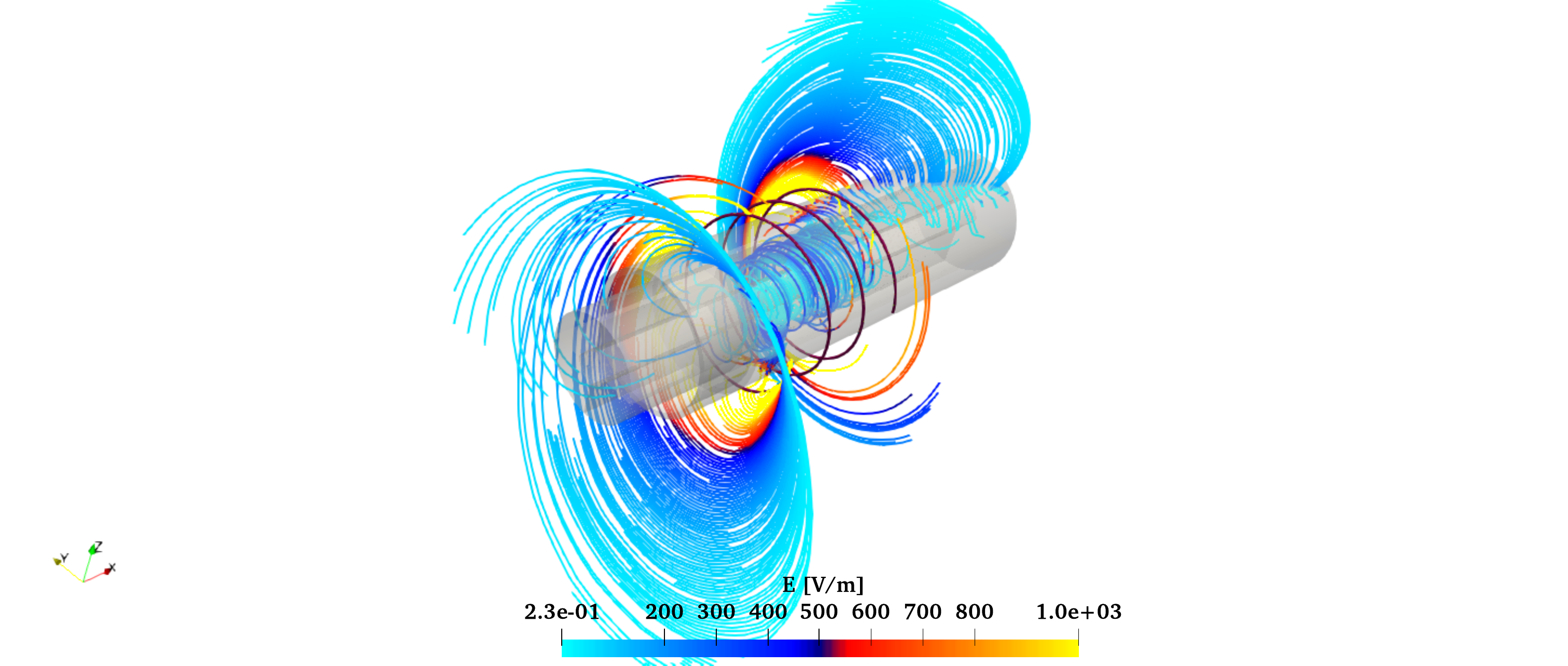}\label{fig:torch_E_B_linesb}}
\caption{Field lines of \protect\subref{fig:torch_E_B_linesa} magnetic induction and \protect\subref{fig:torch_E_B_linesb} electric field generated by the helical coil. ($p_{\mathrm{a}} = \SI{10}{\kilo\pascal}$,  $P = \SI{180}{\kilo\watt}$, $\eta = 55.5\%$).} 
\label{fig:torch_E_B_lines}
\end{figure}

\begin{figure}[hbt!]
\hspace{-1.5cm}
\subfloat[][]{\includegraphics[trim={1.5cm 0 1.5cm 0},clip,scale=0.3]{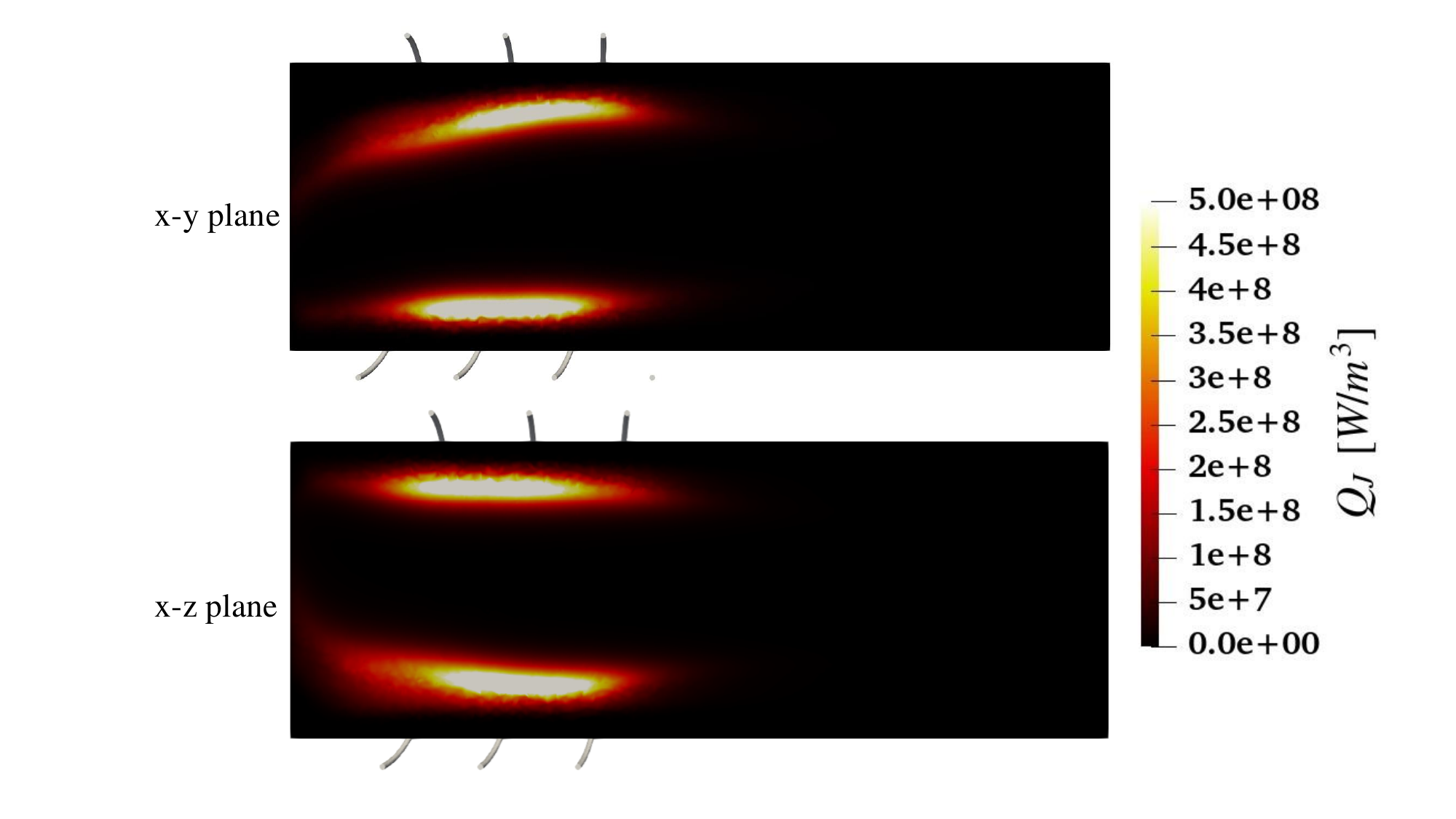}\label{fig:torch_EM_contoursa}} 
\subfloat[][]{\includegraphics[trim={1.5cm 0 1.5cm 0},clip,scale=0.3]{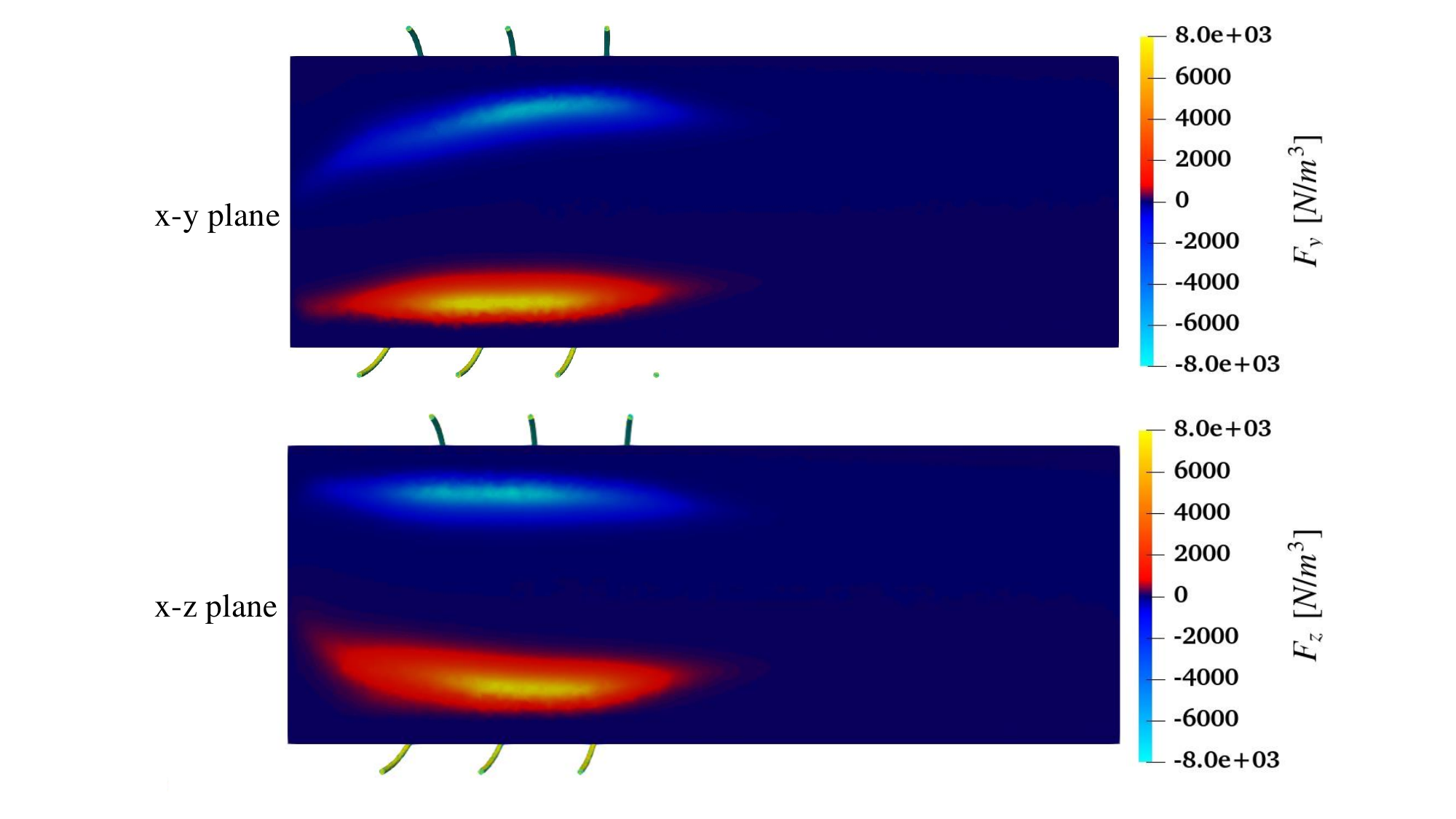}\label{fig:torch_EM_contoursb}}
\caption{Distributions of \protect\subref{fig:torch_EM_contoursa} Joule heating and \protect\subref{fig:torch_EM_contoursb} radial component of Lorentz force inside the torch across the $x$-$y$ and $x$-$z$ planes. ($p_{\mathrm{a}} = \SI{10}{\kilo\pascal}$,  $P = \SI{180}{\kilo\watt}$, $\eta = 55.5\%$).} 
\label{fig:torch_EM_contours}
\end{figure}

\begin{figure}[hbt!]
\centering
\subfloat[][]{\includegraphics[scale=0.5]{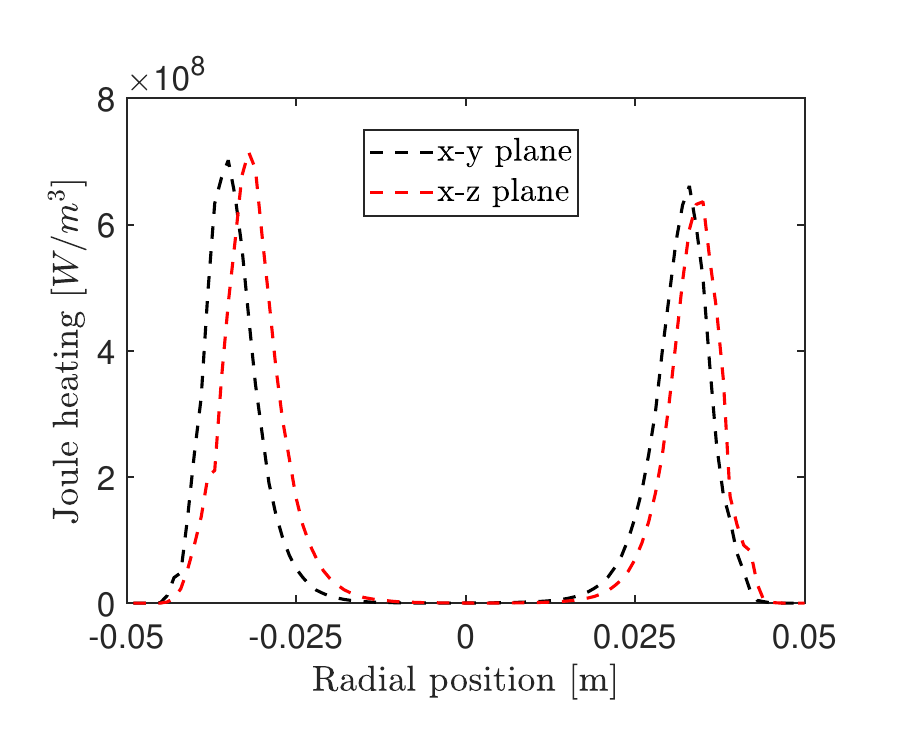}\label{fig:EM_profiles_torcha}} 
\subfloat[][]{\includegraphics[scale=0.5]{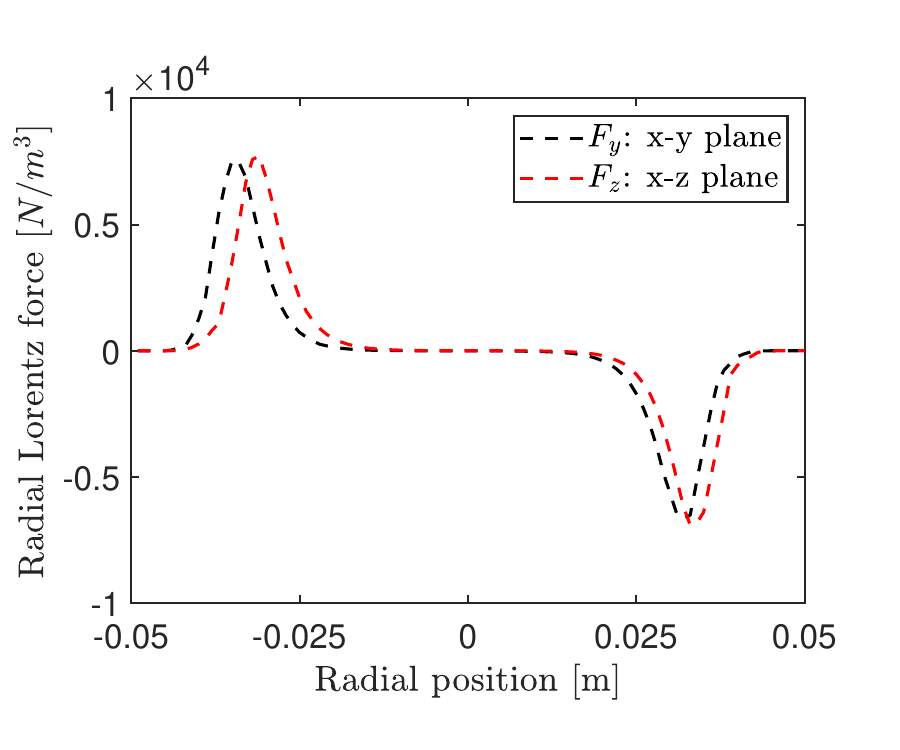}\label{fig:EM_profiles_torchb}}
\caption{Radial profiles across the $x$-$y$ and $x$-$z$ planes (at $x= \SI{0.175}{m}$) of \protect\subref{fig:EM_profiles_torcha} Joule heating and \protect\subref{fig:EM_profiles_torchb} radial component of Lorentz force. ($p_{\mathrm{a}} = \SI{10}{\kilo\pascal}$, $P = \SI{180}{\kilo\watt}$, $\eta = 55.5\%$).} 
\label{fig:EM_profiles_torch}
\end{figure}

\cref{fig:torch_contours} shows the plasma temperature and axial velocity distributions within the torch. The contour maps on the $x$-$y$ and $x$-$z$ planes show large differences, especially in the coil region, indicating a lack of axial symmetry. \cref{fig:profiles_torch} shows the radial profiles of temperature and axial velocity across the $x$-$y$ and $x$-$z$ planes at two axial locations: $x = \SI{0.2}{\meter}$ (coil region) and $x = \SI{0.375}{\meter}$ (nozzle outlet). The temperature profiles in the two planes clearly show a significant difference in the coil region, \emph{i.e.} at $x = \SI{0.2}{\meter}$, whereas at the nozzle outlet, the discrepancy becomes smaller, although non-negligible. The velocity profiles on the two planes show large discrepancies at both axial locations. Moreover, the profiles are not symmetric about the axis (\emph{i.e.} $r = \SI{0}{}$) for both planes. \cref{fig:torch_E_B_lines} shows the magnetic and electric field lines created by the helical coil, highlighting their three-dimensional nature. In a two-dimensional axisymmetric setup with parallel coils, the electric field exhibits only an azimuthal component, and the field lines form toroidal shapes. However, in a three-dimensional configuration, the helical structure of the coil and its end effects introduce axial components to the electric field, though these remain relatively small compared to the dominant azimuthal component. This results in a more complex distribution of the electromagnetic field, which, in turn, has a strong impact on both the Joule heating and Lorentz force. This is put in evidence in \cref{fig:torch_EM_contours}, showing the distribution of the latter quantities inside the torch across the $x$-$y$ and $x$-$z$ planes, which highlights, once more, the lack of axial symmetry. This leads to an unbalanced Lorentz force causing the plasma core to attain an asymmetric configuration, which has also been highlighted in previous works \cite{bernardi2003three,colombo2010three}. \cref{fig:EM_profiles_torch} illustrates the radial profiles of Joule heating and Lorentz force across the $x$-$y$ and $x$-$z$ planes at $x = \SI{0.175}{\meter}$, which marks the center of the coil region. The profiles across the two planes are very different, highlighting the non-axisymmetric distribution. Moreover, even for a given plane, the Lorentz force distribution in one half (\emph{e.g.} $r \leq \SI{0}{m}$) of the torch differs from the other ($r \geq \SI{0}{m}$), thus creating an imbalance in the force acting on the plasma. The non-axisymmetric distribution of the plasma is further influenced by the swirling flow originating from the injector. 

\begin{figure}[hbt!]
\centering
\subfloat[][]{\includegraphics[scale=0.5]{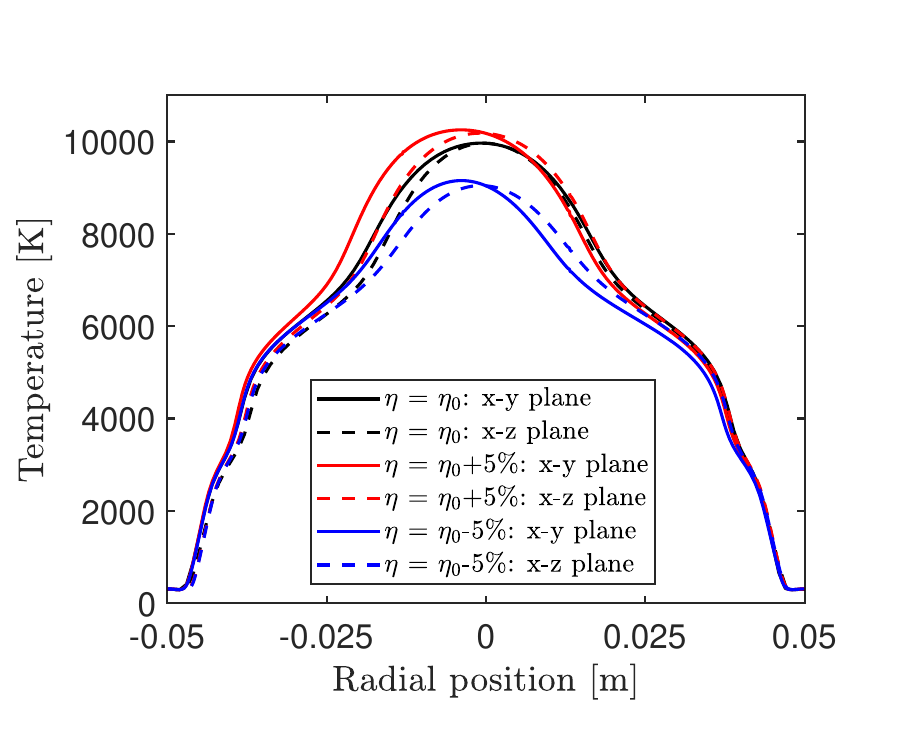}\label{fig:profiles_torcha}} 
\subfloat[][]{\includegraphics[scale=0.5]{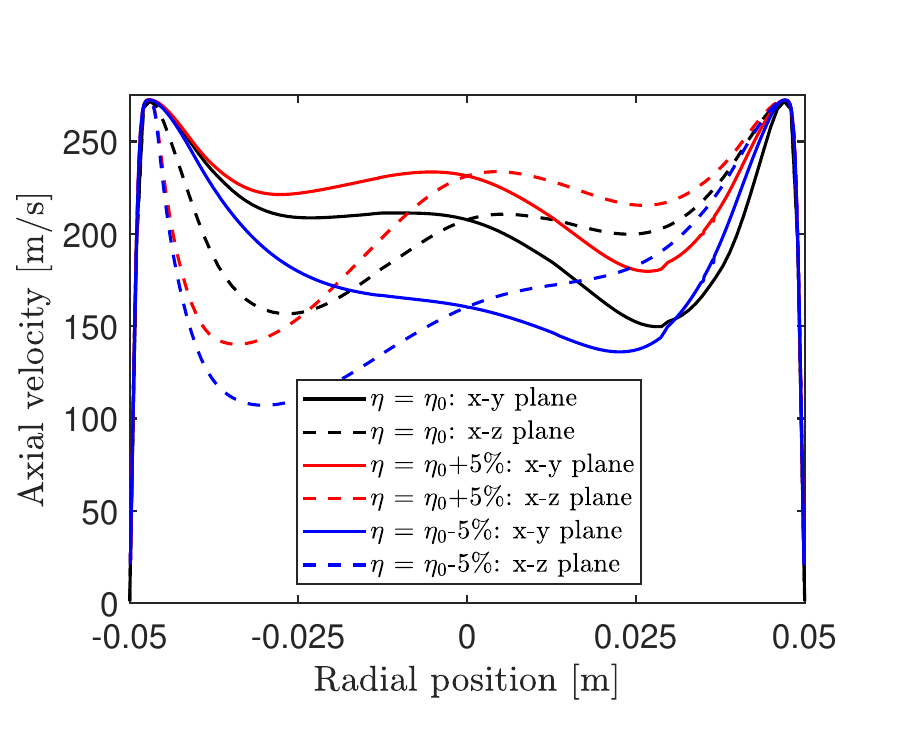}\label{fig:profiles_torchb}}
\caption{Radial profiles across the $x$-$y$ and $x$-$z$ planes for various efficiencies at an axial location of $x = \SI{0.375}{m}$ (nozzle outlet). \protect\subref{fig:profiles_torcha} Temperature, and \protect\subref{fig:profiles_torchb} axial velocity. ($p_{\mathrm{a}} = \SI{10}{\kilo\pascal}$, $P = \SI{180}{\kilo\watt}$, $\eta_0 = 55.5\%$). } 
\label{fig:eta_compare_profiles_torch}
\end{figure}

It is important to note that the calculation of the overall system efficiency ($\eta$) for the Plasmatron X facility is subject to several sources of uncertainty. Typically, the efficiency values range between 50\% and 60\%, and this variability can significantly influence the plasma core behavior. To evaluate the sensitivity of the plasma torch performance to such variations, additional simulations were performed for two $\eta$ values, corresponding to $\pm5\%$ deviations from the nominal efficiency considered in the preceding discussion. \cref{fig:eta_compare_profiles_torch} presents the temperature and axial velocity profiles at the nozzle exit for the three efficiency levels. As expected, the profiles exhibit an increase in peak temperature with higher efficiency and a corresponding decrease with lower efficiency. Specifically, the plasma temperature at the geometric center of the torch increases by only 2.15\% for a 5\% rise in efficiency, which is relatively modest. In contrast, a 5\% reduction in efficiency results in a substantial 9.28\% drop in the centerline temperature. The variations in the axial velocity profiles are even more pronounced. A 5\% increase in efficiency leads to an 11.75\% rise in velocity at the torch centerline, while a 5\% decrease causes a significant reduction of 22.5\%. These findings underscore that the plasma core morphology is highly sensitive to changes in system efficiency, an aspect that must be carefully considered when comparing simulation results against experimental measurements.

Given that the TPM is positioned within the plasma jet region for testing, it is crucial to examine how the three-dimensional characteristics of the plasma in the torch-nozzle combination affect the behavior of the plasma jet within the chamber of the facility. Therefore, the following sections focus on a three-dimensional simulation of the torch-chamber assembly to investigate the spatial dynamics of the plasma jet.

\subsection{\label{sec:torch_chamber} Analysis of the plasma field inside the chamber}
This part of the manuscript investigates the effect of the non-axisymmetric nature of the flowfield inside the torch-nozzle system on the plasma jet in the chamber region. Simulations of the torch-nozzle system along with the chamber region are conducted for two operating conditions: a high-pressure case (\SI{10}{\kilo\pascal}, \SI{100}{\kilo\watt}, $\eta = 57.5\%$), and a low-pressure case (\SI{590}{\pascal}, \SI{300}{\kilo\watt}, $\eta = 53.9\%$) taken from the experiments presented in Ref. \cite{capponi2024multi}. The two pressures roughly characterize the upper and lower limits of the operating pressure considered for testing TPMs in the Plasmatron X facility. The extent of the effect of the coils on the plasma field can be assessed by the magnetic interaction parameter, which is the ratio of the strength of the Lorentz force and the inertia of the plasma given by $I_m = B^2 / \mu_0 \rho u^2$ \cite{david_thesis}. This parameter remains low at low-pressure conditions, where the plasma inertia is large due to higher velocities. However, as the pressure is raised, Lorentz forces begin to dominate over inertia, leading to a rise in the magnetic interaction parameter. Therefore, analyzing the plasma field under these two contrasting operating conditions can provide insights into the degree of three-dimensionality induced by the coils.

\begin{figure}[!htb]
\hspace{-1cm}
\subfloat[][]{\includegraphics[trim={10cm 0 0 0},clip,scale=0.2]{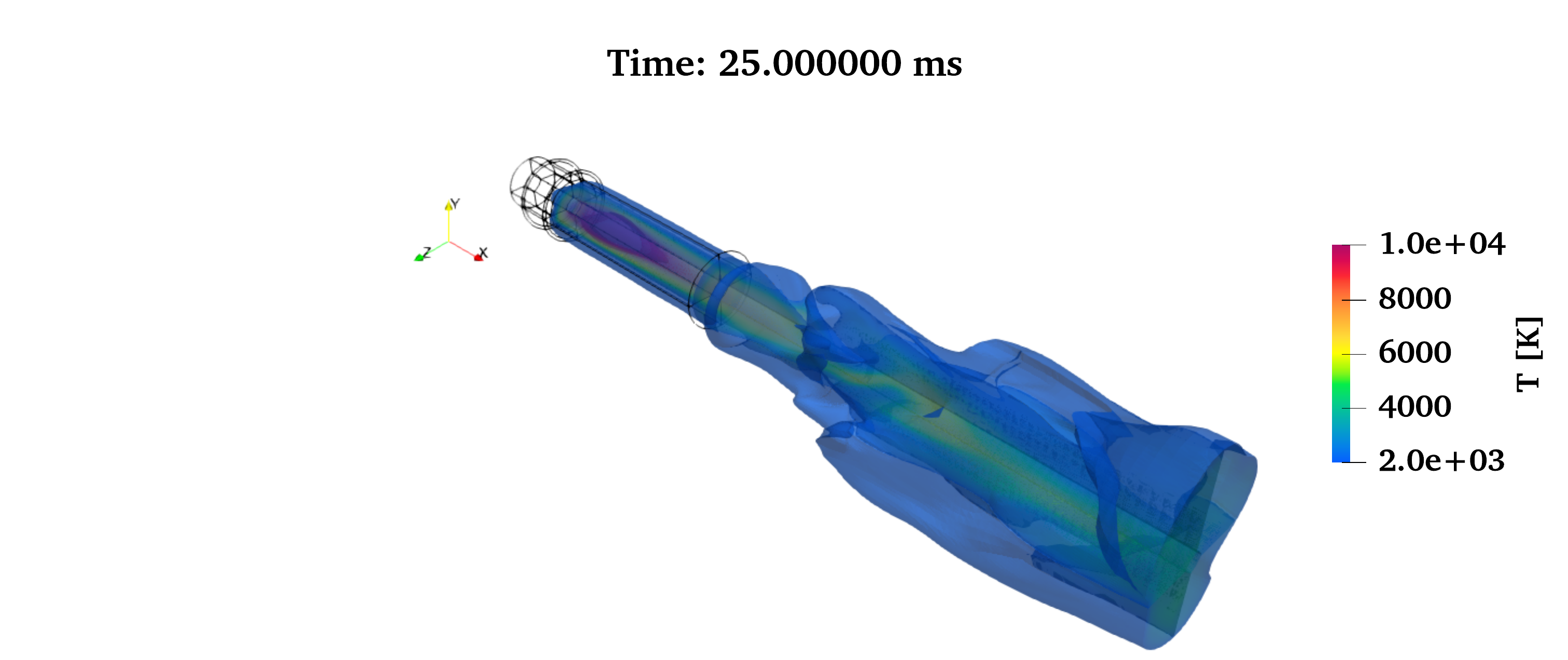}}
\subfloat[][]{\includegraphics[trim={10cm 0 0 0},clip,scale=0.2]{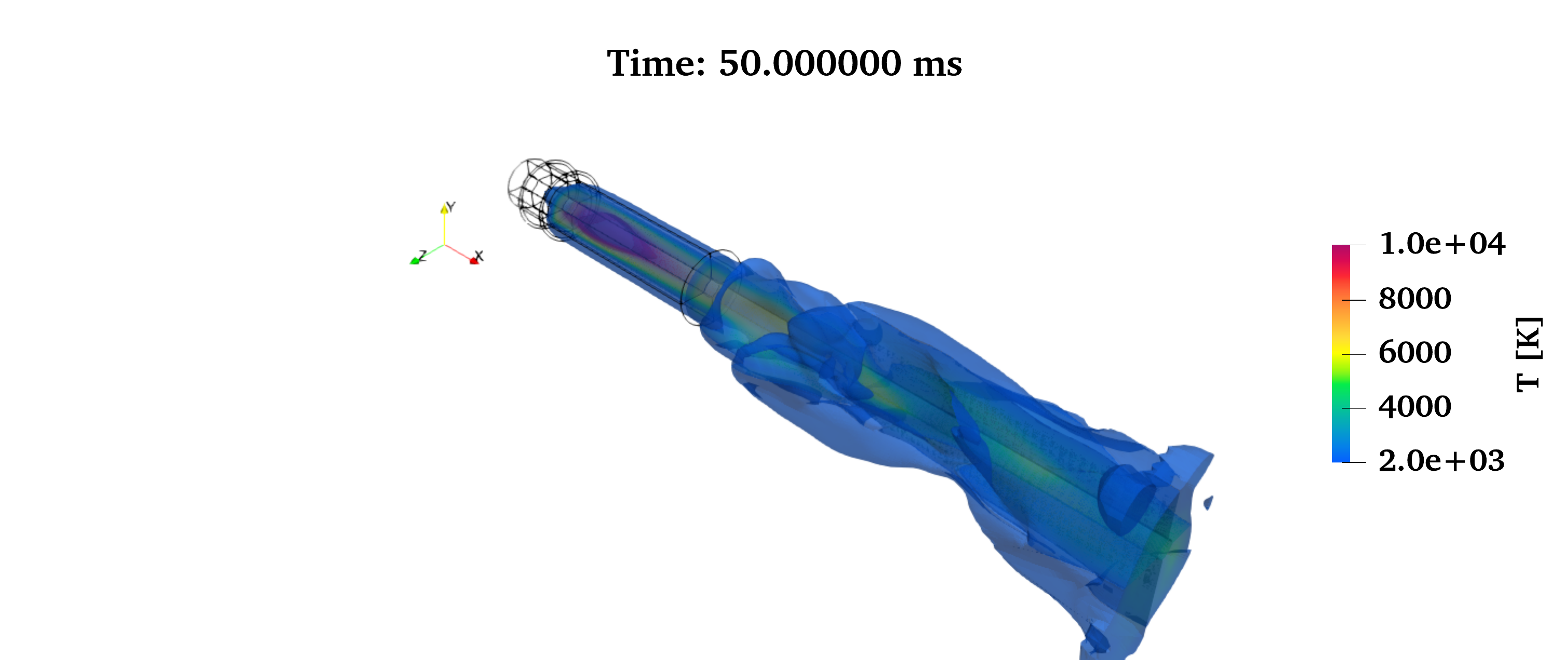}} \\

\hspace{-1cm}
\subfloat[][]{\includegraphics[trim={10cm 0 0 0},clip,scale=0.2]{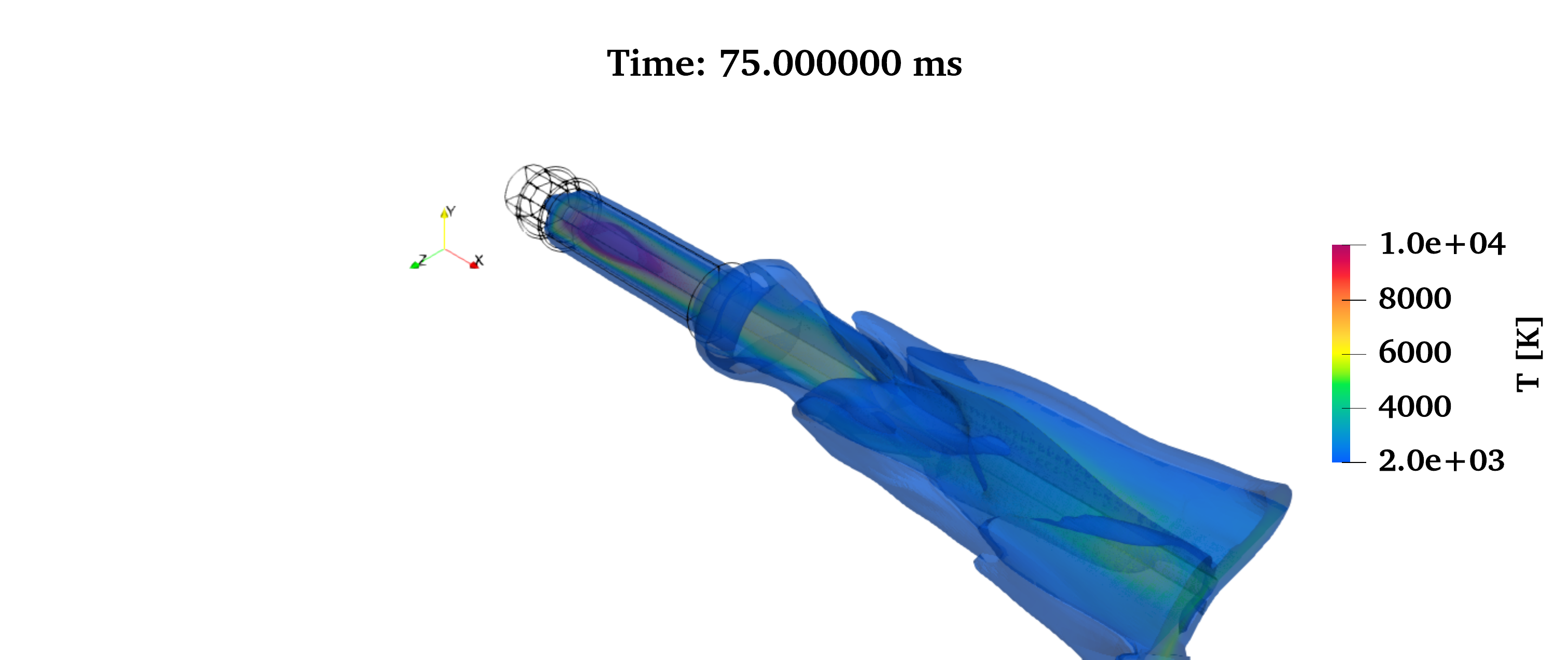}}
\subfloat[][]{\includegraphics[trim={10cm 0 0 0},clip,scale=0.2]{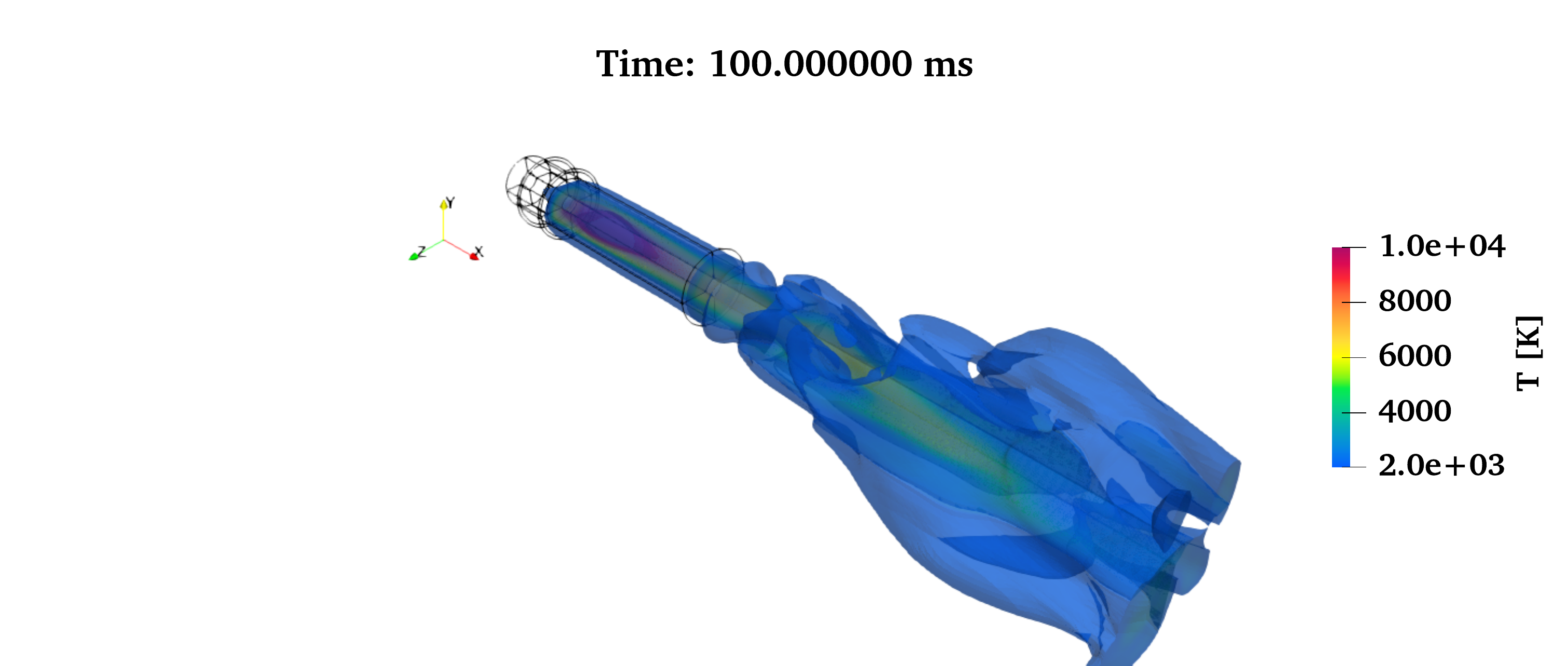}}
\caption{Plasma temperature iso-volumes ranging from \SI{2000}{K} to \SI{10000}{K} at different times. ($p_{\mathrm{a}} = \SI{10}{\kilo\pascal}$, $P = \SI{100}{\kilo\watt}$, $\eta = 57.5\%$).} 
\label{fig:T_iso_dt}
\end{figure}

\begin{figure}[!htb]
\hspace{-1cm}
\subfloat[][]{\includegraphics[trim={3cm 0 6cm 0},clip,scale=0.35]{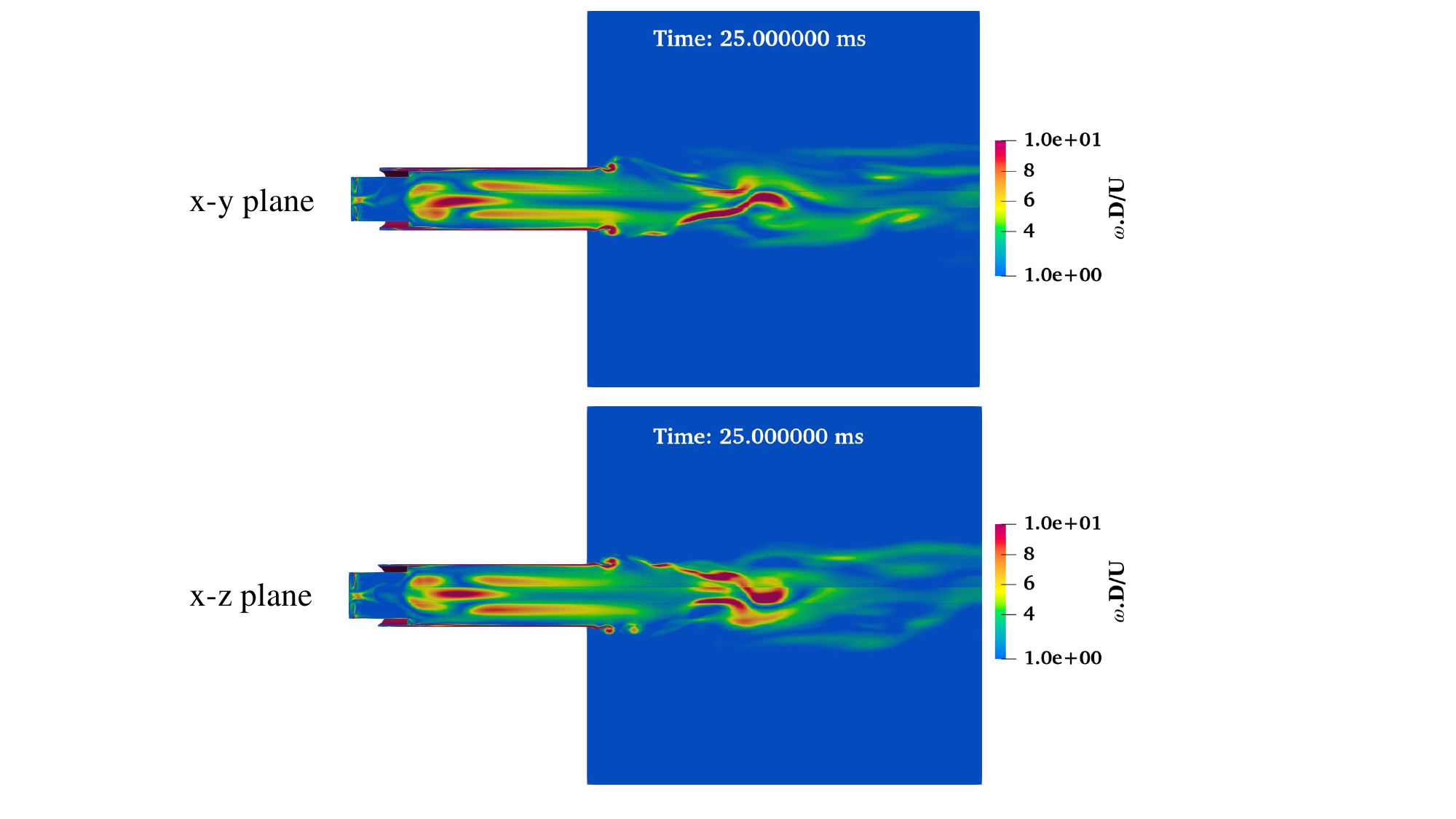}}
\subfloat[][]{\includegraphics[trim={3cm 0 6cm 0},clip,scale=0.35]{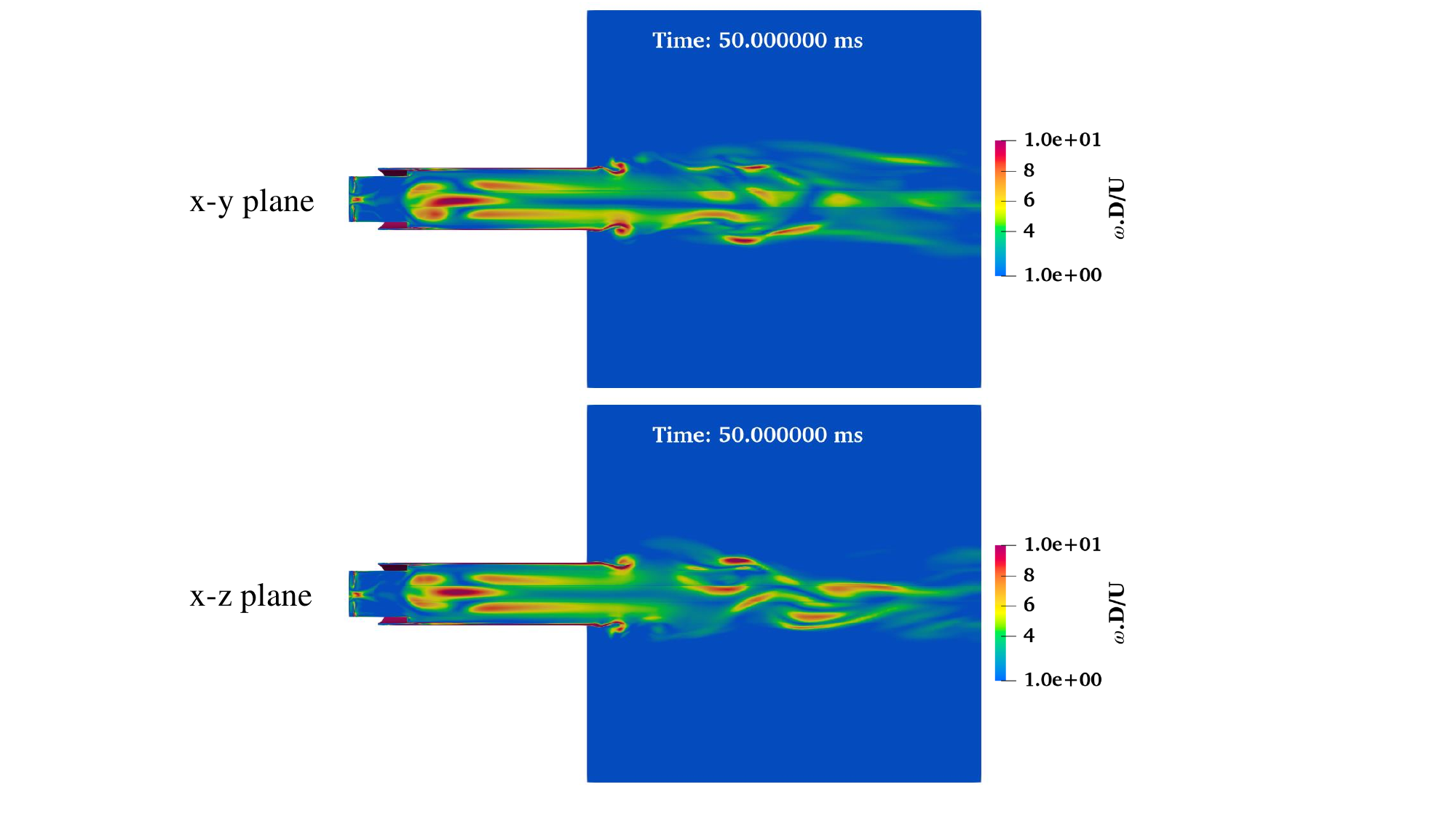}} \\

\hspace{-1cm}
\subfloat[][]{\includegraphics[trim={3cm 0 6cm 0},clip,scale=0.35]{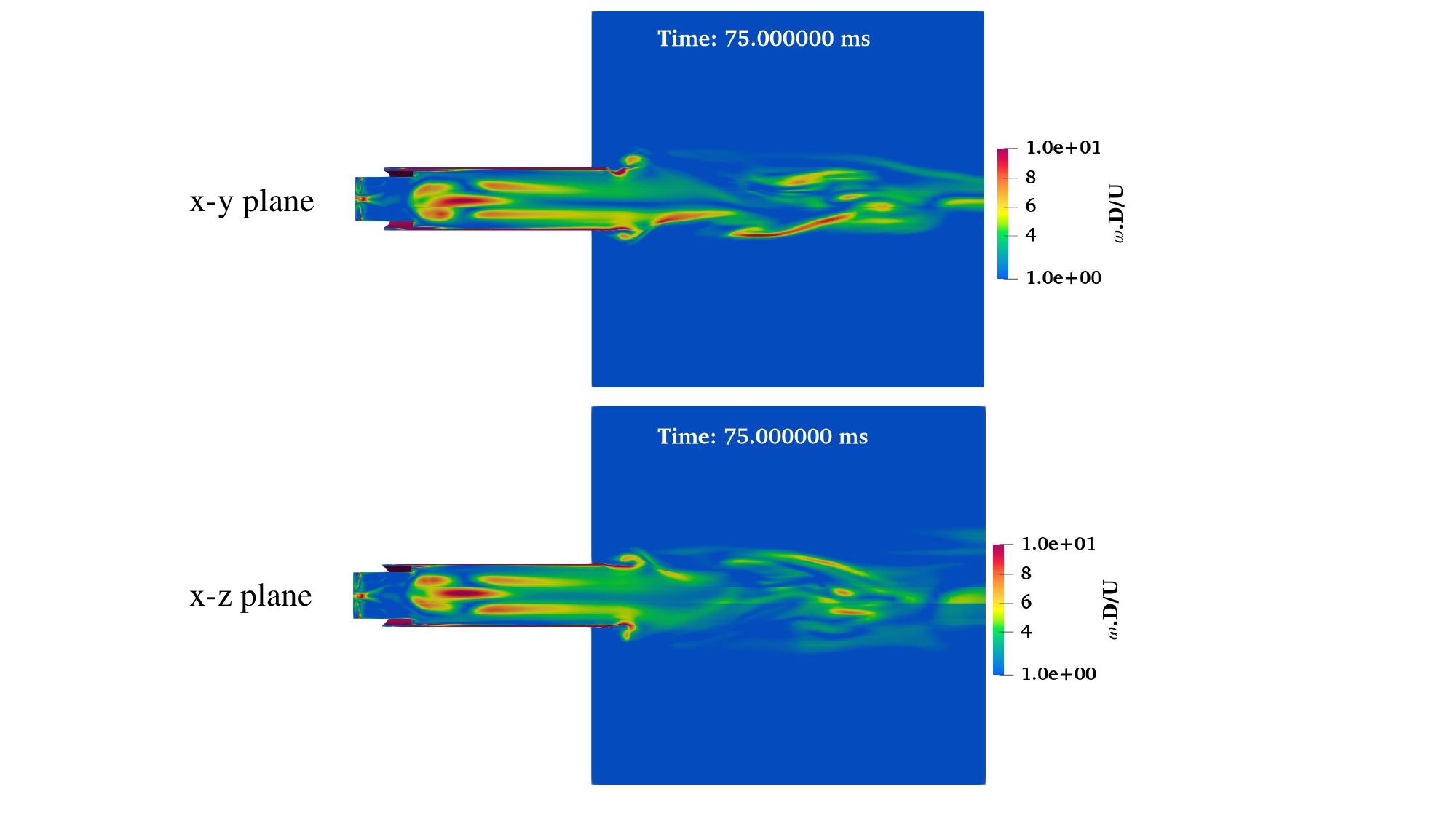}}
\subfloat[][]{\includegraphics[trim={3cm 0 6cm 0},clip,scale=0.35]{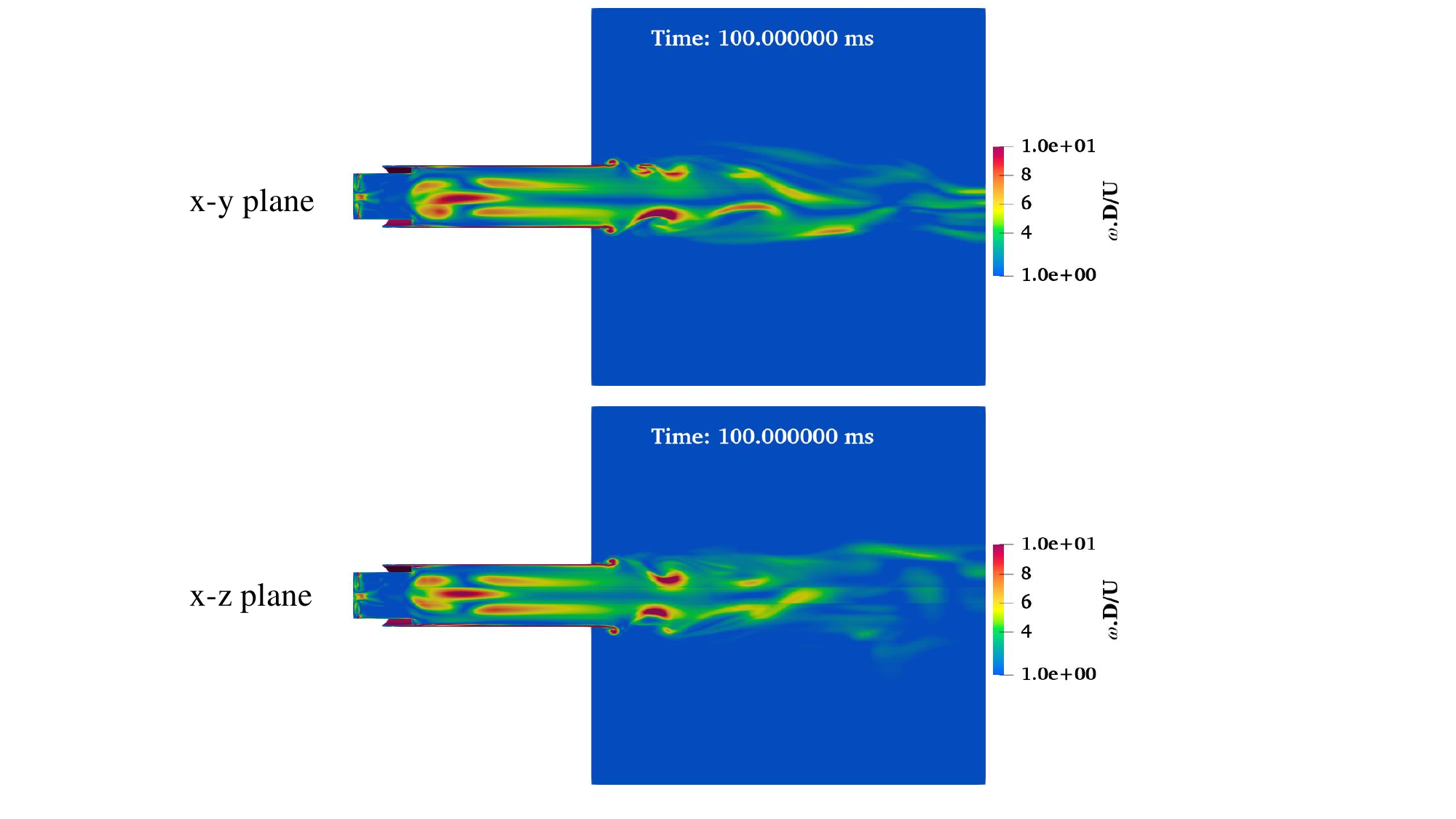}}
\caption{Vorticity contours (normalized by the torch diameter of \SI{0.1}{\meter} and a velocity of \SI{100}{\meter/\second}) across the $x$-$y$ and $x$-$z$ planes at different time instances. ($p_{\mathrm{a}} = \SI{10}{\kilo\pascal}$, $P = \SI{100}{\kilo\watt}$, $\eta = 57.5\%$).} 
\label{fig:W_contours_dt}
\end{figure}

\subsubsection{\label{sec:torch_chamber_lte} High-pressure case}
A time-accurate three-dimensional LTE simulation is performed for the high-pressure case (\SI{10}{\kilo\pascal}, \SI{100}{\kilo\watt}, $\eta = 57.5\%$). For this operating condition, the plasma jet in the chamber region is reported to be highly unsteady \cite{capponi2024multi}, and thus necessitates a time-accurate simulation to capture its nature. The simulation is conducted using Crank-Nicolson time integration with dual-time stepping. The global time step selected for the simulation is \SI{1}{\micro s}, with the coupling between the CFD and the EM solvers occurring every \SI{100}{\micro s} (as mentioned in \cref{sec:numerical_framework}). This coupling frequency is sufficient, as the plasma in the torch is nearly steady, and therefore, the plasma distribution (and hence the electrical conductivity distribution) does not undergo significant changes within this time span in the region where the coupling takes place. Moreover, as discussed later in this section, the dominant frequencies associated with the flow structures are on the order of several milliseconds. Hence, small unsteadiness in the torch (if any) can be well resolved by the coupling frequency chosen in this study. The integration is stopped after \SI{100}{\milli\second}, which is found to be sufficient for the flow to become statistically stationary.

\begin{enumerate}[left=0pt,label=\Alph*.]
  \item \textbf{Qualitative assessment of the flow field:}
  \cref{fig:T_iso_dt} shows the temperature iso-volumes ranging from \SI{2000}{K} to \SI{10000}{K} at various time levels, demonstrating the highly unsteady and three-dimensional nature of the flow. The plasma in the torch region, although non-axisymmetric, is found to be steady. However, the non-axisymmetric profiles at the nozzle outlet, generated by the helical coils, disrupt the symmetry of the plasma jet. Additionally, shear instabilities between the hot plasma core and the surrounding cold gas cause entrainment of the colder gas into the core, resulting in a highly unstable and unsteady jet. To provide a clear visualization of the coherent structures emerging from the jet, \cref{fig:W_contours_dt} illustrates the vorticity contours, normalized by the torch diameter (\SI{0.1}{\meter}) and a velocity of \SI{100}{\meter/\second}, across the $x$-$y$ and $x$-$z$ planes at different time instances. In the vicinity of the nozzle exit (\emph{i.e.}, at the jet inlet), prominent ring-shaped vortex structures are observed, entraining cold ambient gas into the jet core. This behavior is characteristic of the Kelvin-Helmholtz (KH) instability, which arises from the velocity shear between the high-speed jet and the quiescent surrounding medium. Further downstream, the vorticity field exhibits noticeable distortions and undulatory patterns, indicative of vortex interactions and pairing processes that give rise to elongated, stretched vortical formations. Notably, the vorticity field displays a clear, repeating pattern with an orderly downstream progression, wherein the majority of vortices maintain a consistent size. This organized arrangement, coupled with the absence of a broad range of vortex scales, indicates that while the jet displays unsteady behavior, it has not yet fully transitioned to a turbulent state. Additionally, the vorticity contours in the $x$-$y$ and $x$-$z$ planes exhibit significant differences, further highlighting the three-dimensional structure of the jet. In contrast, akin to the temperature contours, the vorticity contours in the torch-nozzle system remain unchanged with time, confirming the steady-state nature of the plasma in that region. 
  \\
  \item \textbf{Modal analysis of the unsteady plasma jet:} 
  To identify the dominant modal structures in the plasma jet, an SPOD (Spectral Proper Orthogonal Decomposition) was performed on the unsteady plasma field in the jet region. SPOD is a linear
data-driven dimensionality reduction approach that combines the advantages of more classic alternatives, such as the POD \cite{lumey2012stochastic,sirovich1987turbulence} and the frequency-based Dynamic Mode Decomposition (DMD) \cite{schmid2010dynamic}. POD identifies the most energetic modes, ensuring that coherent structures with high energy content are effectively captured by its basis functions. However, a key limitation of POD is that its modes can encompass the full spectrum of available frequencies. In contrast, DMD extracts modes as the eigenvectors of the optimal linear dynamical system that approximates the data in a least-squares sense. A drawback of DMD, however, is its slow convergence, as each mode remains strictly harmonic. SPOD effectively isolates spatially and temporally coherent structures, which are either hidden in stochastic turbulent fluctuations or spread over a wide frequency range.

This work uses the SPOD algorithm outlined in \cite{towne2018spectral}. The temperature field fluctuation ($T^{\prime}(\boldsymbol{x}, t) = T(\boldsymbol{x}, t) - \bar{T}(\boldsymbol{x})$) was chosen for the SPOD analysis as the plasma intensity measured during experiments strongly correlates with the temperature field. This makes the temperature field a suitable candidate for the analysis of the plasma jet for facilitating comparison against ICP experiments \cite{anfuso2021multiscale}. The temperature field snapshots were saved every \SI{0.1}{ms}, and the snapshots from \SI{20}{ms} to \SI{100}{ms} have been considered for the analysis. The first \SI{20}{ms} data has been truncated to remove any initial transients. The resulting total time duration for the analysis is \SI{80}{ms}, which gives 800 snapshots in time. As a result, the frequency resolution for the SPOD analysis is \SI{12.5}{Hz} ($\Delta f=1 /$ total time duration) and the maximum number of frequencies available for SPOD is half of the number of snapshots \emph{i.e.} 400, ranging from 0 to Nyquist frequency (5000 Hz in this case). The spatial domain for the analysis was restricted from the nozzle exit to the chamber exit in the axial direction, and 4 radii along the radial direction on either side of the axis. The frequency has been presented in terms of the non-dimensional Strouhal number defined as $St = fD_j/U_j$, where $D_j =$ \SI{0.1}{m} and $U_j =$ \SI{250}{m/s} are the jet diameter and the jet centerline mean velocity.

\cref{fig:energy_spectrum} presents the leading SPOD eigenvalues at each frequency, revealing multiple peaks within the frequency range below 500 Hz. This observation agrees qualitatively with the plasma light intensity short-time FFT data presented in \cite{capponi2024multi} (refer to Fig. 11) for the present operating conditions, where several peaks were reported for the frequency range of 100 to \SI{500}{Hz}. Based on the amplitude of the leading eigenvalues, three SPOD modes, numbered in red in \cref{fig:energy_spectrum}, have been selected for visualization. These modes correspond to Strouhal numbers 0.015, 0.045, and 0.1, respectively. \cref{fig:spod_contours} illustrates the leading SPOD modes at the specified frequencies across the x-y and x-z planes. All three modes exhibit a traveling wave pattern, maintaining their structural strength up to the chamber outlet. Mode 1, associated with the lowest frequencies, primarily features coherent structures within the shear layer, likely linked to periodic vortex shedding at low frequencies. In contrast, modes 2 and 3 predominantly display structures within the core of the plasma jet, emphasizing the periodic activation and deactivation of the entire plasma core due to large-scale entrainment of surrounding cold gas. However, near the nozzle exit, small-scale structures are concentrated in the shear layer, seemingly associated with the periodic ejection of ring vortices induced by the interaction between the hot plasma jet and the surrounding cold gas. Furthermore, the distinct shapes of the coherent structures in the x-y and x-z planes underscore the inherently three-dimensional nature of these modes.
\begin{figure}[hbt!]
\centering
\includegraphics[trim={3cm 3cm 3cm 3cm},clip,scale=0.4]{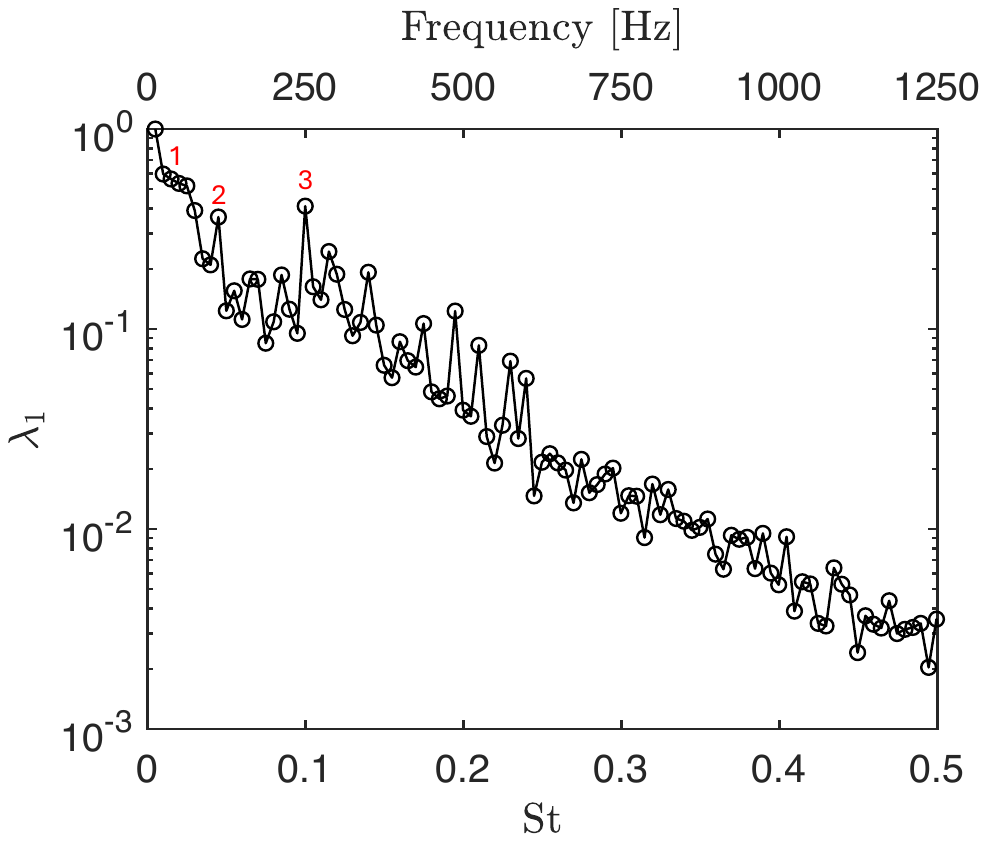}
\caption{Leading SPOD eigenvalues as a function of frequency, normalized by the magnitude of the largest eigenvalue. The SPOD modes numbered in red as 1, 2, and 3 have been considered for subsequent analysis, corresponding to Strouhal numbers 0.015, 0.045, and 0.1, respectively.}
\label{fig:energy_spectrum}
\end{figure}

\begin{figure}[!htb]
\hspace{3cm}
\subfloat{\includegraphics[scale=0.2]{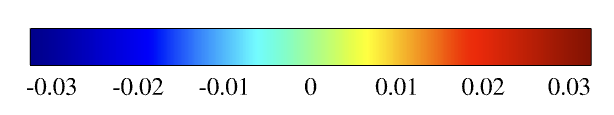}}\\
\setcounter{subfigure}{0}
\centering
\subfloat[][$St = 0.015$]{\includegraphics[trim={10cm 0 10cm 0},clip,scale=0.4]{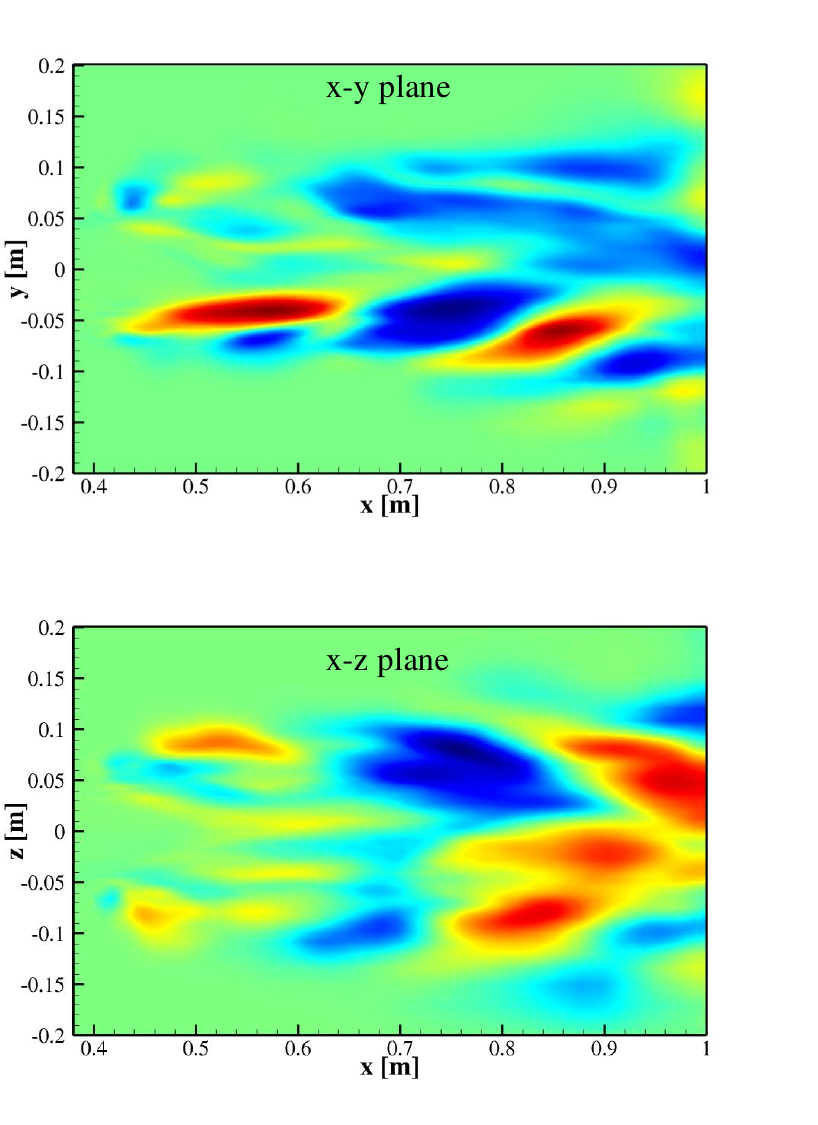}}
\subfloat[][$St = 0.045$]{\includegraphics[trim={10cm 0 10cm 0},clip,scale=0.4]{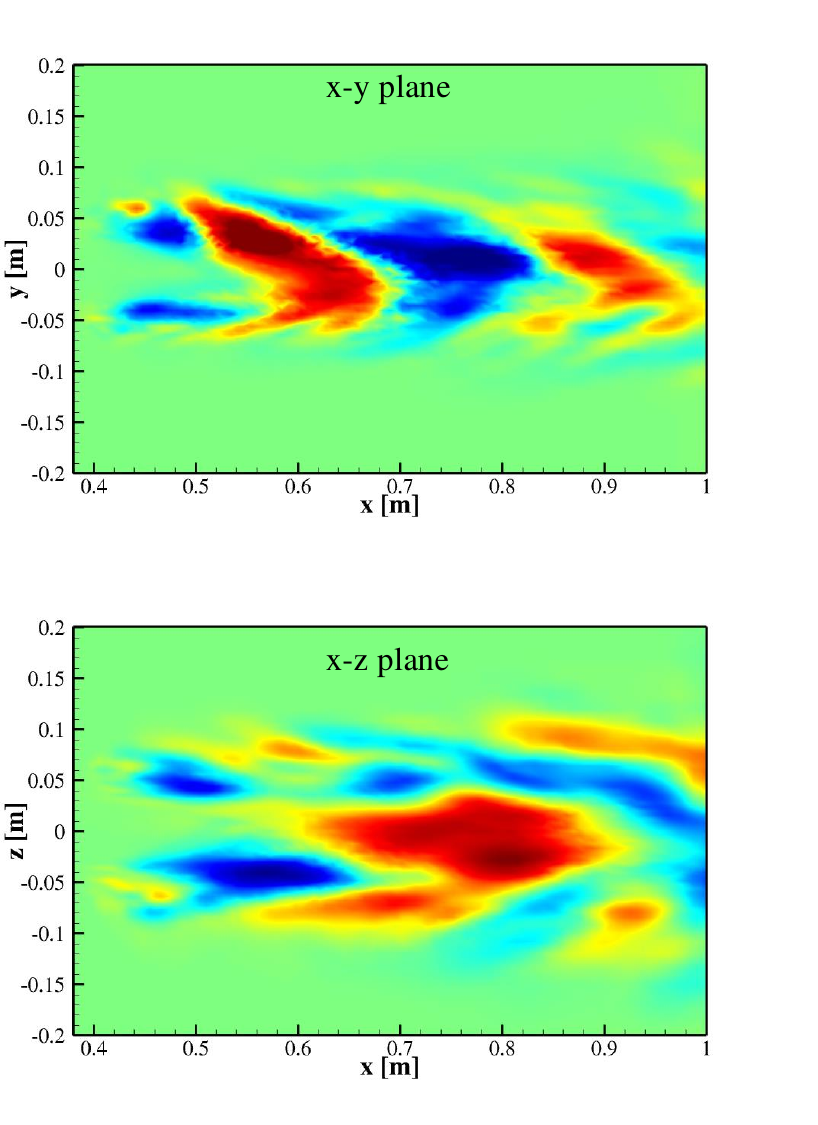}}
\subfloat[][$St = 0.1$]{\includegraphics[trim={10cm 0 10cm 0},clip,scale=0.4]{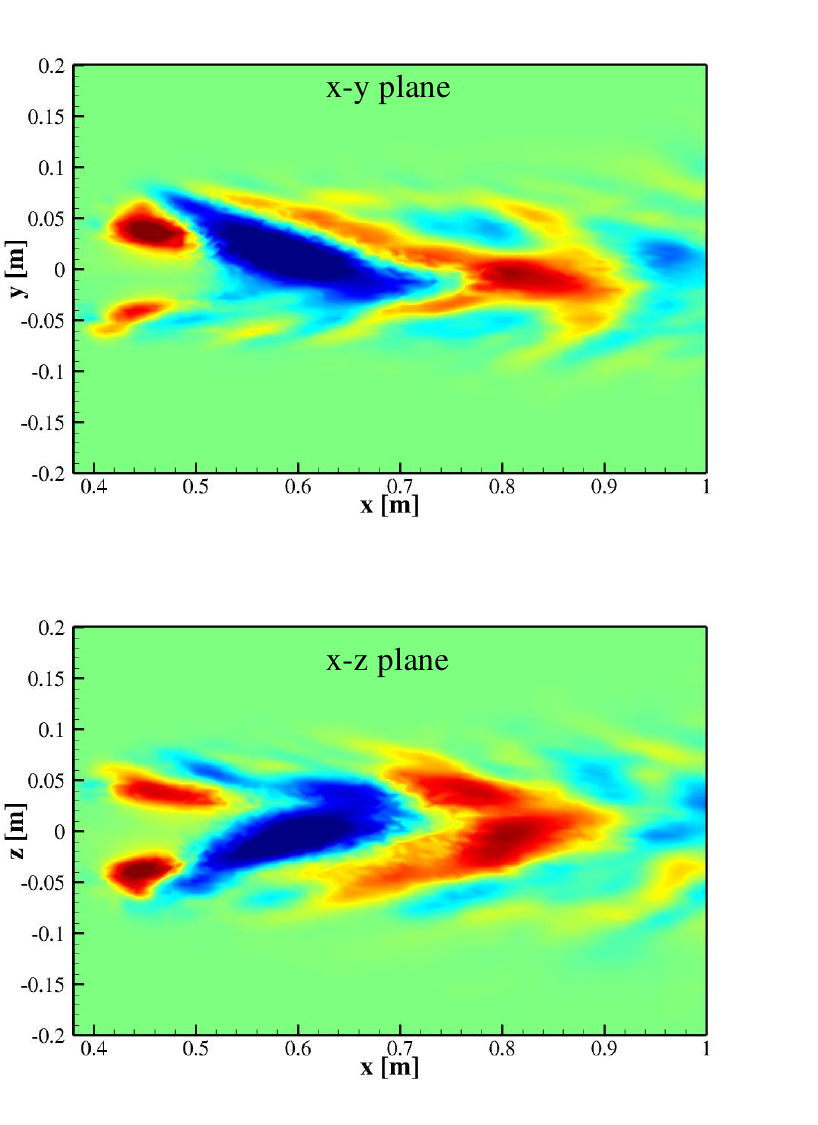}}
\caption{Dominant SPOD modes at the three indicated frequencies across the $x$-$y$ and $x$-$z$ planes. ($p_{\mathrm{a}} = \SI{10}{\kilo\pascal}$, $P = \SI{100}{\kilo\watt}$, $\eta = 57.5\%$).} 
\label{fig:spod_contours}
\end{figure}

  \item \textbf{Analysis of the mean flow field:} 
    The above observations have to be taken into account for a correct interpretation of TPM testing, which may span several seconds. Due to the unsteadiness of the jet, which is on the order of milliseconds, the sample is essentially subjected to a time-averaged flow as far as measurements of quantities of interest are concerned (\emph{e.g.}, heat flux). Therefore, it is more insightful to analyze the mean (\emph{e.g.}, time-averaged) plasma field within the chamber. \cref{fig:Mean_contours} presents the mean plasma temperature and velocity distribution across the $x$-$y$ and $x$-$z$ planes, emphasizing the asymmetric distribution in both the torch-nozzle system and chamber region. This asymmetry is quantified in \cref{fig:mean_profiles_chamber}, where the temperature and axial velocity profiles along the radius are shown for the $x$-$y$ and $x$-$z$ planes. The profiles are taken at axial locations of $x = \SI{0.425}{m}, \, \SI{0.475}{m}$ and \SI{0.525}{m}  (\emph{i.e.,} \SI{50}{\milli\meter}, \SI{100}{\milli\meter}, and \SI{150}{\milli\meter}, respectively, from the nozzle exit) corresponding to typical axial positions of the sample during testing. At all axial locations, both temperature and velocity profiles reveal significant discrepancies between the two orthogonal planes with a maximum relative deviation in temperature of 29\%, 18\%, and 25\% at $x = \SI{0.425}{m}, \, \SI{0.475}{m}$ and \SI{0.525}{m}, respectively and a maximum relative deviation in velocity of 17\%, 25\%, and 28\% at $x = \SI{0.425}{m}, \,  \SI{0.475}{m}$, and \SI{0.525}{m}, respectively. Notably, the profiles in the $x$-$y$ plane exhibit significantly greater asymmetry about the axis (\emph{i.e.}, $r = 0$) compared to those in the $x$-$z$ plane. Additionally, the temperature differences between the two planes are far more pronounced in the plasma jet region than in the torch-nozzle section. This is attributed to the highly three-dimensional nature of the jet instability, which further enhances axisymmetry, whereas the plasma remains essentially steady in the torch-nozzle section, with axisymmetry resulting only from the helical coils.
    
\begin{figure}[!htb]
\hspace{-1cm}
\subfloat[][]{\includegraphics[trim={3cm 0 6cm 0},clip,scale=0.35]{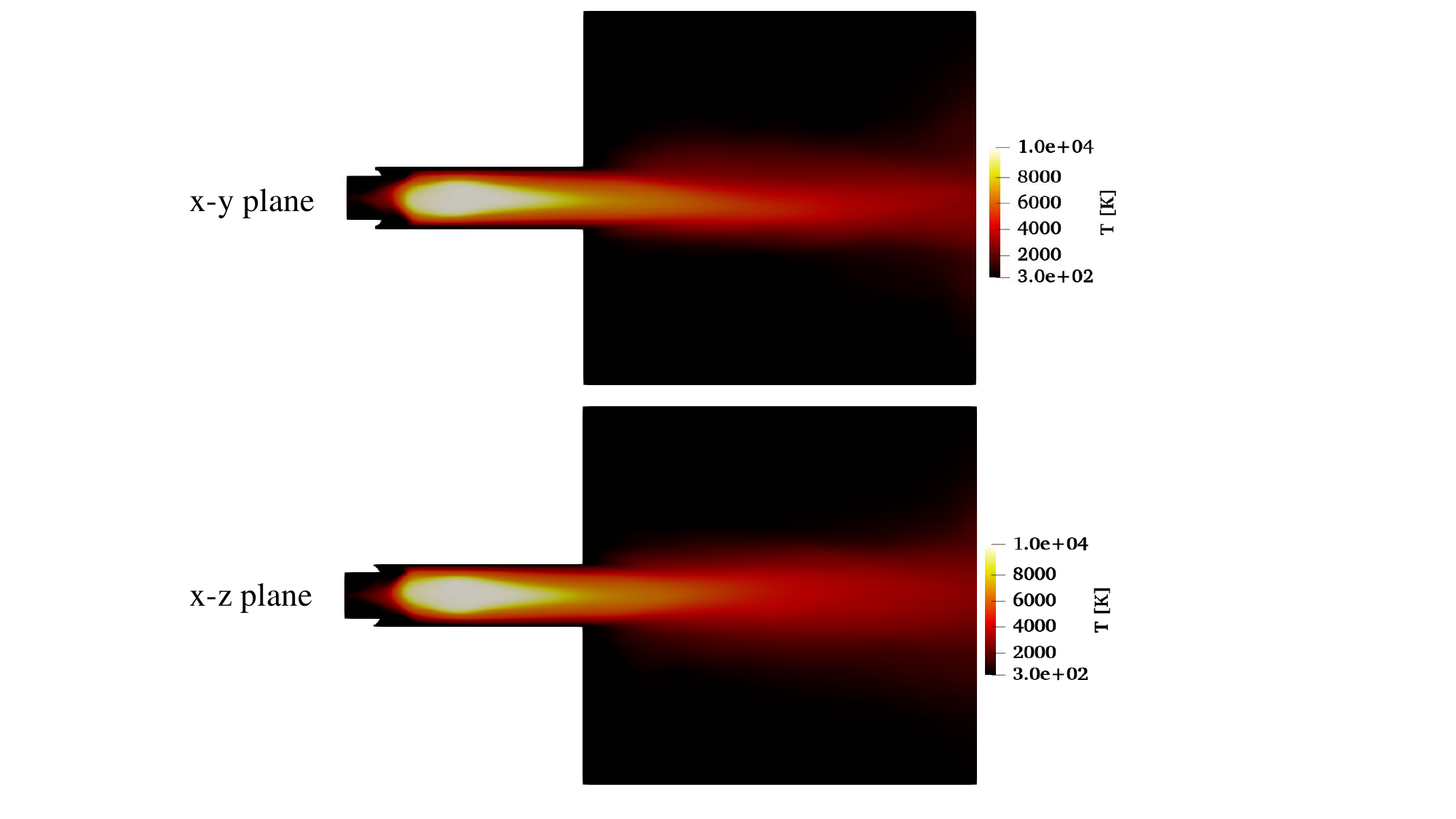}\label{fig:Mean_contoursa}}
\subfloat[][]{\includegraphics[trim={3cm 0 6cm 0},clip,scale=0.35]{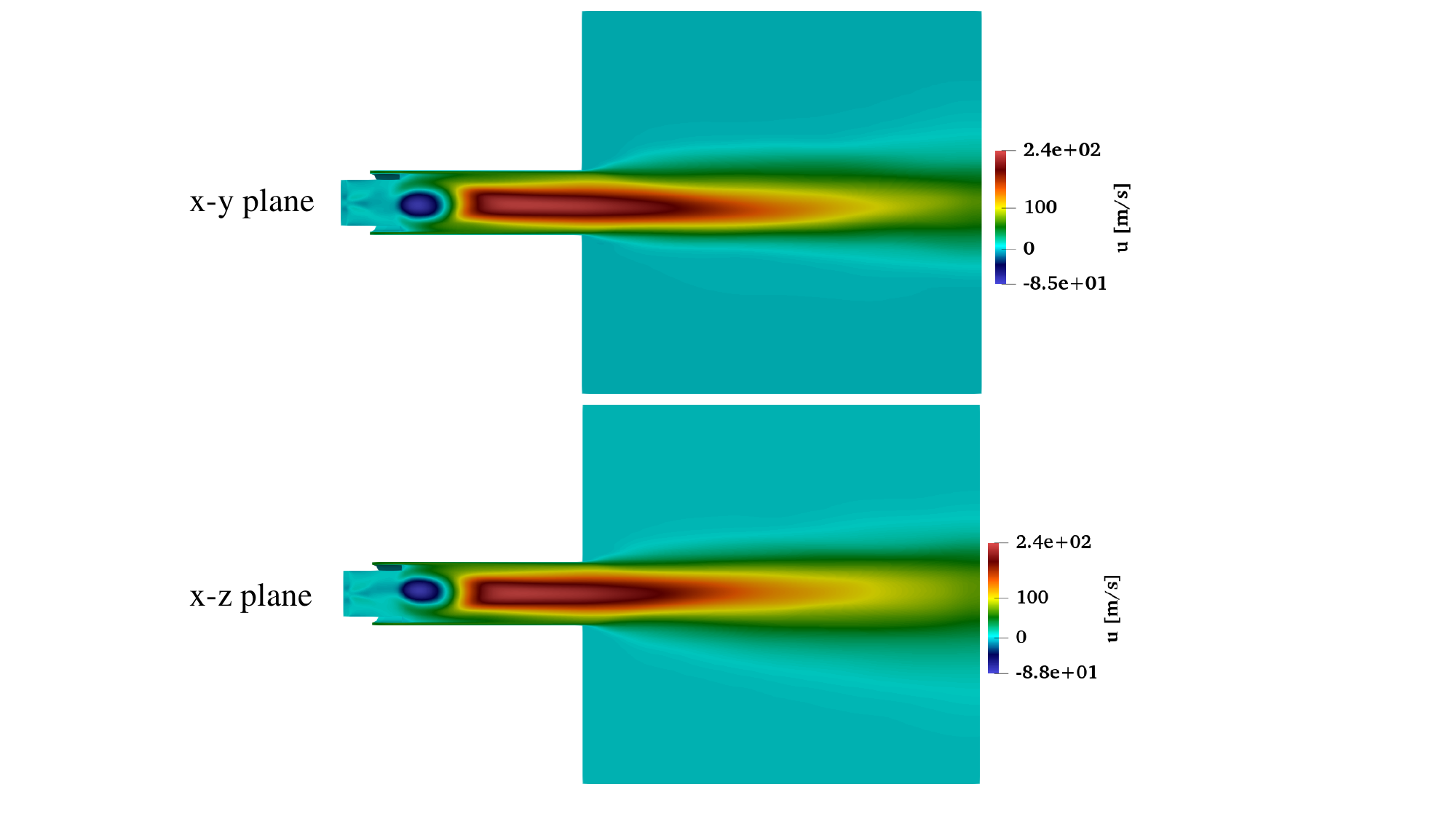}\label{fig:Mean_coutoursb}}
\caption{Mean plasma field across the $x$-$y$ and $x$-$z$ planes. \protect\subref{fig:Mean_contoursa} Temperature, and \protect\subref{fig:Mean_coutoursb} velocity. ($p_{\mathrm{a}} =  \SI{10}{\kilo\pascal}$, $P = \SI{100}{\kilo\watt}$, $\eta = 57.5\%$).} 
\label{fig:Mean_contours}
\end{figure}

\begin{figure}[hbt!]
\centering
\subfloat[][]{\includegraphics[scale=0.5]{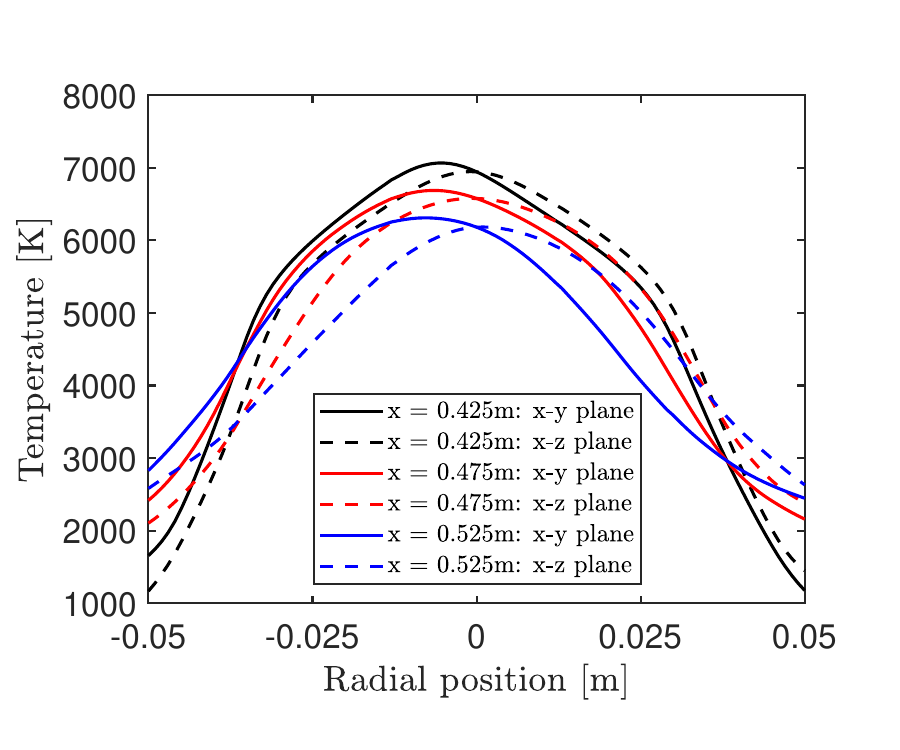}\label{fig:mean_profiles_chambera}} 
\subfloat[][]{\includegraphics[scale=0.5]{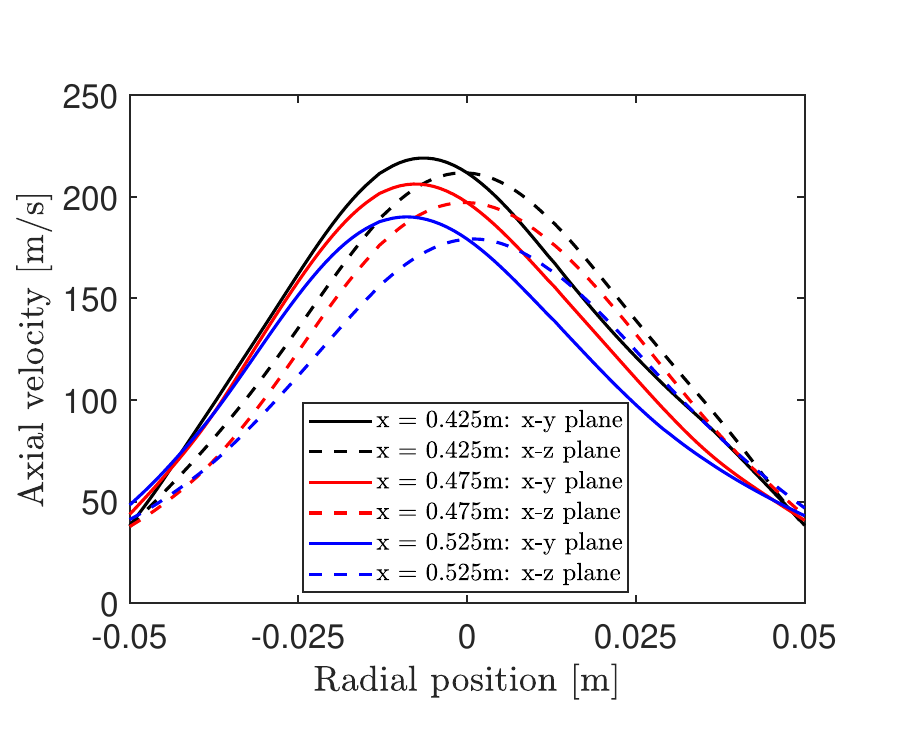}\label{fig:mean_profiles_chamberb}}
\caption{Radial profiles across the $x$-$y$ and $x$-$z$ planes at various axial locations ($x = \SI{0.425}{\meter}, \, \SI{0.475}{\meter}$ and \SI{0.525}{\meter}  \emph{i.e.} \SI{50}{\milli\meter}, \SI{100}{\milli\meter} and \SI{150}{\milli\meter}, respectively, from the nozzle exit). \protect\subref{fig:mean_profiles_chambera} Mean temperature, and \protect\subref{fig:mean_profiles_chamberb} mean axial velocity. ($p_{\mathrm{a}} = \SI{10}{\kilo\pascal}$, $P = \SI{100}{\kilo\watt}$, $\eta = 57.5\%$).} 
\label{fig:mean_profiles_chamber}
\end{figure}

    \item \textbf{Azimuthal FFT of the mean plasma jet:}
    To quantify the asymmetry in the plasma jet, azimuthal FFT (Fast Fourier Transform) was performed at the three previously specified axial locations. This method allows us to decompose the flow field into its azimuthal Fourier modes, helping to quantify asymmetry and identify dominant modes. Flow variables are extracted along a constant radius of \SI{25}{mm} (\emph{i.e.,} half of the torch radius which roughly circumscribes the hot plasma core) at evenly spaced azimuthal angles $\theta$ followed by performing an FFT of the flow variable along $\theta$. Given a flow property $f(\theta)$ (such as velocity or temperature) sampled along a circular contour at a fixed radius, the Fourier series expansion is: 
\begin{equation}
f(\theta)=\sum_{m=-\infty}^{\infty} \hat{f}_m e^{i m \theta}
\end{equation}
where, m is the azimuthal mode number ($m=0, \pm 1, \pm 2, \ldots$), $\hat{f}_m$ are the Fourier coefficients obtained from the azimuthal FFT, and $e^{i m \theta}$ represents the basis functions for periodic decomposition. Mode $m = 0$ corresponds to the axisymmetric mode. Mode $m = 1$ represents a dipole or single-lobed structure, commonly associated with vortex shedding. Mode $m = 2$ signifies a quadrupole or two-lobed pattern, such as vortex pairing. Mode $m = 3$ exhibits a triangular configuration, while higher-order modes ($m>3$) correspond to increasingly intricate structures, often linked to turbulence and fine-scale instabilities. For an axisymmetric flow, the only non-zero mode would be the $m=0$ mode, and all higher modes $(m=1,2,3, \ldots)$ would have zero amplitude. If higher azimuthal modes ($m \geq 1$) have non-zero amplitudes, the flow is non-axisymmetric, meaning variations around the azimuthal direction exist. Furthermore, the greater the relative magnitude of the amplitudes for modes $m \geq 1$ compared to the amplitude of the axisymmetric $m=0$ mode, the stronger the asymmetry. 

\cref{fig:LTE_FFT_amp} presents the amplitudes ($|\hat{f}_m|$) of the first three azimuthal modes (\emph{i.e,} $m=0,1,2$), normalized with respect to the amplitude of the $m=0$ mode, facilitating a clearer interpretation of asymmetry. For the mean temperature, the relative amplitude of the $m=1$ mode ranges from $2.74\%$ at x = \SI{0.425}{m} to $8.33\%$ at x = \SI{0.525}{m}, indicating a substantial asymmetry. For the axial velocity, the asymmetry is even more pronounced with the relative amplitude of the $m=1$ mode varying from $7.07\%$ at x = \SI{0.425}{m} to $11.29\%$ at x = \SI{0.525}{m}. The higher modes ($m \ge 2$) do not have a significant strength as compared to the axisymmetric ($m=0$) mode. Moreover, the relative amplitudes of the non-axisymmetric modes increase along the jet axis, signifying a growing asymmetry downstream in the chamber. This observation suggests that the TPM should be positioned as close as possible to the nozzle exit to minimize asymmetry effects during experiments. \label{sec:azimuthal_fft}
  
\end{enumerate}

\begin{figure}[hbt!]
\centering
\subfloat[][Mean temperature]{\includegraphics[scale=0.5]{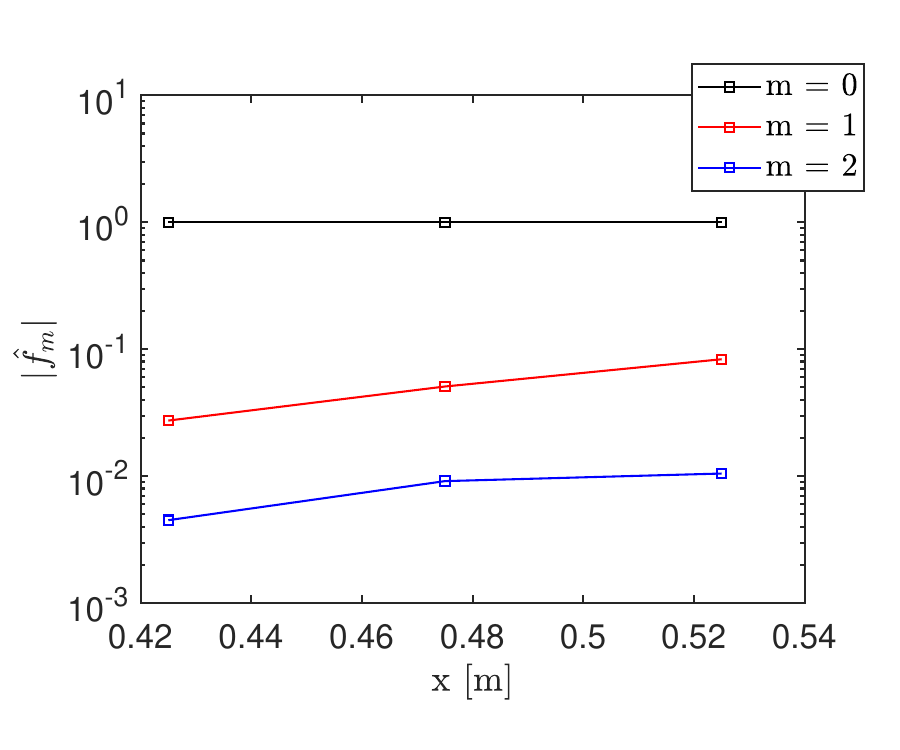}\label{fig:LTE_FFT_ampa}} 
\subfloat[][Mean axial velocity]{\includegraphics[scale=0.5]{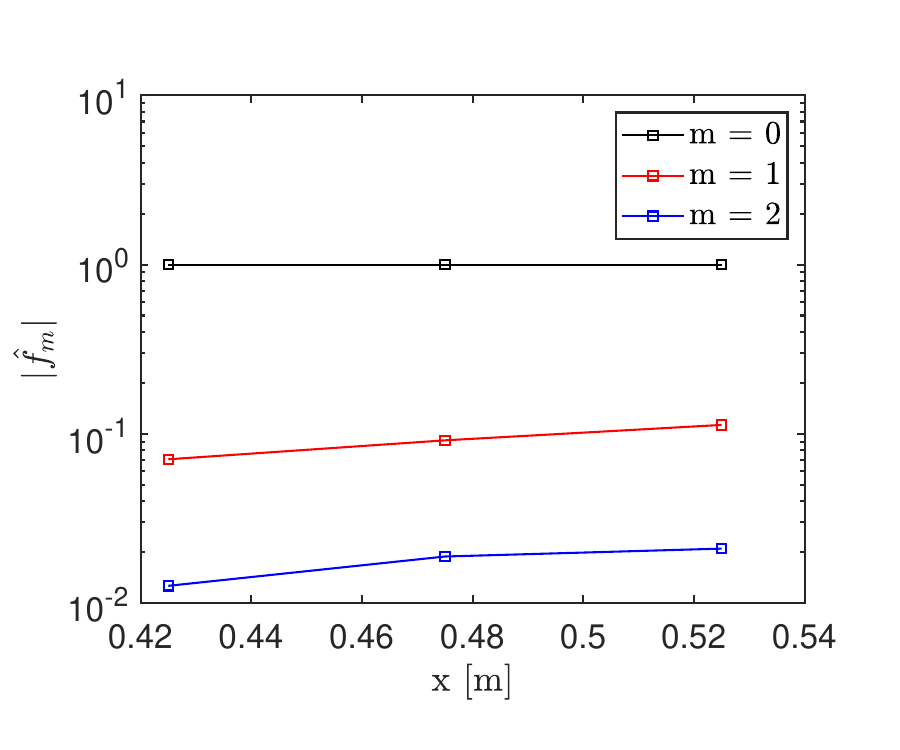}\label{fig:LTE_FFT_ampb}}
\caption{Normalized amplitudes of first three azimuthal modes at various axial locations ($x = \SI{0.425}{\meter}, \, \SI{0.475}{\meter}$ and \SI{0.525}{\meter}  \emph{i.e.} \SI{50}{\milli\meter}, \SI{100}{\milli\meter} and \SI{150}{\milli\meter}, respectively, from the nozzle exit) obtained from the azimuthal FFT along a constant radius of \SI{25}{mm}. ($p_{\mathrm{a}} = \SI{10}{\kilo\pascal}$, $P = \SI{100}{\kilo\watt}$, $\eta = 57.5\%$).} 
\label{fig:LTE_FFT_amp}
\end{figure}

It is to be emphasized that under LTE conditions with negligible demixing \cite{panesi2007analysis}, the plasma composition depends solely on two thermodynamic variables (\emph{e.g.}, temperature, and pressure). Given that the pressure within the facility remains nearly constant, the plasma composition becomes a function of the sole temperature. Consequently, the concentration profiles of the various species will reflect the asymmetric behavior observed for the temperature. Non-axisymmetric distributions of plasma composition, temperature, and velocity, along with unsteadiness, can lead to uneven heating of the sample and affect the key measured quantities such as the time-averaged heat flux, material recession rate, \emph{etc}. In light of this, three-dimensional time-accurate simulations of ICP facilities become crucial for accurate predictions of the TPM behavior and response during testing under these conditions.

\begin{figure}[!htb]
\hspace{-1cm}
\subfloat[][]{\includegraphics[trim={3cm 0 6cm 0},clip,scale=0.35]{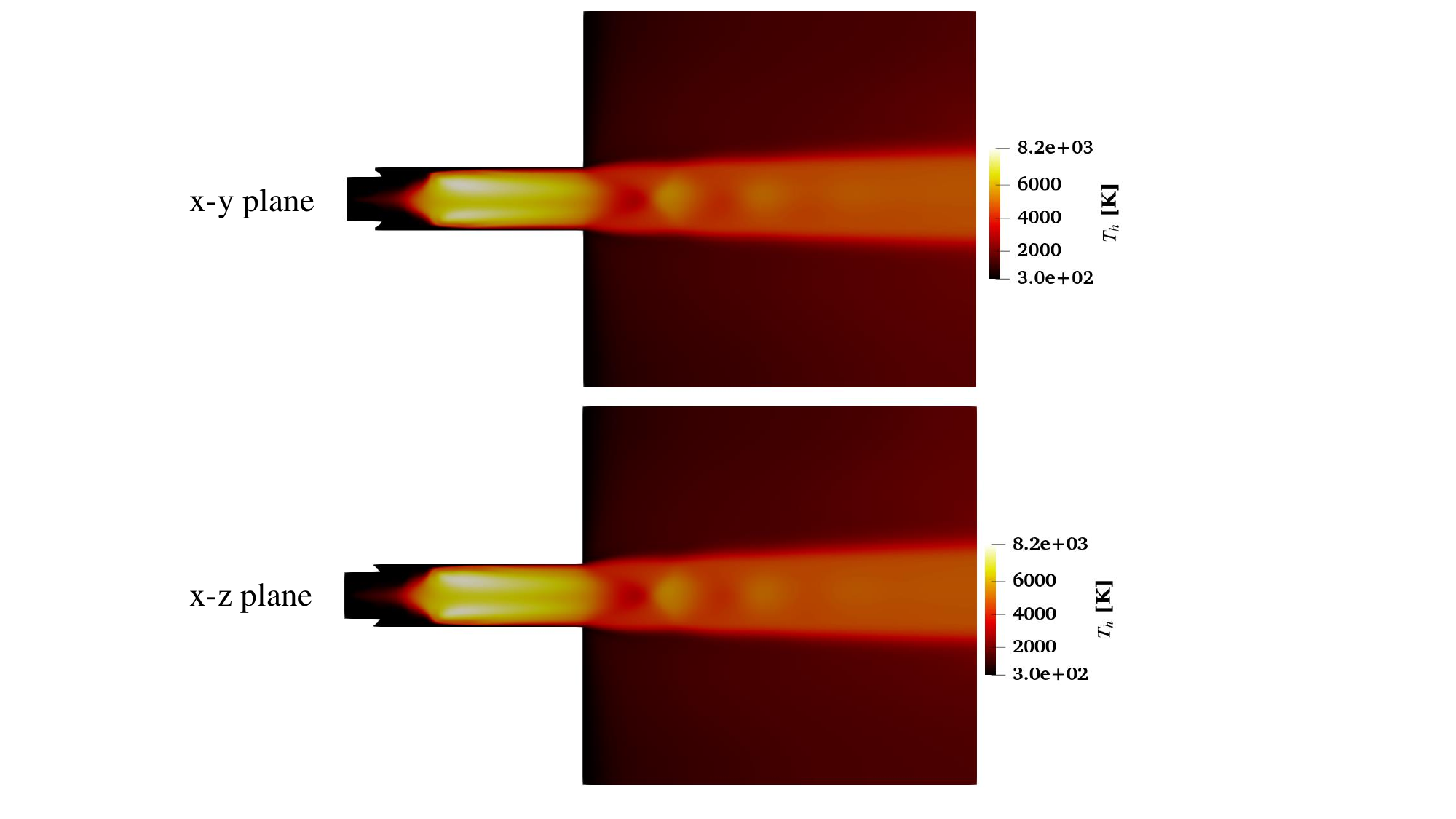}\label{fig:nlte_contoursa}}
\subfloat[][]{\includegraphics[trim={3cm 0 6cm 0},clip,scale=0.35]{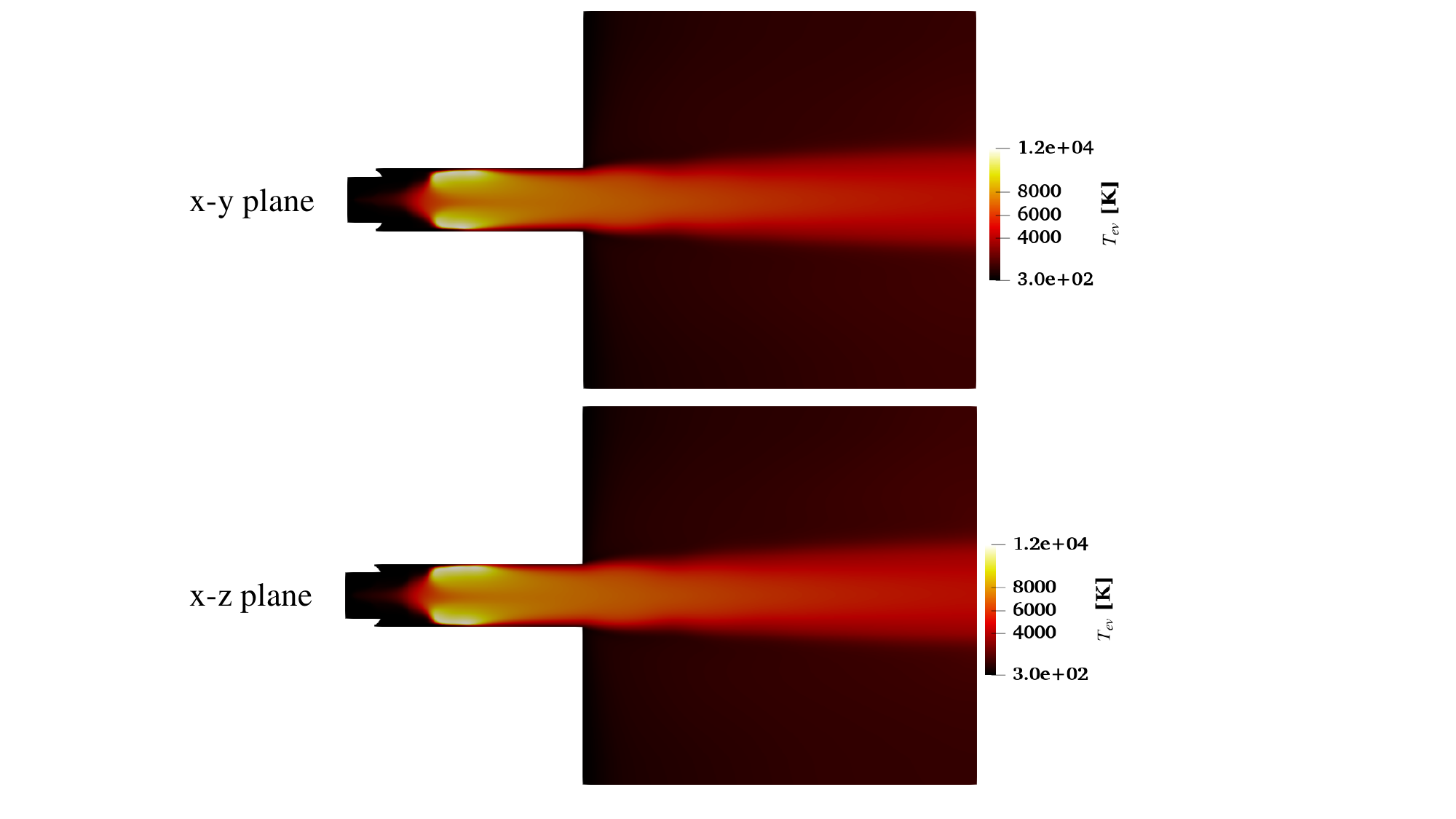}\label{fig:nlte_contoursb}} \\

\hspace{-1cm}
\subfloat[][]{\includegraphics[trim={3cm 0 6cm 0},clip,scale=0.35]{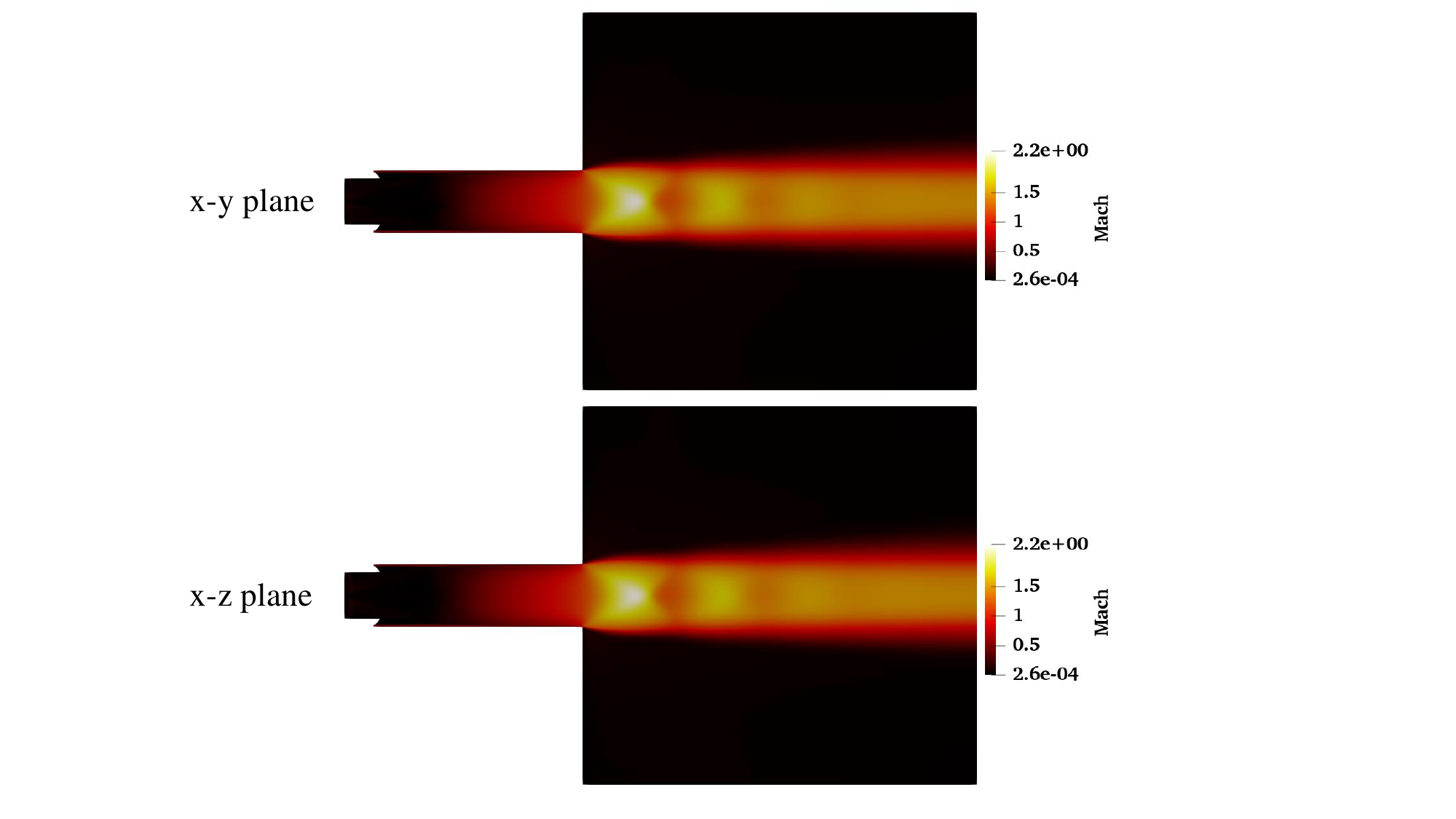}\label{fig:nlte_contoursc}}
\subfloat[][]{\includegraphics[trim={3cm 0 6cm 0},clip,scale=0.35]{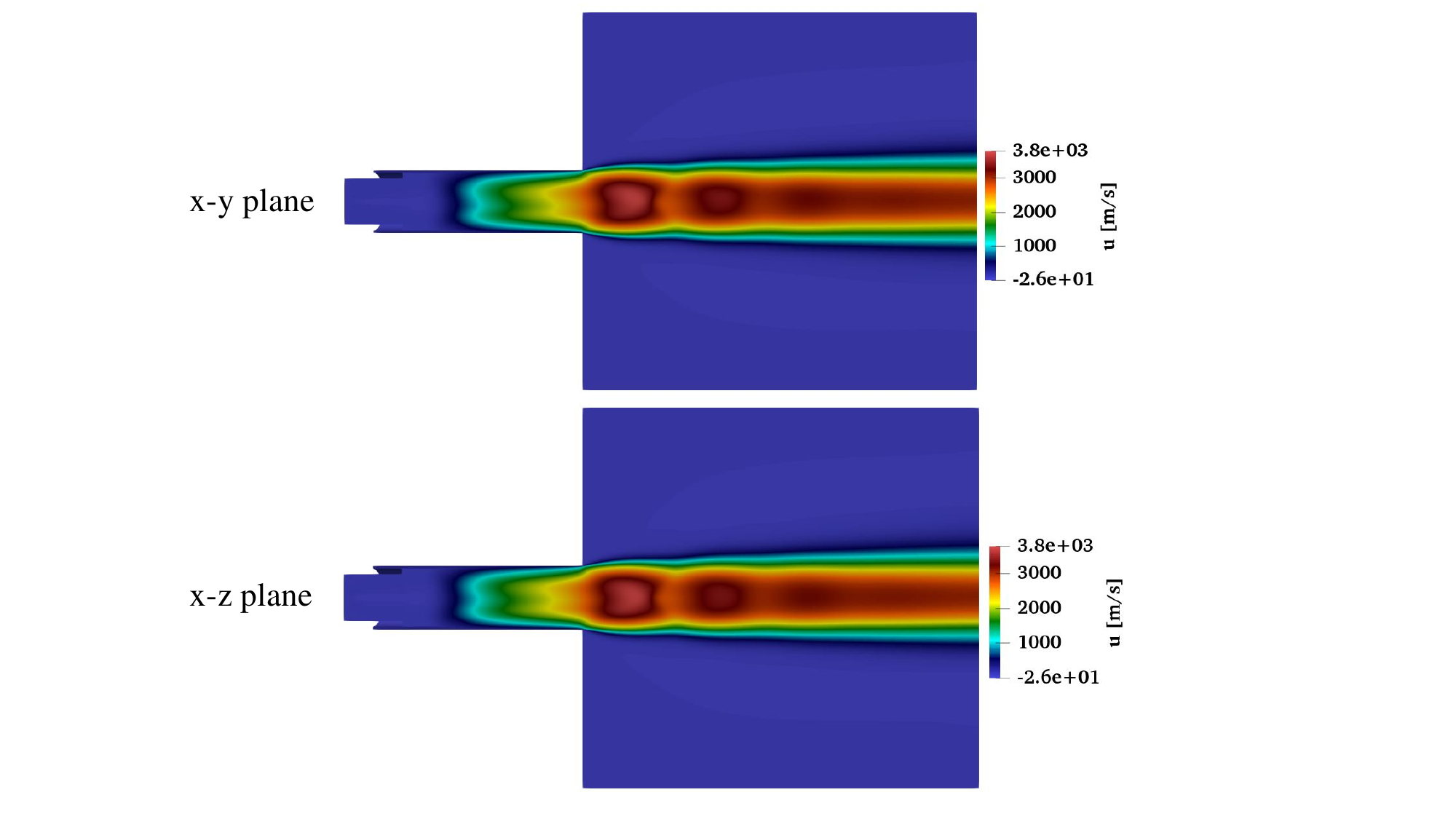}\label{fig:nlte_contoursd}}
\caption{Distributions of flow quantities across the $x$-$y$ and $x$-$z$ planes. \protect\subref{fig:nlte_contoursa} Heavy-particle temperature, \protect\subref{fig:nlte_contoursb} vibronic temperature, \protect\subref{fig:nlte_contoursc} Mach number, and \protect\subref{fig:nlte_contoursd} axial velocity. ($p_{\mathrm{a}} = \SI{590}{\pascal}$, $P = \SI{300}{\kilo\watt}$, $\eta = 53.9\%$).} 
\label{fig:nlte_contours}
\end{figure}

\begin{figure}[hbt!]
\centering
\subfloat[][]{\includegraphics[scale=0.45]{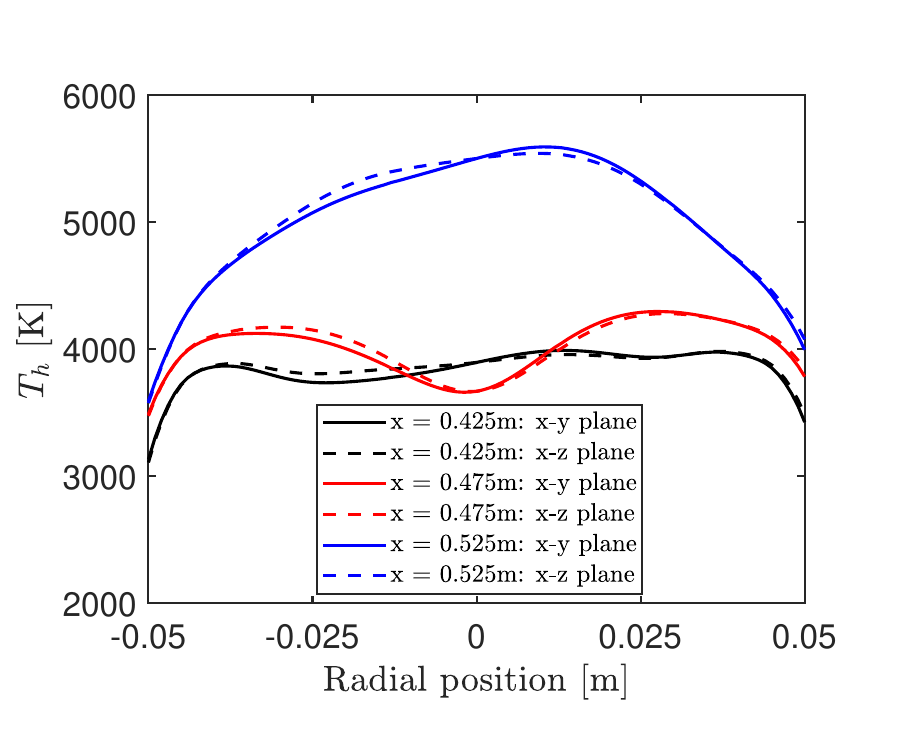}\label{fig:nlte_flow_profilesa}} 
\subfloat[][]{\includegraphics[scale=0.45]{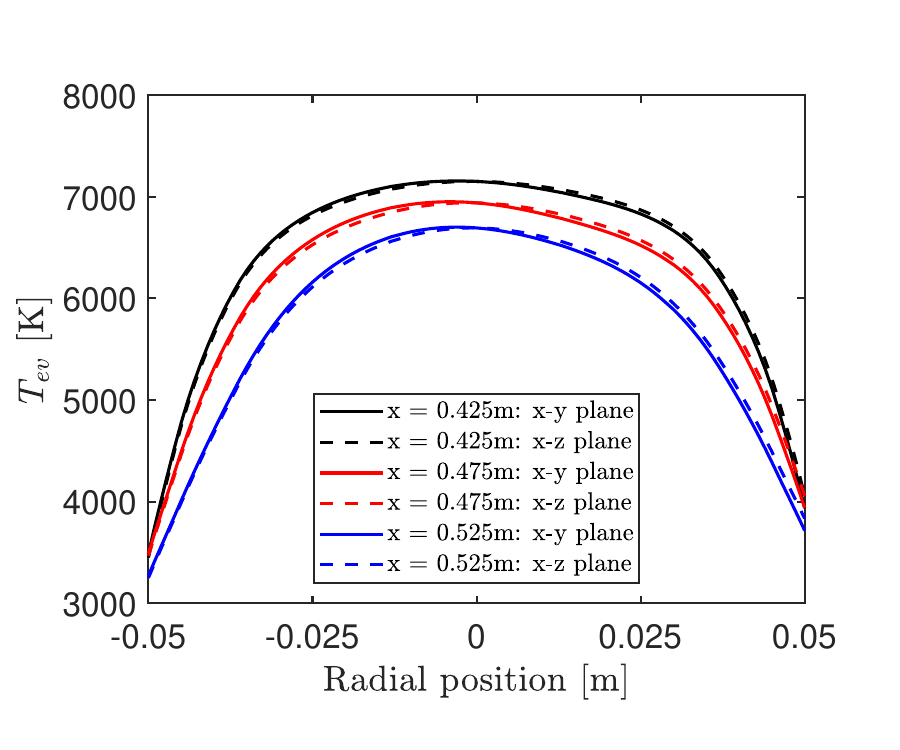}\label{fig:nlte_flow_profilesb}}
\\
\subfloat[][]{\includegraphics[scale=0.45]{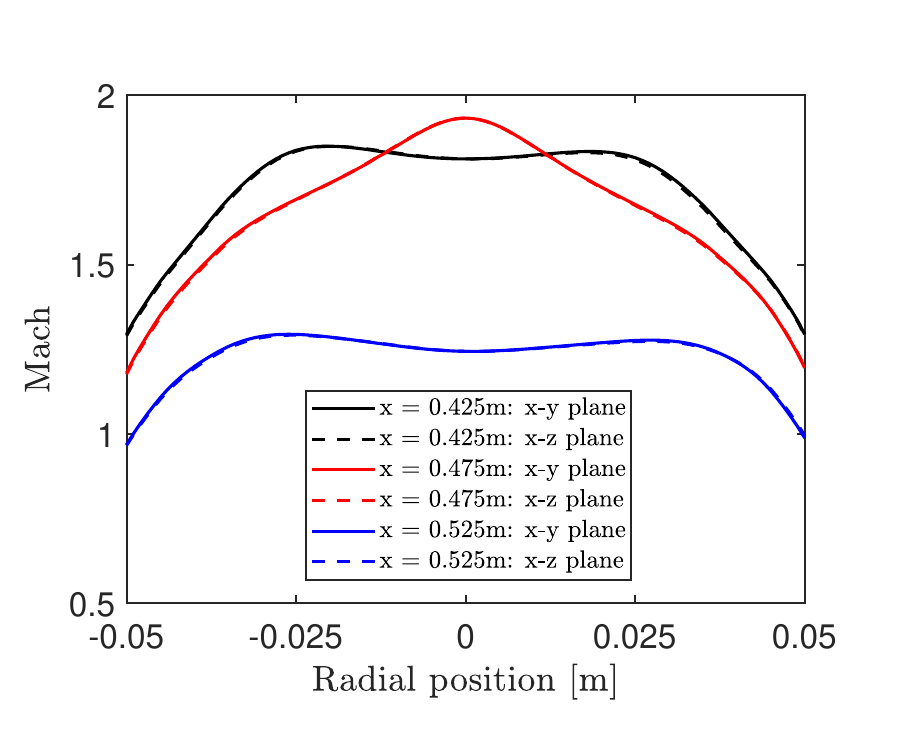}\label{fig:nlte_flow_profilesc}} 
\subfloat[][]{\includegraphics[scale=0.45]{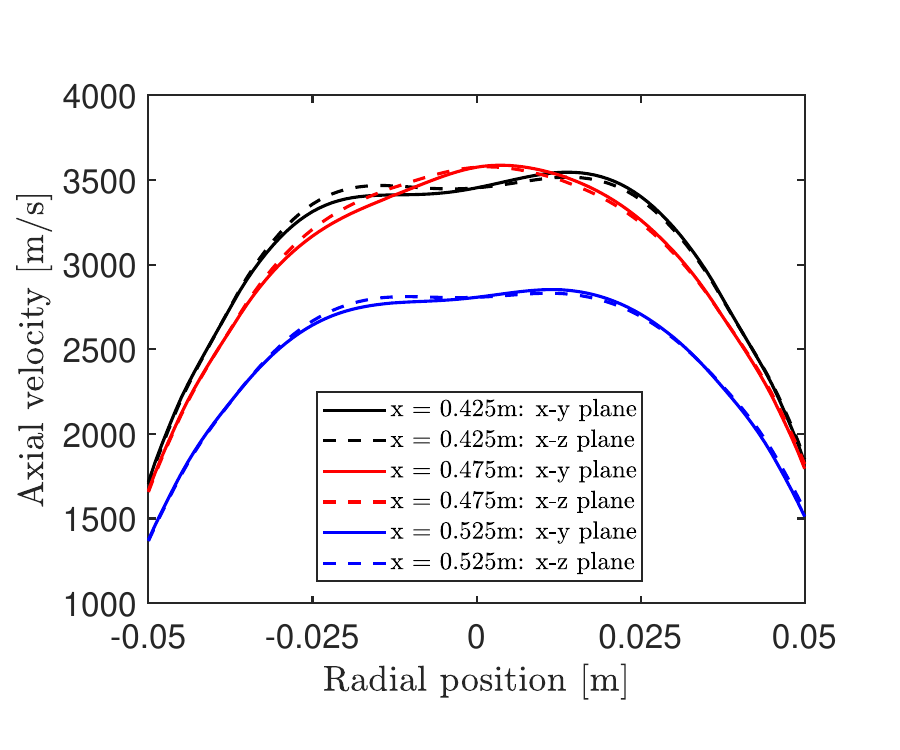}\label{fig:nlte_flow_profilesd}}
\caption{Radial profiles of flow quantities across the $x$-$y$ and $x$-$z$ planes at various axial locations ($x = \SI{0.425}{m}, \, \SI{0.475}{m}$ and \SI{0.525}{m} \emph{i.e.} \SI{50}{mm}, \SI{100}{mm} and \SI{150}{mm}, respectively, from the nozzle exit). \protect\subref{fig:nlte_flow_profilesa} Heavy-particle temperature, \protect\subref{fig:nlte_flow_profilesb} vibronic temperature, \protect\subref{fig:nlte_flow_profilesc} Mach number, and \protect\subref{fig:nlte_flow_profilesd} axial velocity. ($p_{\mathrm{a}} = \SI{590}{\pascal}$, $P = \SI{300}{\kilo\watt}$, $\eta = 53.9\%$).} 
\label{fig:nlte_flow_profiles}
\end{figure}

\begin{figure}[hbt!]
\centering
\subfloat[][Heavy-particle temperature]{\includegraphics[scale=0.45]{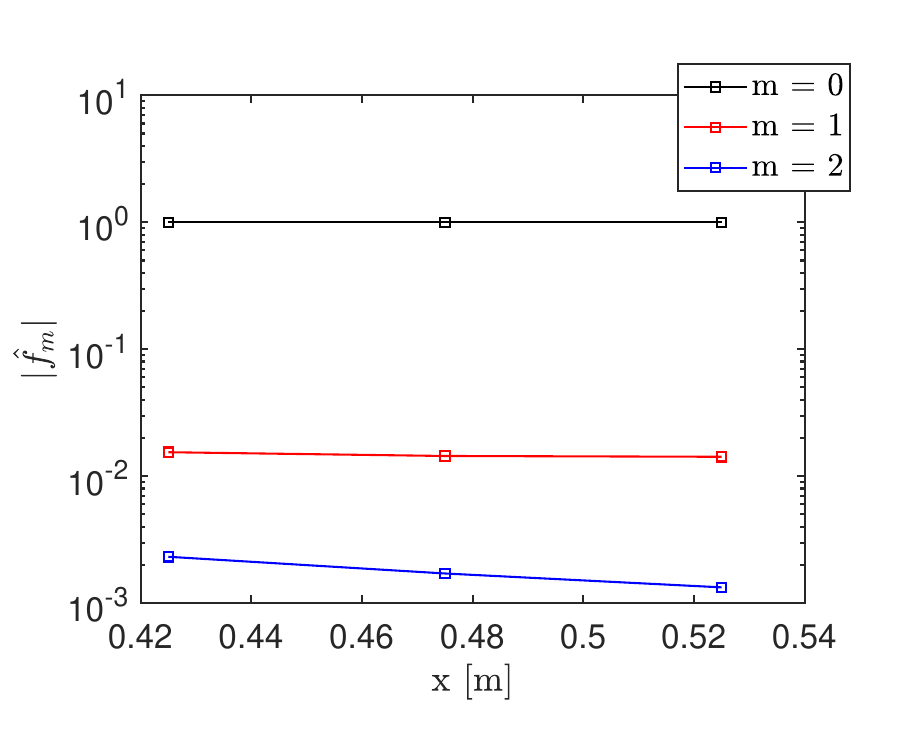}\label{fig:nlte_azimuthal_ampa}} 
\subfloat[][Vibronic temperature]{\includegraphics[scale=0.45]{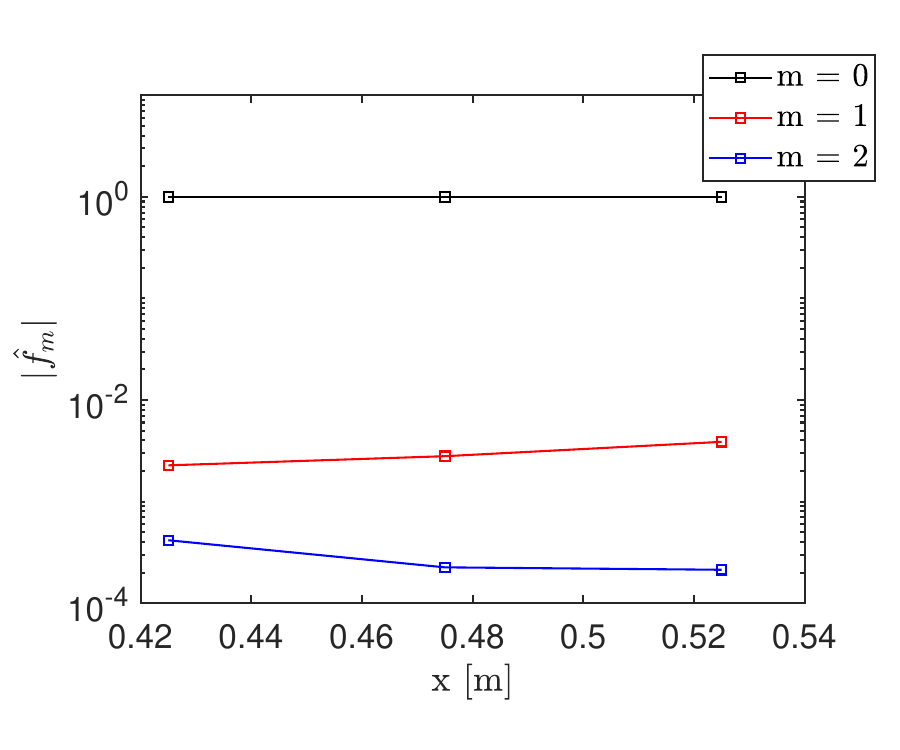}\label{fig:nlte_azimuthal_ampb}}
\\
\subfloat[][Mach number]{\includegraphics[scale=0.45]{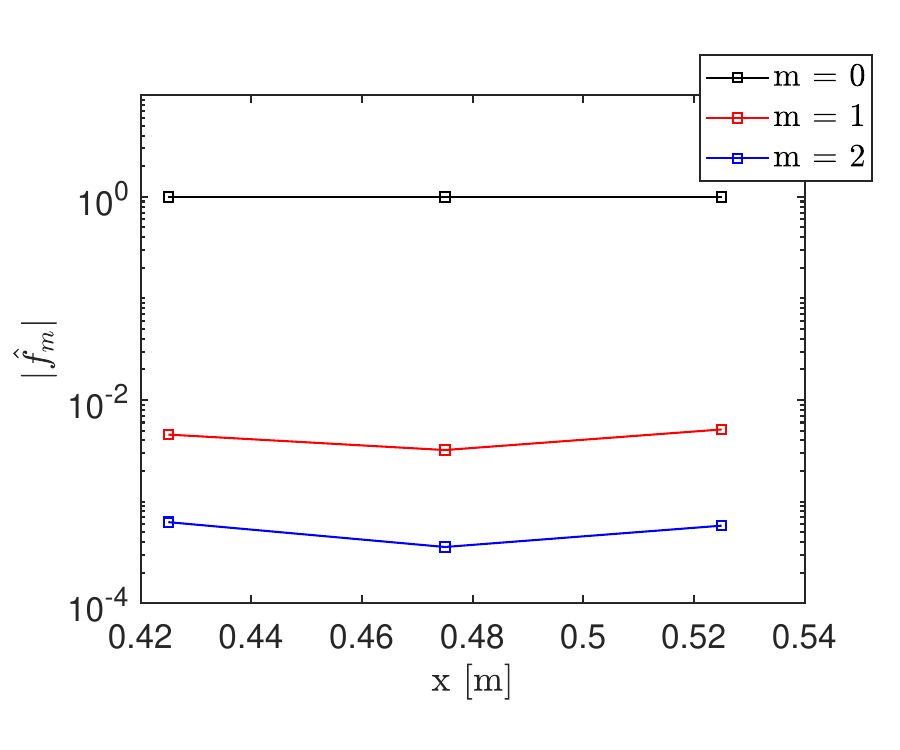}\label{fig:nlte_azimuthal_ampc}} 
\subfloat[][Axial velocity]{\includegraphics[scale=0.45]{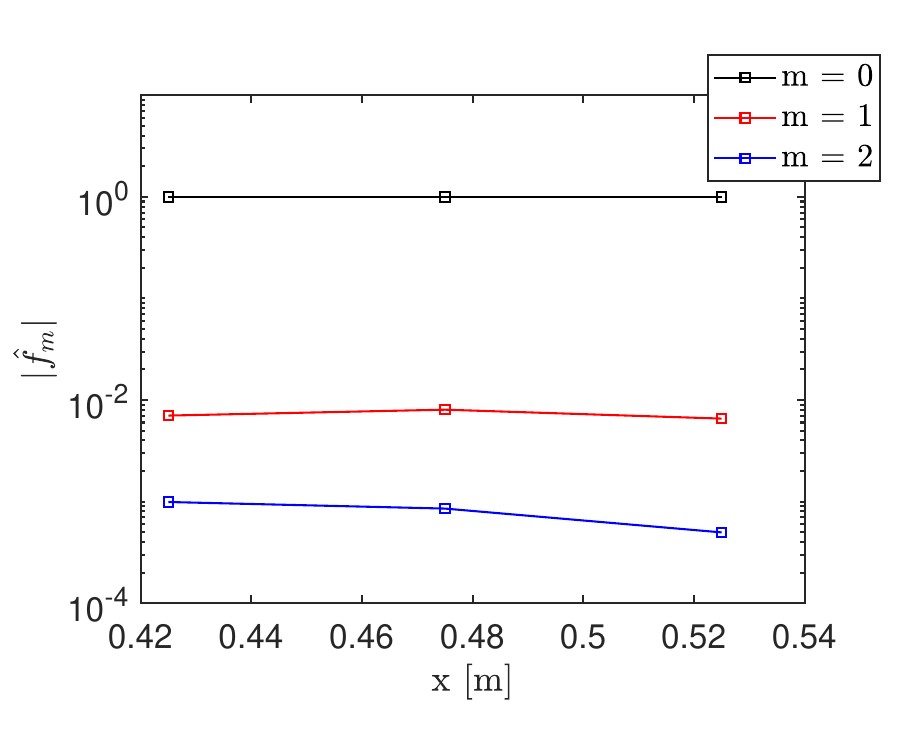}\label{fig:nlte_azimuthal_ampd}}
\caption{Normalized amplitudes of first three azimuthal modes at various axial locations ($x = \SI{0.425}{\meter}, \, \SI{0.475}{\meter}$ and \SI{0.525}{\meter}  \emph{i.e.} \SI{50}{\milli\meter}, \SI{100}{\milli\meter} and \SI{150}{\milli\meter}, respectively, from the nozzle exit) obtained from the azimuthal FFT along a constant radius of \SI{25}{mm}. ($p_{\mathrm{a}} = \SI{590}{\pascal}$, $P = \SI{300}{\kilo\watt}$, $\eta = 53.9\%$).} 
\label{fig:nlte_azimuthal_amp}
\end{figure}

\subsubsection{\label{sec:torch_chamber_nlte} Low-pressure case}
Next, a simulation is conducted for operating conditions of \SI{590}{Pa}, \SI{300}{kW} and $\eta = 53.9\%$. At such low pressures, ICPs are strongly affected by non-equilibrium \cite{kumar2024investigation,zhang2016analysis}. Calculations are thus performed using the 2T NLTE model described in \cref{sec:physical_modeling}, with chemical kinetics parameters taken from Park \emph{et al.} \cite{park2001chemical}. Under conditions of very low pressure and high power, the plasma jet within the facility exhibits supersonic behavior, even when operating with a straight nozzle \cite{capponi2024multi}. This phenomenon arises because the plasma in the torch follows Rayleigh flow dynamics, where heat addition propels a subsonic flow toward sonic velocities. As inductive power is increased, the enhanced heat transfer drives the plasma toward sonic speeds, ultimately resulting in flow choking at the exit of the torch-nozzle system. Moreover, experiments indicate that the supersonic plasma jet is predominantly steady \cite{capponi2024multi}. Hence, a steady-state NLTE simulation is conducted. The governing equations are integrated in time (using a CFL-based local time-stepping) until the residuals of momentum, total energy, and vibronic energy densities decay by four orders of magnitude. 

\begin{figure}[!htb]
\hspace{-1cm}
\subfloat[][]{\includegraphics[trim={3cm 0 6cm 0},clip,scale=0.35]{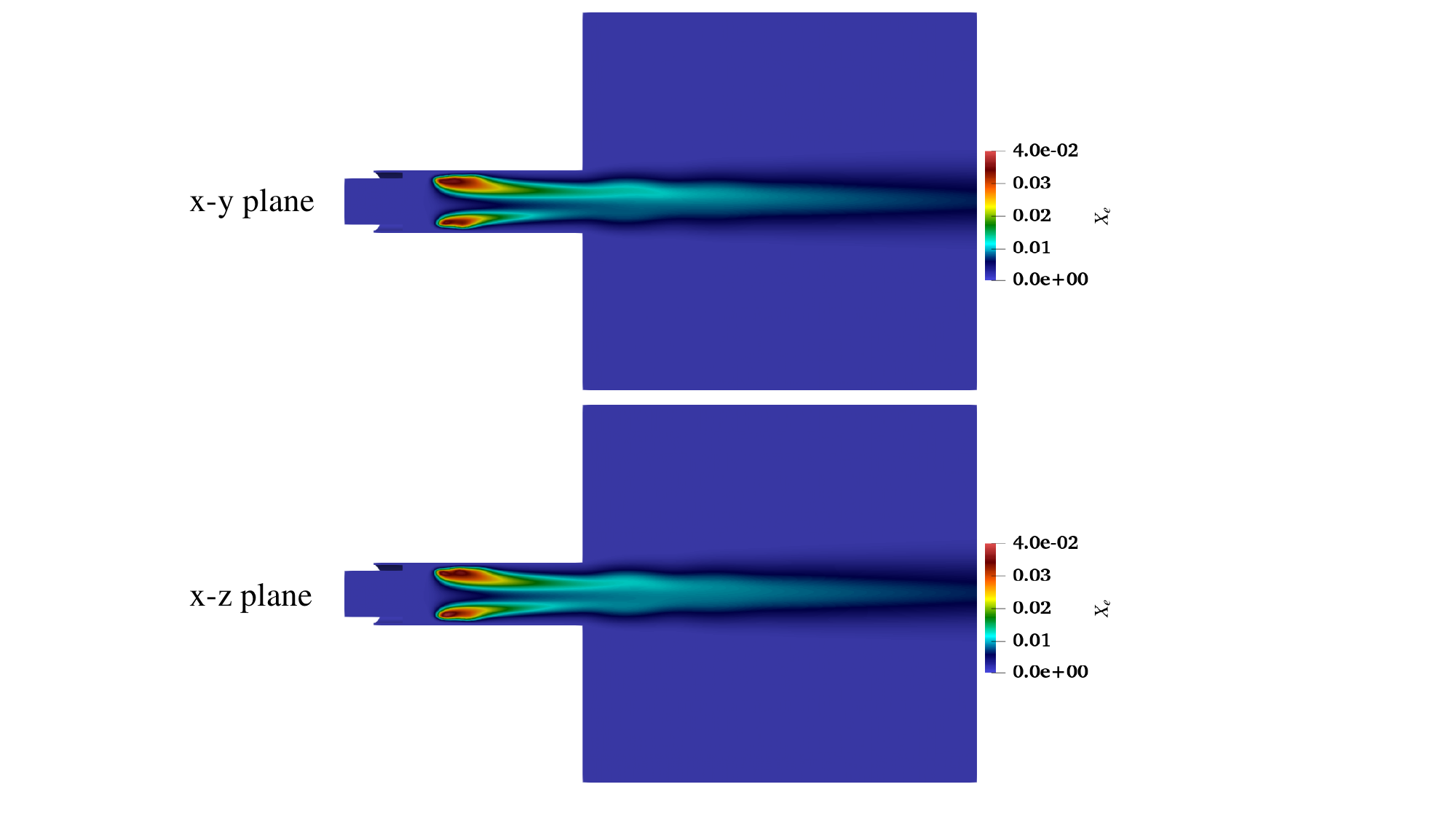}\label{fig:nlte_X_contoursa}}
\subfloat[][]{\includegraphics[trim={3cm 0 6cm 0},clip,scale=0.35]{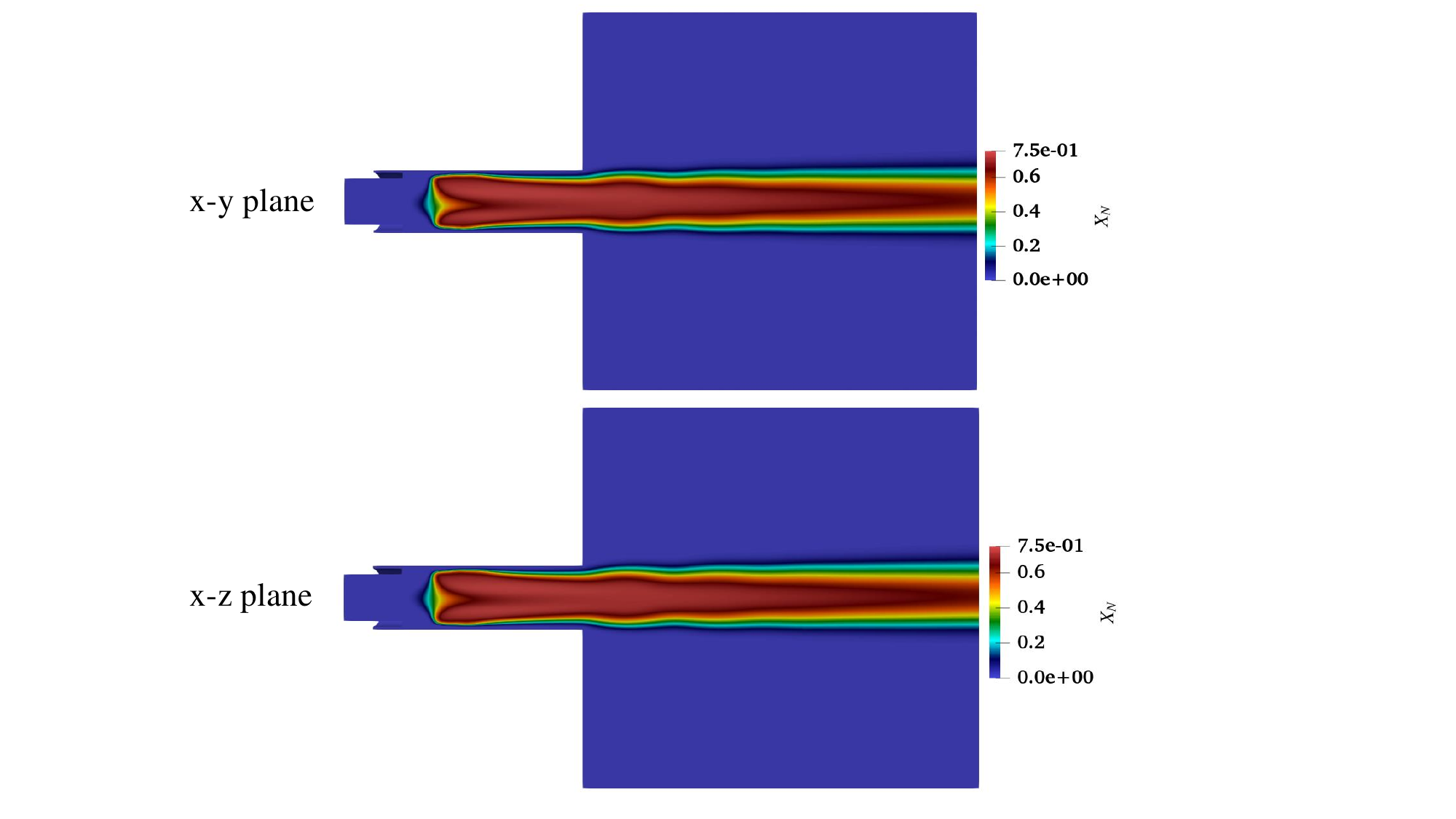}\label{fig:nlte_X_contoursb}} \\

\hspace{-1cm}
\subfloat[][]{\includegraphics[trim={3cm 0 6cm 0},clip,scale=0.35]{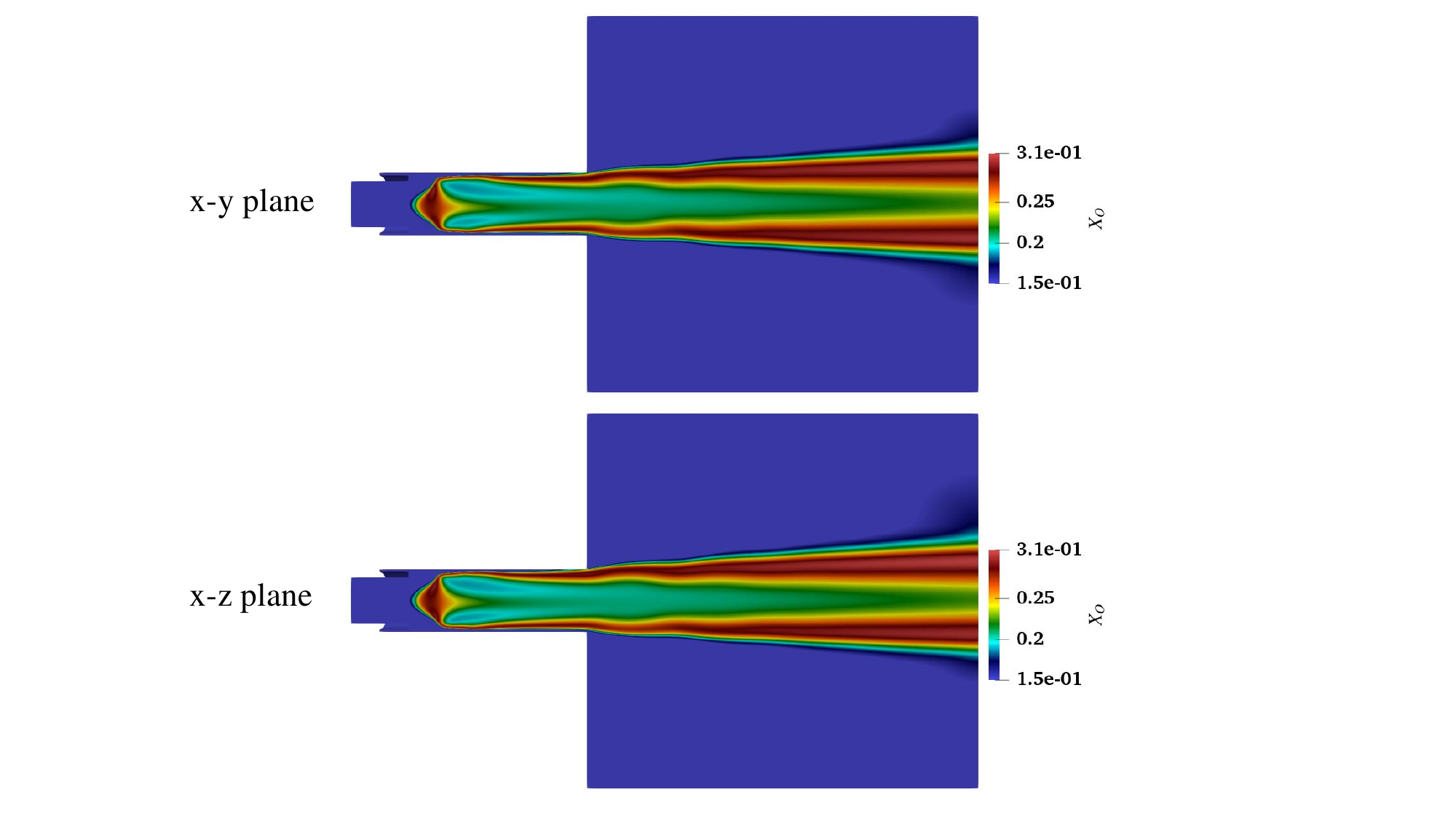}\label{fig:nlte_X_contoursc}}
\subfloat[][]{\includegraphics[trim={3cm 0 6cm 0},clip,scale=0.35]{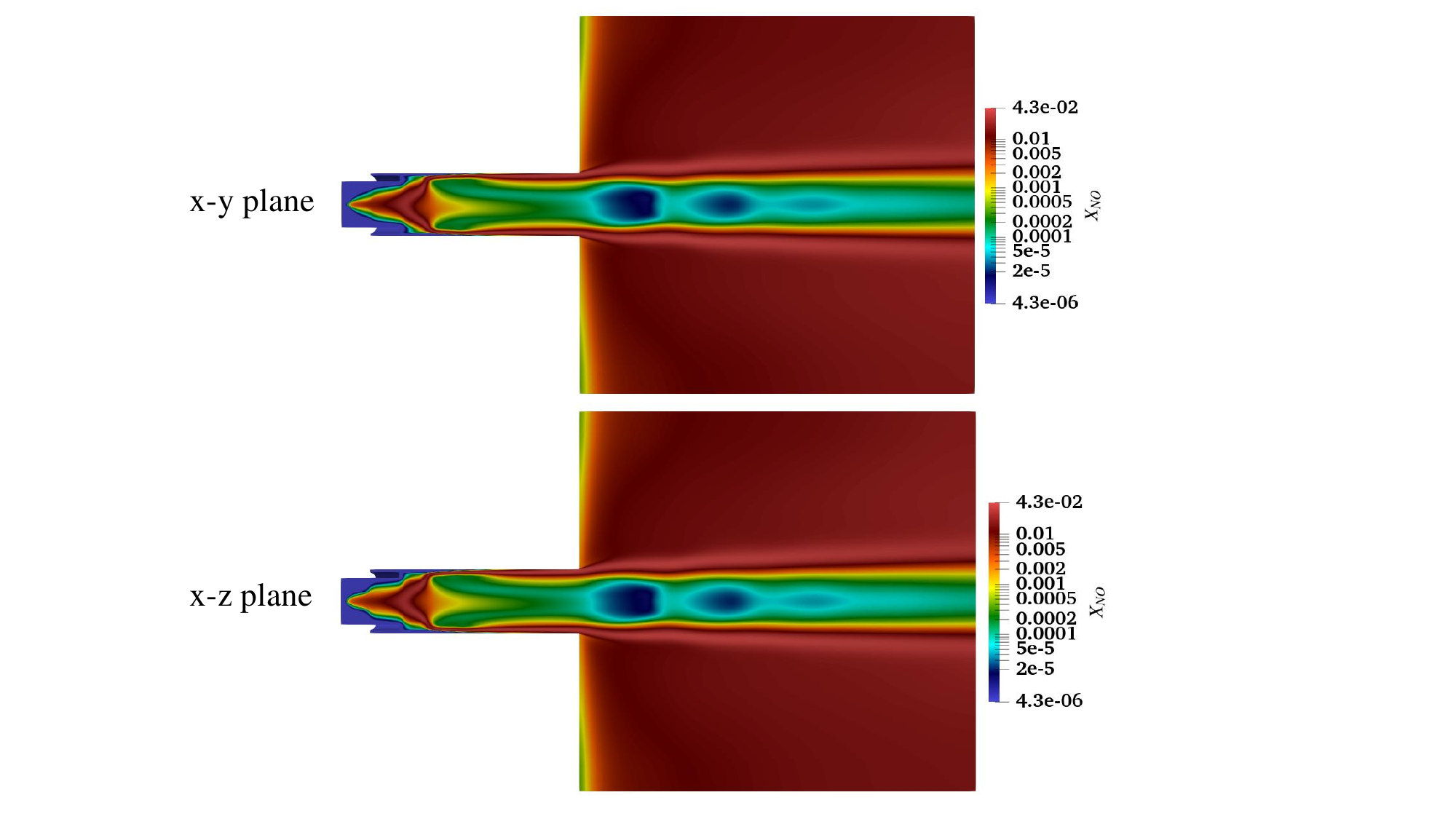}\label{fig:nlte_X_contoursd}}
\caption{Distributions of mole fractions across the $x$-$y$ and $x$-$z$ planes. \protect\subref{fig:nlte_X_contoursa}  Free-electrons, \protect\subref{fig:nlte_X_contoursb} N, \protect\subref{fig:nlte_X_contoursc} O, and \protect\subref{fig:nlte_X_contoursd} NO. ($p_{\mathrm{a}} = \SI{590}{\pascal}$, $P = \SI{300}{\kilo\watt}$, $\eta = 53.9\%$).} 
\label{fig:nlte_X_contours}
\end{figure}

\begin{figure}[hbt!]
\centering
\subfloat[][]{\includegraphics[scale=0.45]{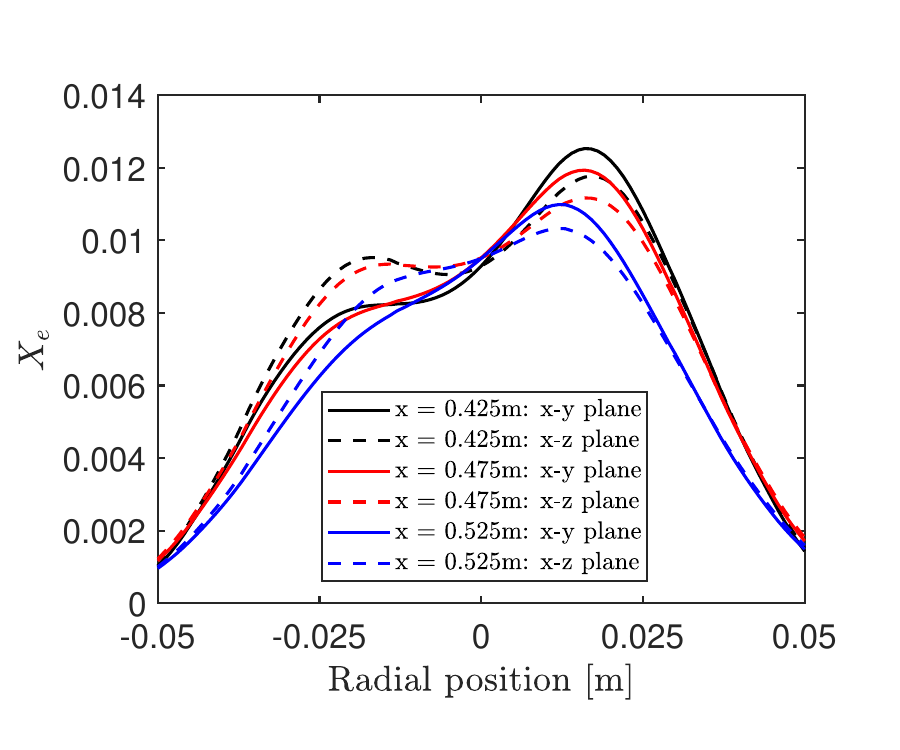}\label{fig:nlte_X_profilesa}} 
\subfloat[][]{\includegraphics[scale=0.45]{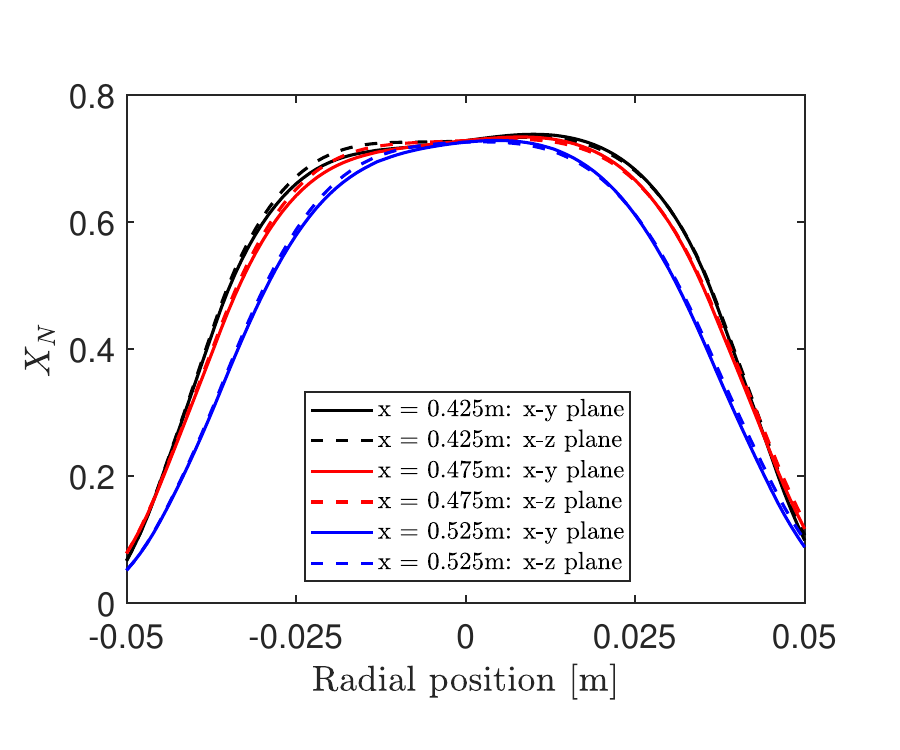}\label{fig:nlte_X_profilesb}}
\\
\subfloat[][]{\includegraphics[scale=0.45]{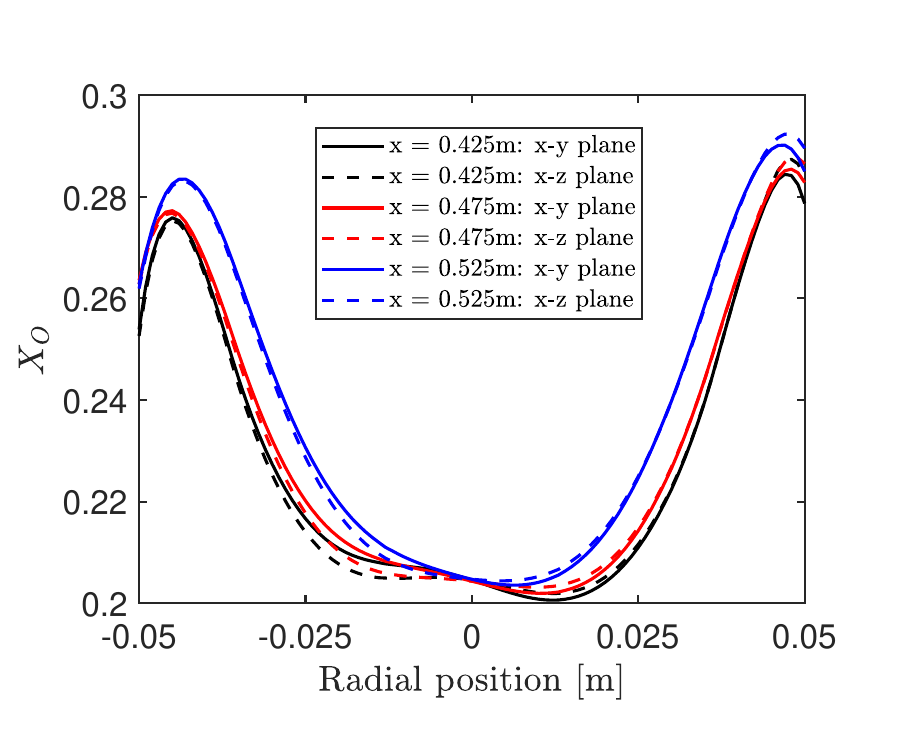}\label{fig:nlte_X_profilesc}} 
\subfloat[][]{\includegraphics[scale=0.45]{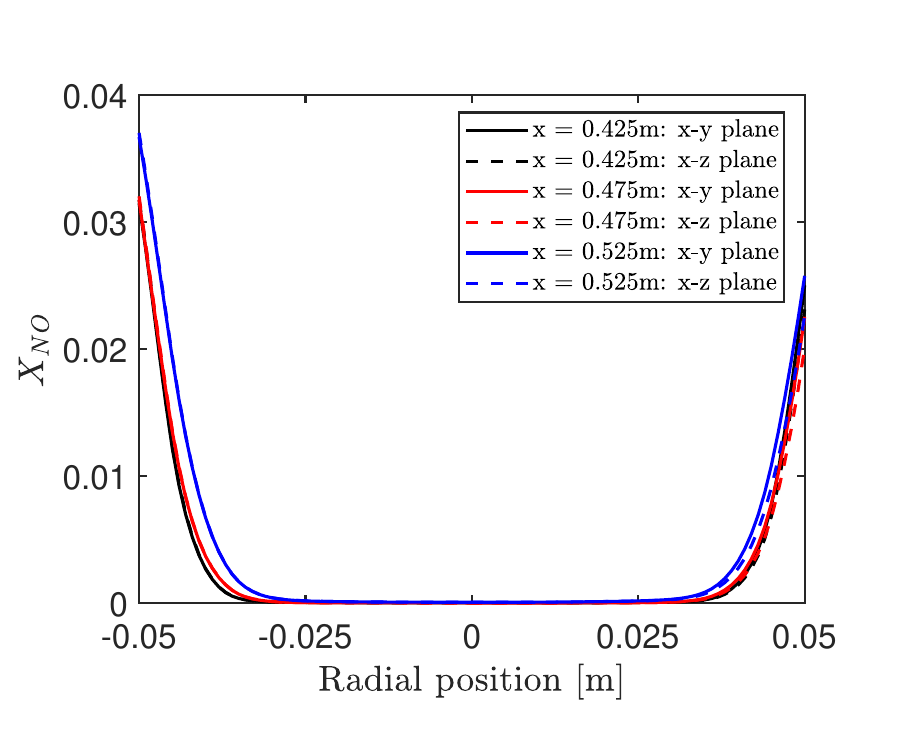}\label{fig:nlte_X_profilesd}}

\caption{Profiles of mole fractions across the $x$-$y$ and $x$-$z$ planes at various locations ($x = \SI{0.425}{\meter}, \, \SI{0.475}{\meter}$ and \SI{0.525}{\meter}  - \emph{i.e.} \SI{50}{\milli\meter}, \SI{100}{\milli\meter}, and \SI{150}{\milli\meter}, respectively, from the nozzle exit). \protect\subref{fig:nlte_X_profilesa} Free-electrons, \protect\subref{fig:nlte_X_profilesb} N, \protect\subref{fig:nlte_X_profilesc} O, and \protect\subref{fig:nlte_X_profilesd} NO. ($p_{\mathrm{a}} = \SI{590}{\pascal}$, $P = \SI{300}{\kilo\watt}$, $\eta = 53.9\%$). } 
\label{fig:nlte_X_profiles}
\end{figure}

\begin{figure}[hbt!]
\centering
\subfloat[][$X_e$]{\includegraphics[scale=0.45]{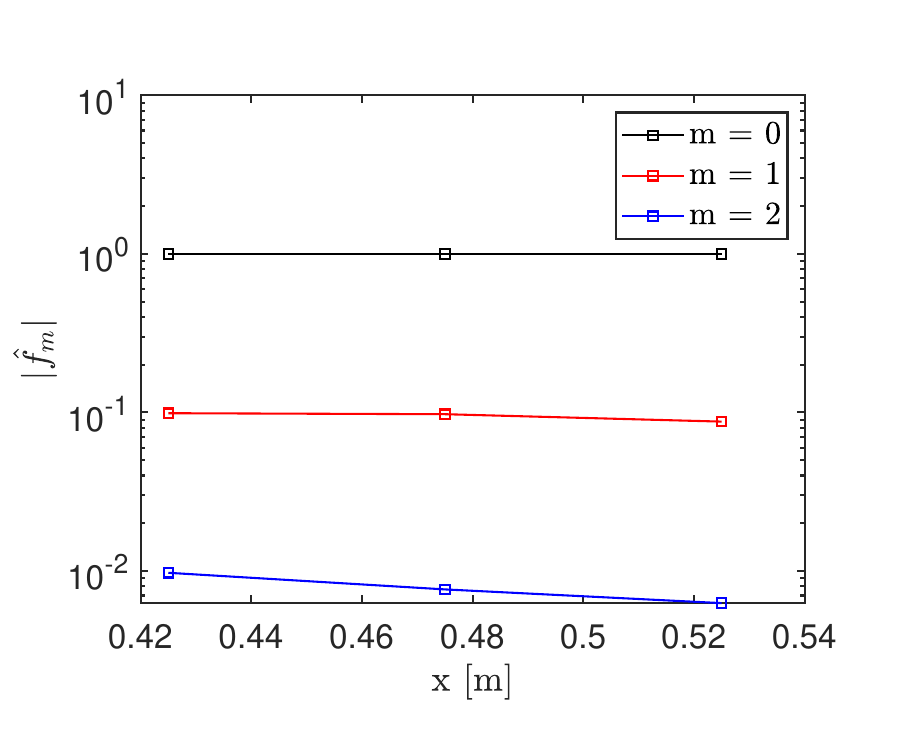}\label{fig:nlte_azimuthal_amp_conca}} 
\subfloat[][$X_N$]{\includegraphics[scale=0.45]{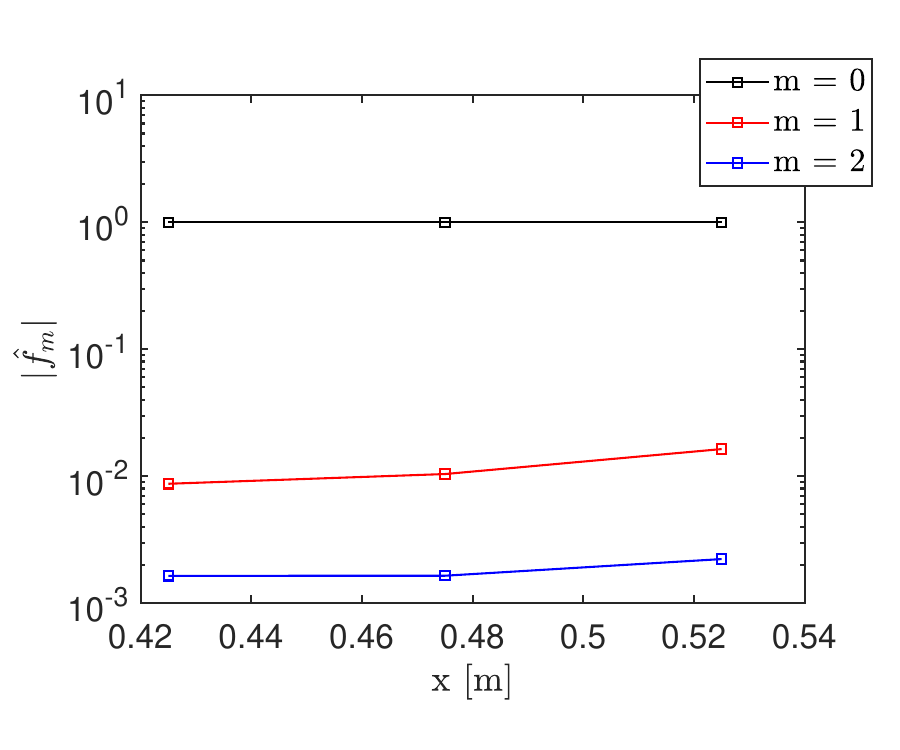}\label{fig:nlte_azimuthal_amp_concb}}
\\
\subfloat[][$X_O$]{\includegraphics[scale=0.45]{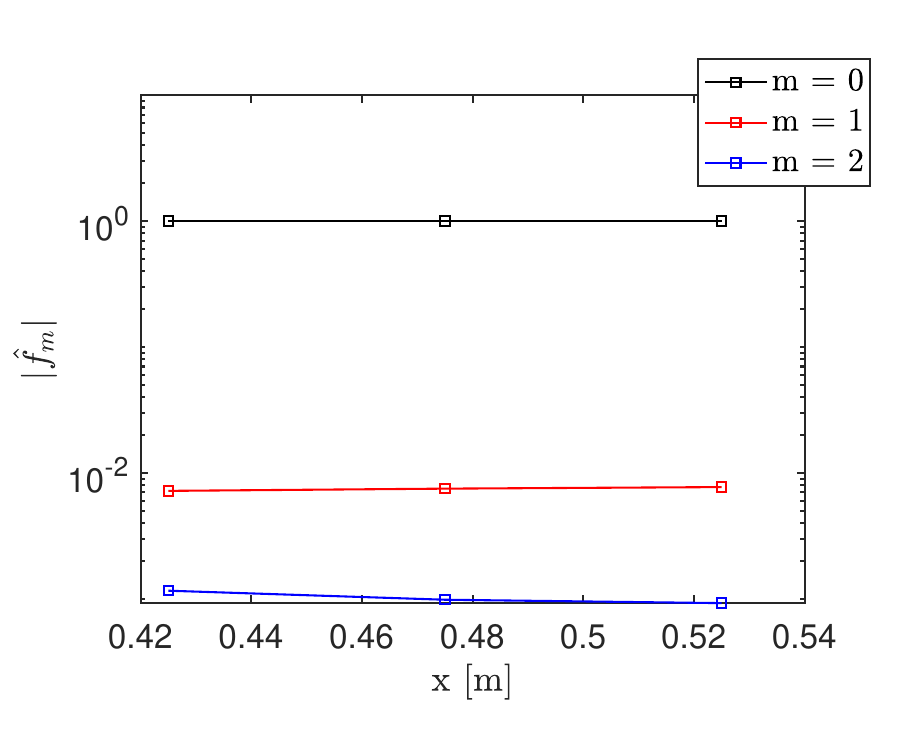}\label{fig:nlte_azimuthal_amp_concc}} 
\subfloat[][$X_{NO}$]{\includegraphics[scale=0.45]{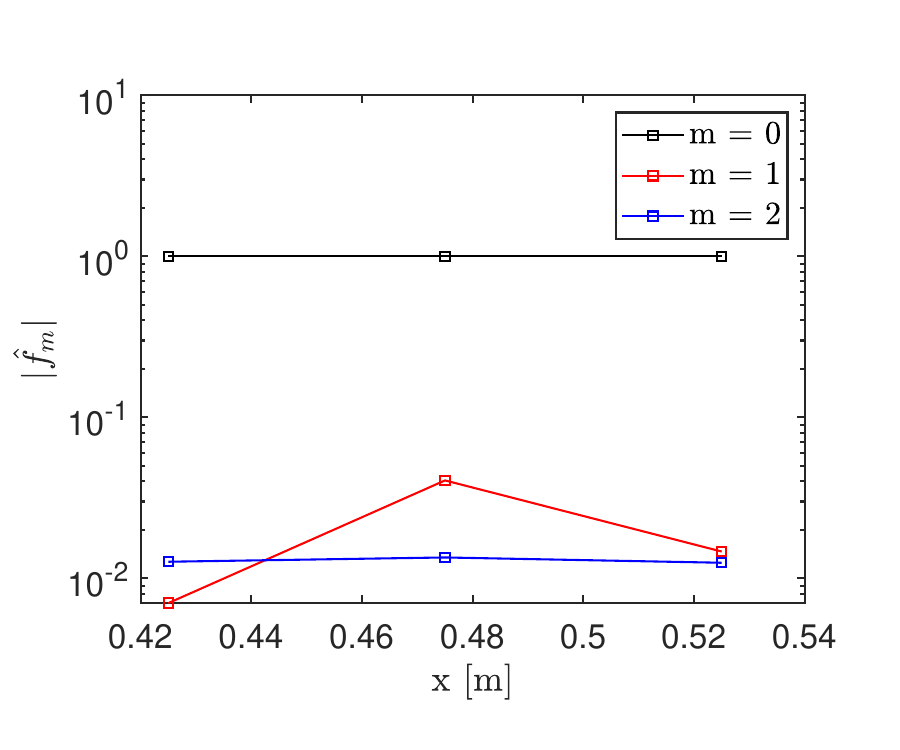}\label{fig:nlte_azimuthal_amp_concd}}
\caption{Normalized amplitudes of first three azimuthal modes at various axial locations ($x = \SI{0.425}{\meter}, \, \SI{0.475}{\meter}$ and \SI{0.525}{\meter}  \emph{i.e.} \SI{50}{\milli\meter}, \SI{100}{\milli\meter} and \SI{150}{\milli\meter}, respectively, from the nozzle exit) obtained from the azimuthal FFT along a constant radius of \SI{25}{mm}. ($p_{\mathrm{a}} = \SI{590}{\pascal}$, $P = \SI{300}{\kilo\watt}$, $\eta = 53.9\%$).} 
\label{fig:nlte_azimuthal_amp_conc}
\end{figure}

\begin{enumerate}[left=0pt,label=\Alph*.]
    \item \textbf{Analysis of the flow quantities distribution:}
     \cref{fig:nlte_contours} shows the distribution of the heavy-particle and vibronic temperatures, frozen Mach number, and axial velocity across the $x$-$y$ and $x$-$z$ planes. The simulation, in alignment with experimental observations, predicts a supersonic plasma jet characterized by distinct shock diamond structures with a peak Mach number of \num{2.2}. Under low-pressure conditions, inertial forces begin to dominate over Lorentz forces, diminishing the impact of the three-dimensional distribution of Joule heating and Lorentz forces. Consequently, the distributions of all the flow quantities exhibited are almost axisymmetric, with minor deviations from axial symmetry in the torch region near the coils.

    \cref{fig:nlte_flow_profiles} further shows the radial profiles of the flow quantities across the $x$-$y$ and $x$-$z$ planes at three axial locations: $x = \SI{0.425}{m}, \, \SI{0.475}{m}$ and \SI{0.525}{m}  (\emph{i.e.,} \SI{50}{mm}, \SI{100}{mm}, and \SI{150}{mm}, respectively, from the torch-nozzle exit). It can be seen that the profiles along the two orthogonal planes overlap at all locations. However, for a given plane, the profiles are not exactly symmetric about the axis (\emph{i.e.,} $r = 0$), especially for the heavy-particle temperature and the axial velocity, indicating a small deviation from axial symmetry. \cref{fig:nlte_azimuthal_amp} shows the normalized amplitudes ($|\hat{f}_m|$) of the first three azimuthal modes (\emph{i.e,} $m=0,1,2$) obtained from the azimuthal FFT (refer \cref{sec:azimuthal_fft} in \cref{sec:torch_chamber_lte} for details) along a constant radius of \SI{25}{mm} for the flow quantities shown in \cref{fig:nlte_contours}. For heavy-particle temperature, the relative amplitudes of the modes with $m\ge 1$ remain below $2\%$ at all the axial locations. For all the other flow quantities, the relative amplitudes of the modes with $m\ge 1$ fall much below $1\%$, indicating an almost axisymmetric configuration.
    \\
    
\item \textbf{Analysis of the species concentration distribution:}
At first, the axisymmetry observed in the flow quantities suggests that two-dimensional axisymmetric simulations provide a good approximation at low-pressure conditions. However, the contours and radial profiles of mole fractions of chemical components of interest, such as e, N, O, and NO, as depicted in \cref{fig:nlte_X_contours,fig:nlte_X_profiles}, respectively, reveal a three-dimensional distribution. The asymmetry in concentrations is further substantiated in \cref{fig:nlte_azimuthal_amp_conc} which presents the normalized amplitudes ($|\hat{f}_m|$) of the first three azimuthal modes (\emph{i.e,} $m=0,1,2$) obtained from the azimuthal FFT along a constant radius of \SI{25}{mm} for the concentration of various species in the plasma. Notably, electron concentrations exhibit the highest degree of asymmetry with the relative amplitude of the mode $m=1$ lying close to $10\%$ at all axial locations. This finding results from the fact that the coil’s energy is first transferred to free-electrons, which then heat the plasma via collisions. Consequently, the asymmetrical electric field distribution induces an asymmetry in electron concentration within the coil region, as illustrated in \cref{fig:nlte_X_contours} (a). This asymmetric electron distribution within the plasma torch extends into the chamber, resulting in a non-axisymmetric distribution even in the chamber region. Notably, the relative amplitudes of modes $m\ge 1$ for free-electrons decrease along the axial direction, unlike the high-pressure unsteady case where they increased with the axial distance. In the high-pressure case, the asymmetry induced by the coils was further amplified by the three-dimensional dynamics of the unsteady jet. However, in the low-pressure case, the asymmetry in the free-electron distribution arises solely from the coils, as the flowfield remains steady and nearly axisymmetric. Hence, as the electrons move further away from the coils, they tend to move towards attaining an axisymmetric distribution, leading to a drop in the relative amplitudes of the higher azimuthal modes. For other species, such as N and O atoms, the distribution remains axisymmetric, with the relative amplitudes of higher azimuthal modes staying below $1\%$. Similarly, for NO, the relative amplitudes are approximately $1\%$, except for mode $m=2$, which exhibits a peak of $5\%$ at x = \SI{0.475}{m}.

\end{enumerate}

\section{Conclusions}\label{sec:conclusions}
A three-dimensional model of the 350 kW inductively coupled plasma (Plasmatron X) facility at UIUC has been developed within a multiphysics-coupled computational framework. Numerical results for sub-atmospheric air plasma within the facility have been obtained using both LTE and NLTE models. The three-dimensional analysis of the plasma discharge in the torch demonstrates pronounced deviations from axisymmetry, attributed to the helical design of the inductor coil. The analysis was further expanded to encompass the torch-chamber geometry, aiming to investigate how the three-dimensional plasma distribution in the torch region influences the plasma jet impinging on the TPM sample. Simulations were conducted for two operating conditions: a low-pressure case (\SI{590}{Pa}) characterized by a low magnetic interaction parameter, and a high-pressure case (\SI{10}{kPa}) characterized by a high magnetic interaction parameter. Since significant deviations from LTE conditions start prevailing at pressures below \SI{5}{kPa}, the low-pressure case was simulated using a 2T NLTE model, whereas the high-pressure case was modeled assuming LTE.

The results demonstrate that the subsonic plasma jet at high pressure is highly unsteady and fundamentally three-dimensional in nature. Spectral Proper Orthogonal Decomposition (SPOD) of the temperature field revealed distinct dominant modes, with peak frequencies occurring below \SI{500}{Hz}, consistent in trend with available experimental evidence. Further, the analysis of the mean plasma jet highlighted significant discrepancies in the temperature and velocity profiles along the two orthogonal planes, with a maximum relative deviation in temperature of 29\%, 18\%, and 25\% and a maximum relative deviation in velocity of 17\%, 25\%, and 28\% at axial distances of \SI{50}{mm}, \SI{100}{mm} and \SI{150}{mm}, respectively from the torch exit. This asymmetry in the profiles was quantified by conducting an azimuthal FFT of the mean temperature and velocity profiles of the plasma jet. The relative amplitude of the m = 1 (non-axisymmetric) mode for temperature increased from 2.74\% at \SI{50}{mm} to 8.33\% at \SI{150}{mm} downstream of the torch exit. For axial velocity, the relative amplitude of the m = 1 mode ranged from 7.07\% to 11.29\% over the same distance, indicating a pronounced asymmetry in the jet structure. This pronounced unsteadiness and asymmetry in the jet at elevated pressures can significantly affect the reliability of Thermal Protection Material (TPM) testing, especially in the measurement of critical quantities such as time-averaged heat flux, recession rates, and surface temperature distributions. Crucially, gaining a deeper understanding of these unsteady flow features is essential for the design and refinement of plasma diagnostic tools, many of which currently rely on assumptions of quasi-steady, axisymmetric jet behavior—assumptions that may no longer hold under such flow conditions.

The low-pressure case exhibits a steady supersonic plasma jet with only minor deviation from axisymmetric distribution in the flow quantities such as temperature and velocity. For all the flow quantities, the relative amplitudes of the non-axisymmetric modes ($m\ge 1$) fall much below $2\%$ at all the axial locations, indicating an almost axisymmetric configuration. Despite the flow properties (e.g., temperatures, velocities, \emph{etc.}) closely approximating an axisymmetric distribution, strong non-equilibrium effects induce a three-dimensional distribution in species concentrations, especially the free-electron concentration which exhibit the highest degree of asymmetry with the relative amplitude of the mode m = 1 lying close to 10\% at all axial locations. Given the influence of species concentrations near the TPM sample on its catalytic activity, any deviation from an axisymmetric distribution could impact the uniformity of material responses. 

 Therefore, three-dimensional simulations are essential for accurately characterizing the plasma state within the facility during TPM testing under all operating conditions, ultimately enhancing the precision of predictions regarding the TPM material’s response. The versatile nature of the current framework enables a broad range of studies in ICP facilities, as three-dimensional modeling of ICPs is also critical for various other applications such as powder spheroidization and nanoparticle production. Future research will focus on exploring key effects such as turbulence within the facility, the effect of the secondary vortices created due to discrete hole injectors, \emph{etc.}, and validating results with experimental three-dimensional measurement data. Moreover, the simulations presented in this paper provide critical plasma property data that can serve as valuable reference data, helping guide the design and calibration of future diagnostics in the Plasmatron X facility and providing experimentalists with prior insight into the expected ranges of these quantities.

\section*{Acknowledgments}
         This work is funded by the Vannevar Bush Faculty Fellowship OUSD(RE) Grant No: N00014-21-1-295 with M. Panesi as the Principal Investigator. The work is also supported by the Center for Hypersonics and Entry Systems Studies
(CHESS) at UIUC. Computations were performed on Frontera, an HPC resource provided by the Texas Advanced
Computing Center (TACC) at The University of Texas at Austin, on allocation CTS20006, with D. Bodony as the Principal Investigator.
    
\section*{Competing interest}
The authors declare no competing interests.

\section*{References}
\bibliographystyle{aiaa}
\bibliography{bibliography}

\end{document}